\documentclass[11pt]{article}

    \usepackage[dvipsnames]{xcolor}
    \usepackage[T1]{fontenc}
	\usepackage[utf8]{inputenc}
	\usepackage[english]{babel}
	\usepackage{amsfonts}
	\usepackage{amssymb}
	    \usepackage{amsmath}
	
	\usepackage{enumerate}
	\usepackage{bbm}
	\usepackage{mathtools}
	\usepackage{subfig}
	\usepackage{booktabs}	
	\usepackage{setspace}
	\usepackage{graphicx}
	\usepackage{fullpage}
	
	\usepackage{soul}
	\usepackage{xspace}
	\usepackage{csquotes}
	\usepackage{placeins}
	\usepackage[footfont=normalsize,font=normalsize]{floatrow}
	\usepackage{floatrow}
	\usepackage{lscape}
	\usepackage{longtable}
	\usepackage{mathabx}
    \usepackage{accents}
    \usepackage{float}

    \usepackage{tikz}
    \usetikzlibrary{calc,intersections}

	\usepackage[]{natbib}
	\usepackage{bibunits}
	\defaultbibliography{AugI_Bibliography.bib}  
	\defaultbibliographystyle{aer}

\usepackage{geometry}
\usepackage{caption}
\usepackage{pgfplotstable}

\usepackage{pgfplotstable}
	
	\usepackage{hyperref}
	\hypersetup{
    colorlinks=true,
    linkcolor=Red,
    filecolor=magenta,      
    urlcolor=blue,
    citecolor = Blue
}
	\usepackage{cleveref}

    \usepackage{rotating}
	\usepackage[framemethod=tikz]{mdframed}
	\usepackage{todonotes}
	\presetkeys{todonotes}{color=black!5,inline}{} 

    \usepackage{tcolorbox}
    \newtcbox{\feedback}{nobeforeafter,colframe=black,colback=white,boxrule=0.5pt,arc=2pt,
      boxsep=0pt,left=2pt,right=2pt,top=2pt,bottom=2pt,tcbox raise base}
      
    \usepackage{amsthm}

    \theoremstyle{definition}

\usepackage{array}
\newcolumntype{L}[1]{>{\raggedright\let\newline\\\arraybackslash}m{#1}}
\newcolumntype{C}[1]{>{\centering\let\newline\\\arraybackslash\hspace{0pt}}m{#1}}
\newcolumntype{R}[1]{>{\raggedleft\let\newline\\\arraybackslash\hspace{0pt}}m{#1}}

\usepackage{ragged2e}
\newlength\ubwidth

\def\eea{\end{eqnarray*}}
\def\bea{\begin{eqnarray*}}
\renewcommand{\[}{\begin{equation}}
\renewcommand{\]}{\end{equation}}
\newcommand{\bi}{\begin{itemize}}
\newcommand{\ei}{\end{itemize}}

\graphicspath{{../Empirics/Output/}{./}}

\usepackage{pgfplotstable}

\usepackage{geometry}
\usepackage{caption}
\begin{document}
\thispagestyle{empty}

\vspace{20pt}
\begin{center}
\textcolor{white}{fill}\\
\vspace{20pt}
\LARGE{Generative AI at Work$^{*}$}\\\mbox{ }
\end{center}

\begin{singlespace}
\begin{center}
\begin{tabular}{ccccccc}
\large{Erik Brynjolfsson} & \mbox{ } & \mbox{ } & \large{Danielle Li} & \mbox{ } & \mbox{ } & \large{Lindsey Raymond} \\
\large{Stanford \& NBER} & \mbox{ } & \mbox{ } & \large{MIT \& NBER} & \mbox{ } & \mbox{ } & \large{MIT} \\
\end{tabular}
\end{center}

\end{singlespace}

\begin{center}
\vspace{.20in}
\normalsize{\today}

\vspace{10pt}
First Draft: 23 April 2023\\
\end{center}
\vspace{10pt}

\begin{abstract}
\vspace{-11pt}
\singlespacing
\noindent

We study the staggered introduction of a generative AI-based conversational assistant using data from 5,172 customer support agents.  Access to AI assistance increases worker productivity, as measured by issues resolved per hour, by 15\% on average, with substantial heterogeneity across workers.  Less experienced and lower-skilled workers improve both the speed and quality of their output while the most experienced and highest-skilled workers see small gains in speed and small declines in quality.  We also find evidence that AI assistance facilitates worker learning and improves English fluency, particularly among international agents. While AI systems improve with more training data, we find that the gains from AI adoption are largest for relatively rare problems, where human agents have less baseline training and experience. Finally, we provide evidence that AI assistance improves the experience of work along two key dimensions: customers are more polite and less likely to ask to speak to a manager.

\end{abstract}


\bigskip

\hspace{-25pt}\rule[0.1ex]{0.33\textwidth}{0.2mm}\vspace{0.05in}\\\indent
\begin{singlespace}
\vspace{-25pt}
\noindent \scriptsize{\mbox{  }\mbox{  }\mbox{  }\mbox{  }\mbox{ }$^{*}$Correspondence to erikb@stanford.edu, d\_li@mit.edu, and lraymond@mit.edu. 
We are grateful to Daron Acemoglu, David Autor, Amittai Axelrod, Eleanor Dillon, Zayd Enam, Luis Garicano, Alex Frankel, Sam Manning, Sendhil Mullainathan, Emma Pierson, Scott Stern, Ashesh Rambachan, John Van Reenen, Raffaella Sadun, Kathryn Shaw, Christopher Stanton, Sebastian Thrun, and various seminar participants for helpful comments and suggestions. We thank Max Feng for providing excellent research assistance and the Stanford Digital Economy Lab for funding. The content is solely the responsibility of the authors and does not necessarily represent the official views of Stanford University, MIT, or the NBER.}
\end{singlespace}

\thispagestyle{empty}

 \setcounter{page}{0}

\clearpage
\newpage

The emergence of generative artificial intelligence (AI) has attracted significant attention, but few studies have examined its economic impact.  Although various generative AI tools have performed well in laboratory settings, excitement about their potential has been tempered by concerns that these tools may be less effective in real-world settings, where they may encounter unfamiliar problems, face organizational resistance, or provide misleading information in a consequential environment \citep{peng2023check, rooseConversationBingChatbot2023}.

In this paper, we study the adoption of a generative AI tool that provides conversational guidance to customer support agents.\footnote{A note on terminology. There are many definitions of artificial intelligence and of intelligence itself---\citet{legg2007collection} list over 70 of them.  In this paper, we define ``artificial intelligence'' (AI) as an umbrella term that refers to systems that exhibit intelligent behavior, such as learning reasoning and problem-solving.  ``Machine learning'' (ML) is a branch of AI that uses algorithms to learn from data, identify patterns, and make predictions or decisions without being explicitly programmed \citep{aivsml}. Large language models (LLMs) and tools built around LLMs such as ChatGPT are an increasingly important application of machine learning. LLMs generate new content, making them a form of ``generative AI.'' ``Generative AI'' is a type of artificial intelligence that can create new content, such as text, images, or music, by learning patterns from existing data.}  This is, to our knowledge, the first study of the impact of generative AI deployed at scale in the workplace.  We find that access to AI assistance increases the productivity of agents by 15\%, as measured by the number of customer issues they are able to resolve per hour. We find that these gains accrue disproportionately to less-experienced and lower-skill customer support workers.  This finding suggests that generative AI systems may be capable of capturing and disseminating the behaviors of the most productive agents. 

Computers and software have transformed the economy with their ability to perform certain tasks with far more precision, speed, and consistency than humans.  To be effective, these systems typically require explicit and detailed instructions for how to transform inputs into outputs: a software engineer must literally program the computer.  Yet, despite significant advancements in traditional computing, many workplace activities---such as writing emails, analyzing data, or creating presentations---are difficult to ``codify''---and have therefore defied automation.

Machine learning (ML) algorithms work differently from traditional computer programs: instead of requiring explicit instructions to function, these systems infer instructions from examples. Given a training set of images, for instance, ML systems can learn to recognize specific individuals even though one cannot fully explain what physical features characterize a given person's identity. This ability highlights a key distinguishing aspect of ML systems: they can learn to perform tasks even when no instructions exist---including tasks requiring tacit knowledge that could previously only be gained through lived experience \citep{polanyi_tacit_1966, autorPolanyiParadoxShape2014, brynjolfssonmitchell2017}.\footnote{As \citet{meijer2018behind} puts it ``where the Software 1.0 Engineer formally specifies their problem, carefully designs algorithms, composes systems out of subsystems or decomposes complex systems into smaller components, the Software 2.0 Engineer amasses training data and simply feeds it into an ML algorithm...''}

In addition, ML systems are often trained on data from human workers, who naturally vary in their abilities.  By seeing many examples of tasks---making sales pitches, driving a truck, or diagnosing a patient, to name a few---performed well and poorly, these models can implicitly learn what specific behaviors and characteristics set high-performing workers apart from their less effective counterparts.  That is, not only are generative AI models capable of performing complex tasks, they might also be capable of capturing the skills that distinguish top workers.  The use of ML tools may therefore differentially expose lower-skill workers to new skills and techniques, leading to disparate changes in productivity even among workers performing the same task. 

We study the impact of generative AI on productivity and worker experience in the customer service sector, an industry with one of the highest rates of AI adoption \citep{chui_global_2021}. We examine the staggered deployment of a chat assistant using data from 5,000 agents working for a Fortune 500 software firm that provides business process software. The tool we study is built on a recent version of the Generative Pre-trained Transformer (GPT) family of large language models developed by OpenAI \citep{GPT4TechnicalReport2023}.  It monitors customer chats and provides agents with real-time suggestions for how to respond. It is designed to augment agents, who remain responsible for the conversation and are free to ignore or edit the AI's suggestions.

We have three sets of findings.  

First, AI assistance increases worker productivity, resulting in a 15\% increase in the number of chats that an agent successfully resolves per hour.  This increase reflects shifts in three components of productivity: a decline in the time it takes an agent to handle an individual chat, an increase in the number of chats that an agent handles per hour (agents may handle multiple chats at once), and a small increase in the share of chats that are successfully resolved.  

Within customer support workers, the productivity impacts of AI assistance are highly uneven.  We find that less-skilled and less-experienced workers improve significantly across all productivity measures we consider, including approximate increase of 30\% in the number of issues they are able to resolve per hour.  Access to the AI tool helps newer agents move more quickly down the experience curve: treated agents with two months of tenure perform just as well as untreated agents with more than six months of tenure.  In contrast, we find minimal impacts on the productivity of more-experienced or more-skilled workers.  Indeed, we find evidence that AI assistance leads to a small \textit{decrease} in the quality of conversations generated by the most skilled agents.  These results contrast, in spirit, with studies that find evidence of skill-biased technical change for earlier waves of computer technology and robotics \citep{bres2002, bartel2007, dixon2020}.  

Our second set of results investigates the mechanism underlying our main findings. We show that AI recommendations appear useful to workers: agents who follow recommendations more closely see larger gains in productivity, and adherence rates increase over time, particularly those who were initially more skeptical.  We also find that engagement with AI recommendations can generate durable learning.  Using data on software outages---periods in which the AI software fails to provide any suggestions---we show that workers see productivity gains relative to their pre-AI baseline even when recommendations are unavailable.  These outage-period gains are more pronounced for workers with more prior exposure to AI assistance, and especially those who had been following AI recommendations more closely.  In addition, we examine the heterogeneous impact of AI access by the technical support conversation topics that agents encounter.  While AI systems improve with access to more training data, we find that the gains to AI adoption---when used to complement human workers---are largest for relatively rare problems, perhaps because agents are already capable of addressing the problems they encounter most frequently.  We further analyze the text of agents' chats and provide evidence that access to AI improves their English language fluency, especially among international agents.  Finally, we compare the text of conversations before and after AI and provide suggestive evidence that AI adoption drives convergence in communication patterns: low-skill agents begin communicating more like high-skill agents. 

Our third set of results focus on agents' experience of work.  Work in contact centers\footnote{The term ``contact center'' updates the term ``call center,'' to reflect the fact that a growing proportion of customer service contacts no longer involve phone calls.} is often difficult.  Agents are regularly exposed to hostile treatment from upset (and anonymous) customers, and because much work is outsourced, many agents work overnight shifts in order to service US business hours.  AI assistance may help agents communicate more effectively but could also increase the likelihood that agents are perceived as mechanical or inauthentic. We show that access to AI assistance markedly improves how customers treat agents, as measured by the sentiment of their chat messages.  We also find that customers are less likely to question the competence of agents by requesting to speak to a supervisor.  These changes come alongside a decrease in worker attrition, which is driven by the retention of newer workers.  

Our overall findings show that access to generative AI can increase the productivity of individual workers, and improve their experience of work.  We emphasize, however, that these findings capture medium run impacts within a single firm: our paper is not designed to shed light on the aggregate employment or wage effects of generative AI tools. In the longer run, firms may respond to increasing productivity among novice workers by hiring more of them, or by seeking to develop more powerful AI systems that replace their use of labor altogether. This latter scenario highlights the possibility that while the introduction of generative AI may increase demand for lower-skill labor \textit{within} an occupation, the equilibrium response to AI assistance may lead to \textit{across} occupation shifts in labor demand that instead benefit higher skill workers \citep{autorSkillContentRecent, acemogluLowSkillHighSkillAutomation2018, acemogluSimpleMacroeconomicsAI}.\footnote{For example, transcripts from a fully automated customer service division could be sent to human analysts who mine this data for information on how to improve a product.}  Unfortunately, our data do not allow us to observe changes in wages, overall labor demand, or the skill composition of workers hired at the firm.

Our results also highlight the longer-term incentive challenges raised by the adoption of AI systems.  In our data, top workers increase their adherence to AI recommendations, even though those recommendations marginally decrease the quality of their conversations.  Yet, with fewer original contributions from the most skilled workers, future iterations of the AI model may be less effective in solving new problems.  Our work therefore raises questions about whether and how workers should be compensated for the data they provide for training AI systems.

Our paper is related to a large literature on the impact of technological adoption on worker productivity and the organization of work \citep[e.g.][]{rosen1981, autorComputingInequalityHave1998, atheystern2002, bres2002, bartel2007, acemoglu2007, hoffman2018, bloom2014, michaelsHasIctPolarized2014, garicano2015, acemogluRobotsJobsEvidence2020, feltenOccupationalHeterogeneityExposure2023}.  Many of these studies, particularly those focused on information technologies, find evidence that IT complements higher-skill or more-educated workers \citep{akermanSkillComplementarityBroadband2015,taniguchiICTCapitalSkillComplementarity2022}. For instance, \citet{bartel2007} finds that firms that adopt IT tend to use more skilled labor and is associated with increased skill requirements for machine operators in valve manufacturing. Other research compares workers with different degrees of educational attainment between occupations; \citet{acemogluRobotsJobsEvidence2020} study the diffusion of robots and finds that the negative effects of robots on employment are most pronounced for workers in blue-collar occupations and those with fewer than a college education. 

There have been substantially fewer studies involving AI-based technologies, generative or not. \citet{acemoglu2022, zolas2020, calvino2023} examine economy-wide data from the US and OECD and show that the adoption of AI tools is concentrated among larger and younger firms with relatively high productivity.  So far, evidence on the productivity impacts of these technologies is mixed: for example, \citet{acemoglu2022} finds no detectable relationship between investments in AI-specific tools, while \citet{babina2022} finds evidence of a positive relationship between firms' AI investments and their subsequent growth and valuations.\footnote{\citet{oecd2023} reports that when surveyed firms are asked directly, 57\% of employers in finance and 63\% in manufacturing reported that AI positively impacted productivity and 80\% of surveyed workers who work with AI report higher job performance.}  These studies all caution that the productivity effects of AI technologies may be challenging to identify at the macro-level because AI-adopting firms differ substantially from non-adopters. 

Even at the decision level, the effects of AI on decision quality are mixed, often revealing unexpected challenges in human-AI collaboration. Some studies on AI decision support tools report positive impacts. For instance, \citet{kanazawa2022} examine a non-generative AI tool for taxi drivers that suggests routes with the highest likelihood of finding customers. This tool reduces driver search time by 5\%, with low-skill drivers seeing the largest reductions in search. On the other hand, several studies find that the combination of AI and human decisions performs worse than either AI or humans alone \citep{hoffman2018, angelova2023, agrawal2023, poursabzisangdeh2021}. In fact, a meta-analysis of more than 100 experimental studies concludes that, on average, human-AI collaborations underperform both the AI alone and the best human decision-makers 
 \citep{vaccaro2024combinationshumansaiuseful}. These results underscore the particular challenges introduced when using AI-based tools designed to augment human decision making. 

In this paper, we provide micro-level evidence on the adoption of a generative AI tool across thousands of workers employed by a given firm and its subcontractors.  Our work is closely related to several other studies examining the impacts of generative AI in lab-like settings.  \citet{peng2023impact} recruit software engineers for a specific coding task (writing an HTTP server in JavaScript) and show that those given access to GitHub Copilot complete this task twice as quickly. Similarly, \citet{noyExperimentalEvidenceProductivity2023} conduct an online experiment showing that subjects given access to ChatGPT complete professional writing tasks more quickly. In the legal domain, \citet{choiAIAssistanceLegal2023} provided law students with AI assistance on a law school exam, while in management consulting, \citet{dellacquaNavigatingJaggedTechnological2023} find that access to GPT-4 suggestions improves the quality of responses on management consulting tasks within its capabilities, but can negatively impact performance on tasks outside its capabilities. Consistent with our findings, \citet{noyExperimentalEvidenceProductivity2023}, 
\citet{choiAIAssistanceLegal2023}, \citet{peng2023check} and \citet{dellacquaNavigatingJaggedTechnological2023} find that generative AI assistance compresses the productivity distribution, with lower-skill workers driving the bulk of improvements.  Our paper, however, is the first to examine longer-term effects in a real-world workplace where we can also track patterns of learning, customer-side effects, and changes in the experience of work.

\section{Generative AI and Large Language Models}

In recent years, the rapid pace of AI development and public release tools such as ChatGPT, GitHub Copilot, and DALL-E have attracted widespread attention, optimism, and alarm \citep{whitehouse2022}.  These technologies are all examples of ``generative AI,'' a class of machine learning technologies that can generate new content---such as text, images, music, or video---by analyzing patterns in existing data.  In this section, we provide background on generative AI as a technology and discuss its potential economic implications.

\subsection{Technical Primer}

This paper focuses on an important class of generative AI, large language models (LLMs). LLMs are neural network models designed to process sequential data \citep{bubeck2023sparks}.  An LLM is trained by learning to predict the next word in a sequence, given what has come before, using a large corpus of text (such as Wikipedia, digitized books, or portions of the Internet).  This knowledge of the statistical co-occurrence of words allows it to generate new text that is grammatically correct and semantically meaningful.  Although ``large language model'' implies human language, the same techniques can be used to produce other forms of sequential data (``text'') such as protein sequences, audio, computer code or chess moves \citep{el2023}. 

Recent progress in generative AI has been driven by four factors: computing scale, earlier innovations in model architecture, the ability to ``pre-train'' using large amounts of unlabeled data and refinements in training techniques.\footnote{For a more detailed technical review of progress, see \citet{radfordImprovingLanguageUnderstanding2018, radfordLanguageModelsAre2019, liuSummaryChatGPTGPT42023, ouyangTrainingLanguageModels2022}.}  

First, the quality of LLMs is strongly dependent on scale: the amount of computing power used for training, the number of model parameters, and dataset size \citep{kaplan2020scaling}. Firms are increasingly devoting more resources to increasing this scale.  The GPT-3 model included 175 billion parameters, was trained on 300 billion tokens, and generated approximately \$5 million dollars in computing costs alone; the GPT-4 model, meanwhile, is estimated to include 1.8 trillion parameters, trained on 13 trillion tokens, at a rumored computing-only cost of \$65 million \citep{liOpenAIGPT3Language2020, brownLanguageModelsAre2020, gpt4scale}

In terms of model architecture, modern LLMs use two earlier key innovations: positional encoding and self-attention. Positional encodings keep track of the order in which a word occurs in a given input.\footnote{For instance, a model would keep track of ``the, 1'' instead of only ``the'' (if ``the'' was the first word in the sentence).}  Meanwhile, self-attention assigns importance weights to each word in the context of the entire input text. Together, this approach enables models to capture long-range semantic relationships within an input text, even when that text is broken up into smaller segments and processed in parallel \citep{vaswaniAttentionAllYou2017, bahdanau2016neural}.

Next, LLMs can be pre-trained on large amounts of unlabeled data from sources such as Reddit or Wikipedia. Because unlabeled data are far more prevalent than labeled data, LLMs can learn about natural language on a much larger training corpus \citep{brownLanguageModelsAre2020}.  By seeing, for example, that the word ``yellow'' is more likely to be observed with ``banana'' or ``sun'' or ``rubber duckie,'' the model can learn about semantic and grammatical relationships even without explicit guidance \citep{radfordImprovingLanguageUnderstanding2018}.  The resulting model can be used in multiple applications because its training is not specific to a particular set of tasks.

Finally, general-purpose LLMs can be further ``fine-tuned'' to generate output that matches the priorities of any specific setting \citep{ouyangTrainingLanguageModels2022, liuSummaryChatGPTGPT42023}. For example, a model trained to generate social media content would benefit from receiving labeled data that contain not just the content of a post or tweet, but also information on the amount of user engagement it received.  Similarly, an LLM may generate several potential responses to a given query, but some of them may be factually incorrect or contain toxic language.  To discipline this model, human evaluators can rank these outputs to train a reward function that prioritizes desirable responses. These types of refinements can significantly improve model quality by making a general-purpose model better suited to its specific application \citep{ouyangTrainingLanguageModels2022}. 

Together, these innovations have generated meaningful improvements in model performance. The Generative Pre-trained Transformer (GPT) family of models, in particular, has attracted considerable media attention for their rapidly expanding capabilities.\footnote{For instance, GPT-4 has recently been shown to outperform humans in taking the US legal bar exam \citep{liuSummaryChatGPTGPT42023, bubeck2023sparks, GPT4TechnicalReport2023}.}

\subsection{The Economic Impacts of Generative AI}\label{sec:aibackground}

Computers have historically excelled at executing pre-programmed instructions, making them particularly effective at tasks that can be reduced to explicit rules \citep{autorPolanyiParadoxShape2014}. Consequently, computerization has disproportionately reduced the demand for workers performing ``routine'' tasks such as data entry, bookkeeping, and assembly line work, reducing wages in these jobs \citep{acemoglu2011skills}. At the same time, computerization has also increased the demand for workers who possess complementary skills such as programming, data analysis, and research.  Along with changes in supply of skilled workers, these changes have contributed to the increase in wage inequality in the United States and have been linked to a variety of organizational changes \citep{katzChangesRelativeWages1992c, autorSkillContentRecent, michaelsHasIctPolarized2014, bres2002, bakerMakeBuyTrucking2003, oecd2023}.

In contrast, generative AI tools do not require explicit instructions to perform tasks.  If asked to write an email denying an employee a raise, generative AI tools will likely respond with a professional and conciliatory note.  This occurs because the model will have seen many examples of workplace communication in which requests are declined in this manner.  Importantly, the model produces such an output even though no programmer has explicitly specified what tone would be appropriate for what context, nor even defined what a tone like ``professional'' or ``conciliatory'' means.  Indeed, the ability to behave ``appropriately'' is one that cannot be fully articulated even by those who possess it.  This type of ``tacit knowledge'' underlies most tasks humans perform, both on and off the job \citep{polanyi_tacit_1966, autorPolanyiParadoxShape2014}. 

The fact that generative AI models display such skills suggests that they can acquire tacit knowledge that is embedded in the training examples they encounter.  This ability expands the types of tasks that computers may be capable of performing to include non-routine tasks that rely on judgment and experience.  For example, Github Copilot, an AI tool that generates code suggestions for programmers, has achieved impressive performance on technical coding questions and, if asked, can provide natural language explanations of how the code it produces works \citep{nguyen2022a, zhao2023}.  These capabilities extend to tasks traditionally dominated by highly skilled professionals, including complex mathematics, scientific analysis, coding, and financial modeling. As these tasks have typically been performed by workers who were previously insulated from or benefited from earlier waves of technological adoption, the rise of generative AI has the potential to significantly alter the established relationships between technology, labor productivity, and economic inequality \citep{whitehouse2022}.

Generative AI tools can not only expand the types of tasks that machines can perform, they may also reveal valuable information about how the most productive human workers differ from others.  This is because the ML models underlying generative AI systems are commonly trained on data generated by human workers and, consequently, encounter many examples of people performing tasks both well and poorly.  In learning to predict good outcomes from human-generated data, ML models can implicitly identify characteristics or patterns of behavior that distinguish high and low performers, including subtleties rooted in tacit knowledge.  Generative AI systems then take this knowledge and use it to produce new behaviors that embody what top performers might do.  This ability could be used in different ways: firms may choose to replace lower-skill workers with AI-based tools, such tools could be used to demonstrate best practices to help lower-skill workers improve or help less experienced workers get up to speed more quickly.  In either case, generative AI tools may have differential impacts by worker ability, even amongst workers performing the same tasks. 

Despite their potential, generative AI tools face significant challenges in real-world applications.  At a technical level, popular LLM-based tools, such as ChatGPT, have been shown to produce false or misleading information unpredictably, raising concerns about their reliability in high-stakes situations.  Second, while LLM models often perform well on specific tasks in the lab \citep{GPT4TechnicalReport2023, peng2023impact, noyExperimentalEvidenceProductivity2023}, the types of problem that workers encounter in real-world settings are likely to be broader and less predictable.  Furthermore, generative AI tools often require prompts from human operators, yet finding ways to effectively combine human and AI expertise may be challenging: for instance, earlier research indicates that decision-support systems integrating AI with human judgment often perform worse than those that rely on the human or the AI alone \citep{vaccaro2024combinationshumansaiuseful}. These challenges raise concerns about the ability of AI systems to provide accurate assistance in every circumstance and---perhaps more importantly---workers' capacity to distinguish cases where AI suggestions are effective from those where they are not.  Finally, the efficacy of new technologies is likely to depend on how they interact with existing workplace structures.  Promising technologies may have more limited effects in practice due to the need for complementary organizational investments, skill development, or business process redesign.  Because generative AI technologies are only beginning to be used in the workplace, little is currently known about their impacts.  

\section{Our Setting: LLMs for Customer Support}

\subsection{Customer Support and Generative AI}

We study the impact of generative AI in the customer service industry, an area with one of the highest surveyed rates of AI adoption.\footnote{For instance, of the businesses that report using AI, 22\% use AI in their customer service centers \citep{chui_global_2021}.}  Customer support interactions play a crucial role in building strong customer relationships and company reputation; however, as in many industries, there is substantial variation in worker productivity \citep{berg2018, syverson2011}.

Newer workers require significant training and take time to become more productive.  At the same time, turnover is high: industry estimates suggest that 60\% of agents in contact centers leave each year, costing firms \$10,000 to \$20,000 dollars per agent \citep{buesing_customer_2020, gretzBoostingContactcenterPerformance2018}. To address these workforce challenges, the average supervisor spends at least 20 hours per week coaching agents with lower performance \citep{berg2018}.  Faced with variable productivity, high turnover, and high training costs, firms are increasingly turning to AI tools \citep{chui_global_2021}.

At a technical level, customer support is well-suited for current generative AI tools.  From an AI's perspective, customer-agent conversations can be thought of as a series of pattern-matching problems in which one is looking for an optimal sequence of actions.  When confronted with an issue such as ``I can't login,'' an AI/agent must identify which types of underlying problems are most likely to lead a customer to be unable to log in and think about which solutions typically resolve these problems (``Can you check that caps lock is not on?'').  At the same time, they must be attuned to a customer's emotional response, making sure to use language that increases the likelihood that a customer will respond positively (``that wasn't stupid of you at all!  I always forget to check that too!''). Because customer service conversations are widely recorded and digitized, pre-trained LLMs can be fine-tuned for customer service using many examples of both successfully and unsuccessfully resolved conversations. 

Customer service is also a setting where there is high variability in the abilities of individual agents.  For example, in top-performing customer support agents are often more effective at diagnosing the underlying technical issue given a customer's problem description.  These workers often ask more questions before settling on a diagnosis of the problem; this initially takes longer but reduces the likelihood that agents waste time trying to solve the wrong problem.  Such differences in agent behavior can often be inferred from the large amounts of training data to which customer-service-specific AI models have access.  As a result, customer service is also a setting in which generative AI models can potentially encode some of the ``best practices'' that top-performing agents use.

In the remainder of this section, we provide details about the firm we study and the AI tool they adopt.  

\subsection{Data Firm Background} 

We work with a company that provides AI-based customer service support software (hereafter, the ``AI firm'') to study the deployment of their tool in one of their client firms (hereafter, the ``data firm''). 

Our data firm is a Fortune 500 enterprise software company that specializes in business process software for small and medium-sized businesses in the United States.  It employs a variety of chat-based technical support agents, both directly and by contracting with third-party outsourcing firms.  The majority of agents in our sample work from offices located in the Philippines, with a smaller group working in the United States and in other countries.  Across locations, agents are engaged in a fairly uniform job: answering technical support questions from US-based small business owners.

Within this group of agents, customer chats were assigned on the basis of agent availability, with no additional pre-screening.\footnote{Our data firm employed other groups of agents to provide chat-based support for different customer segments, such as self-employed individuals or larger businesses.  Within chats from US-based small businesses, there was no additional sorting.}  Support sessions are relatively long, averaging 40 minutes, with much of the conversation spent trying to diagnose the underlying technical problem. 

The job requires a combination of detailed product knowledge, problem solving skills, and the ability to deal with frustrated customers. 

Our firm measures productivity using three metrics that are standard in the customer service industry: ``average handle time,'' the average time an agent takes to finish a chat; ``resolution rate,'' the share of conversations that the agent successfully resolves; and ``net promoter score,'' (customer satisfaction), which is calculated by randomly surveying customers after a chat and calculating the percentage of customers who would recommend an agent minus the percentage who would not.  A productive agent is able to field customer chats quickly while maintaining a high resolution rate and net promoter score.  

Across locations, agents are organized into teams with a manager who provides feedback and training to agents. Once a week, managers hold one-on-one feedback sessions with each agent. For example, a manager might share the solution to a new software bug, explain the implication of a tax change, or suggest how to better manage customer frustration with technical issues. Agents work individually and the quality of their output does not directly affect others. 

Agents in our data are typically paid a baseline hourly wage, and receive bonuses for hitting specific performance targets (usually as they relate to chats per hour and chat resolution rate).  While we lack data on individual pay, the managers we interviewed reported that performance bonuses were moderate, accounting for an estimated 20 to 40 percent of total take home pay for the typical customer service worker.

\subsection{AI System Design}

The AI system we study is designed to identify conversational patterns that predict efficient call resolution.  It builds on a recent version of GPT and is fine-tuned on a large dataset of customer-agent conversations labeled with various outcomes, such as call resolution success and handling time.  Our AI firm also flags whether the agent in charge of the conversation was considered a ``top'' performer by the data firm for AI system training.  Many aspects of successful agent behavior are difficult to quantify, including asking clarifying questions, being attentive to customer concerns, deescalating tense situations, adapting communication styles, and explaining complex topics in simple terms. By explicitly training the AI model on text from top performers, the system places value on these subtle behaviors and tone.  The AI firm further trains its model using a process similar in spirit to \citet{ouyangTrainingLanguageModels2022} to prioritize agent responses that express empathy, provide appropriate technical documentation, and limit unprofessional language.  This additional training mitigates some of the concerns associated with relying on LLMs to generate suggestions.  

Once deployed, the AI system generates two main types of output: 1) real-time suggestions for how agents should respond to customers and 2) links to the data firm's internal documentation for relevant technical issues.  In both cases, recommendations are based on a history of the conversation.\footnote{For example, the correct response when a customer says ``I can't track my employee's hours during business trips'' depends on what version of the data firm's software the customer uses. Suppose that the customer has previously mentioned that they are using the premium version. In that case, they should have access to remote mobile device timekeeping, meaning that the support agents need to diagnose and resolve a technical issue that prevents the software from working. If, however, the customer stated that they are using the standard version, then the correct solution is for the customer to upgrade to the premium version in order to access this feature.  For more on context tracking, see, for instance, \citet{dunn2021}.} 

Appendix Figure \ref{afig:AI-sample} illustrates an example of AI assistance. In the chat window (Panel A), Alex, the customer, describes their problem to the agent.  Here, the AI assistant generates two suggested responses (Panel B). In this example, it has learned that phrases like ``I can definitely assist you with this!'' and ``Happy to help you get this fixed asap'' are associated with positive outcomes. Panel C of Appendix Figure \ref{afig:AI-sample} shows an example of a technical recommendation from the AI system, which occurs when it recommends a link to the data firm's internal technical documentation.

Importantly, the AI system we study is designed to augment, rather than replace, human agents. The output is shown only to the agent, who has full discretion over whether to incorporate (fully or partially) the AI suggestions. Giving the agent final decision rights over messages to customers reduces the likelihood that off-topic or incorrect outputs make their way into customer conversations.  Furthermore, the system does not provide suggestions when it has insufficient training data for that situation.  In these situations, the agent must respond on their own.

\section{Deployment, Data, and Empirical Strategy}

\subsection{AI Rollout}\label{sec:rollout}  
AI assistance was introduced after a small randomized pilot involving a small number of agents.\footnote{The RCT is discussed in Section \ref{sec:rct}.} Appendix Figure \ref{afig:deploymenttimeline} illustrates the timing of the rollout, which primarily took place during the Fall of 2020 and Winter of 2021. The variation in access and timing was influenced by two key factors: bottlenecks in the onboarding process due to limited training resources, and the firm's overall budget for AI assistance.

Agents gained access to the AI tool only after completing a three-hour online onboarding session conducted by the AI firm. To maintain quality and consistency, these training sessions were kept small and were exclusively led by one of two employees from our AI firm, both of whom had prior contact center experience and deep knowledge of the AI system. Since these individuals had other full-time responsibilities, the AI firm could only offer limited training sessions each week, with session timings adjusted to accommodate the time zones of the global workforce of the data firm.

Additionally, because generative AI was costly and relatively untested at that time, the data firm allocated a limited budget for its deployment. Once the total number of onboarded agents reached the predefined contractual limit, onboarding ceased. However, when AI-enabled agents left, new agents were added to these existing licenses. These constraints on licenses and training sessions created the variation in AI adoption that we analyze in our study.

Managers at each office oversaw the selection and allocation of agents to training sessions.  In interviews, employees of our AI firm reported that managers made these decisions with the goal of minimizing customer service disruptions.  If all agents on a team were assigned to the same training session, customers would experience a large increase in wait time.  To avoid this, managers tried to assign workers on the same team to different training sessions in order to maintain coverage.  After their initial onboarding session, workers were not given any additional training on how to use the AI software.  This is because the AI firm only employed a small product management team and were not able to provide ongoing support or guidance to the thousands of agents using the tool.

These considerations meant that our AI rollout effectively occurred at the individual level: within the same team, same office, individuals would be onboarded to AI assistance at different times.  In October 2020, within a team, the average share of active workers with access to AI assistance was only 5\%, growing to about 70\% in January of 2021.  While our analysis primarily focuses on the individual adoption dates, we also provide results in Appendix Table \ref{tab:ivdd_companylocteam} that instrument individual adoption dates with team-level adoption patterns.

\subsection{Summary Statistics}\label{sec:sumstats}

Table \ref{sumstat_agents} provides details on the sample characteristics, divided into four groups: all agents (`all''), agents who never have access to the AI tool during our sample period (``never treated''), pre-AI observations for those who eventually get access (``treated, pre'') and post-AI observations (``treated, post'').  In total, we observe the conversation text and outcomes associated with 3 million chats by 5,172 agents.  Within this, we observe 1.2 million chats by 1,636 agents in the post-AI period.  Most of the agents in our sample, 89\%, are located outside of the United States, mainly in the Philippines.  For each agent, we observe their assigned manager, tenure, geographic location, and which firm employs the agent (the data firm or a subcontractor). 

To assess the impacts of this deployment, we rely on several key outcome variables, all aggregated at the agent-month level, the most granular level with complete data.  Our primary productivity measure is resolutions per hour (RPH), which reflects the number of chats a worker successfully resolves per hour and serves as our summary of a worker's productivity.  RPH is influenced by several factors, which we also measure individually: the average time it takes to handle an individual chat (AHT), the number of chats an agent is able to handle per hour, which accounts for multitasking (CPH), and resolution rate (RR), the share of chats that are successfully resolved.  In addition, we measure customer satisfaction using the net promoter score (NPS) from post-call surveys.  While our main outcome measures are at the agent-month level, some data, like chat duration, are available at a more granular chat level.  We also construct additional measures of sentiment, topics, and language fluency from chat text.

Our dataset includes average handle time (AHT) and chats per hour (CPH) for all agents in our sample. However, subcontractors fail to consistently collect call quality metrics for all agents. 
As a result, we only observe our omnibus productivity measure---resolutions per hour---for this smaller subset of agents with call quality outcomes. Workers may work only for a portion of the year or part-time, so we calculate year-month observations based solely on the periods that an agent is assigned to chats.
Appendix Section \ref{asec:key_vars} includes a more extensive discussion of our sample construction and key variables. 

\begin{figure}[ht!]
\begin{center}
\captionsetup{justification=centering}
\caption{\textsc{Figure \ref{prod-distributions}: Raw Productivity Distributions, by AI Treatment}}
\makebox[\linewidth]{
\begin{tabular}{c}
\textsc{A. Resolutions Per Hour} \\
\includegraphics[scale=0.5]{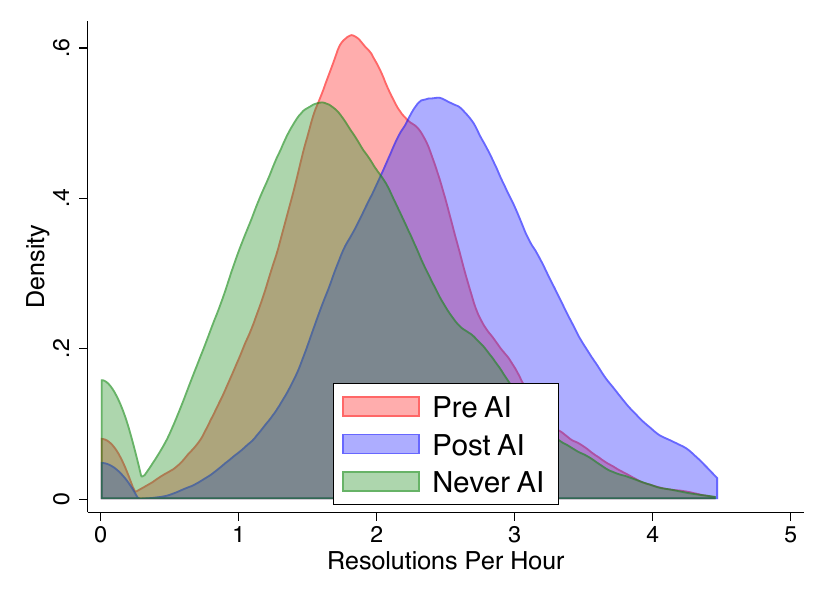} \\
\end{tabular}
}
\makebox[\linewidth]{
\begin{tabular}{cc}
\textsc{B. Average Handle Time} & \textsc{C.  Chats Per Hour}\\
\includegraphics[scale=0.5]{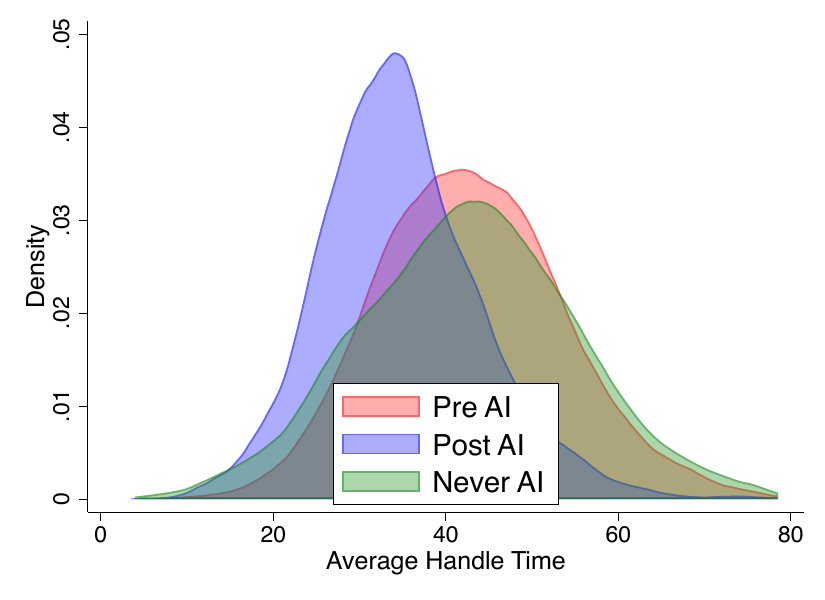} &\includegraphics[scale=0.5]{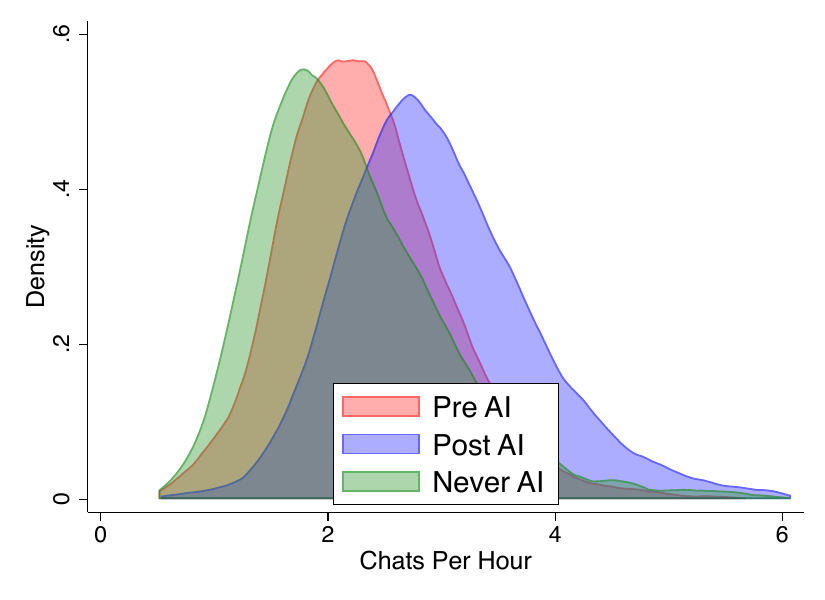} 
\\
\textsc{D. Resolution Rate} & \textsc{E.  Customer Satisfaction (NPS)}\\
\includegraphics[scale=0.5]{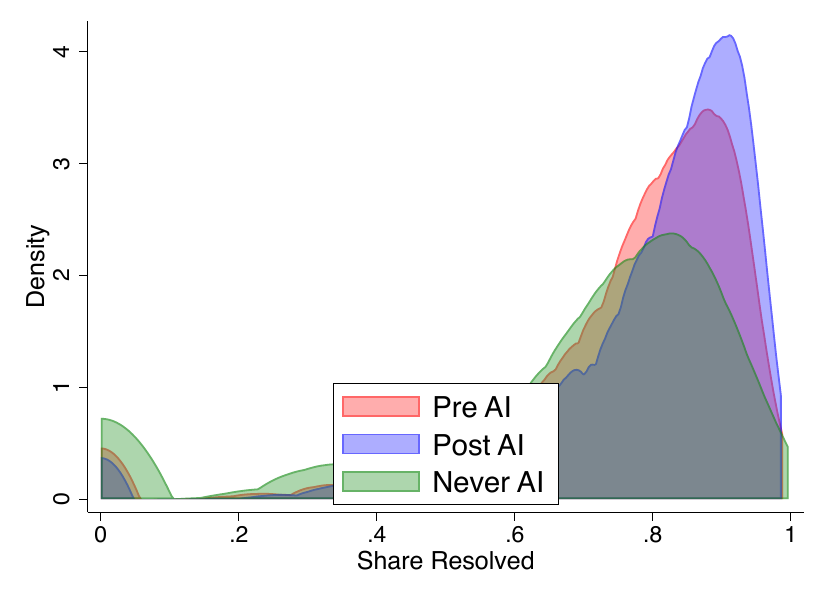} &\includegraphics[scale=0.5]{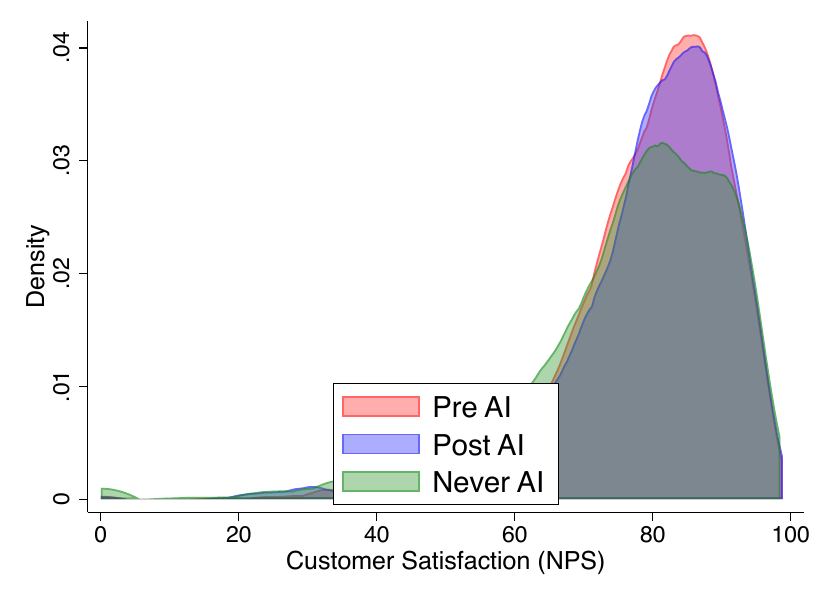} 
\\
\end{tabular}
}
\label{prod-distributions}
\end{center}
\end{figure}

Figure \ref{prod-distributions} plots the raw distributions of our outcomes for each of the never, pre-, and post-treatment subgroups.  Several of our main results are readily visible in these raw data.  In Panels A through D, we see that post-treatment agents do better along a range of outcomes, relative to both never-treated agents and pre-treatment agents.  In Panel E, we see no discernible differences in surveyed customer satisfaction among treated and non-treated groups. 

Focusing on our main productivity measure, Panel A of Figure \ref{prod-distributions} and Table \ref{sumstat_agents} show that never-treated agents resolve an average of 1.7 chats per hour, whereas post-treatment agents resolve 2.5 chats per hour.  Some of this difference may be due to initial selection: treated agents have higher resolutions per hour prior to AI model deployment (2.0 chats) relative to never treated agents (1.7).  This same pattern appears for chats per hour (Panel C) and resolution rates (Panel D): while ever-treated agents appear to be stronger performers at the outset than agents who are never treated, post-treatment agents perform substantially better.  Looking instead at average handle times (Panel B), we see a starker pattern: pre-treatment and never-treated agents have similar distributions of average handle times, centered at 40 minutes, but post-treatment agents have a lower average handle time of 35 minutes.  These figures, of course, reflect raw differences that do not account for potential confounding factors such as differences in agent experience or differences in selection into treatment.  In the next section, we will more precisely attribute these raw differences to the impact of AI model deployment.

\subsection{Empirical Strategy}

We isolate the causal impact of access to AI recommendations using a standard difference-in-differences regression:
\begin{equation} \label{eq:dd_main}
y_{it}= \delta_{t} + \alpha_{i} + \beta AI_{it} + \gamma X_{it}+ \epsilon_{it}
\end{equation}

Our outcome variables, $y_{it}$, is performance measures for agent $i$ in year-month $t$, with resolutions per hour as our main measure of productivity.  We measure these outcomes in levels, and report percentage changes off the baseline pre-treatment means.  Our main variable of interest is $AI_{it}$, an indicator that equals one if AI assistance is activated for agent $i$ at time $t$.  All regressions include year-month fixed effects, $\delta_{t}$, to control for common, time-varying factors such as tax season or the end of the business quarter.  In our preferred specifications, we also include agent fixed effects $\alpha_{i}$ to control for time-invariant differences in productivity across agents and time-varying tenure controls $X_{it}$ (specifically, fixed effects for agent tenure in months).  In our main specifications, we weight each agent-month equally and cluster standard errors at the agent level to reflect the fact that AI access is rolled out individually, but Appendix Tables \ref{tab:robustness_clustering} and \ref{tab:robustdd_nchat_weight} show that our results are robust to alternative weightings and clustering.

A rapidly growing literature has shown that two-way fixed effects regressions deliver consistent estimates only with strong assumptions about the homogeneity of treatment effects and may be biased when treatment effects vary over time or by adoption cohort \citep{cengiz_effect_2019,dechaisemartin2020,sun_estimating_2020,goodman-bacon_2021,callaway_2021, borusyak2022revisiting}. For example, workers may take time to adjust to using the AI system, in which case its impact in the first month may be smaller.  Alternatively, the onboarding of later cohorts of agents may be smoother, so that their treatment effects may be larger.

We study the dynamics of treatment effects using the interaction weighted (IW) estimator proposed in \citet{sun_estimating_2020}.  \citet{sun_estimating_2020} show that this estimator is consistent assuming parallel trends, no anticipatory behavior, and cohort-specific treatment effects that follow the same dynamic profile.\footnote{This last assumption means that treatment effects are allowed to vary over event-time and that average treatment effects can vary across adoption-cohorts (because even if they follow the same event-time profile, we observe different cohorts for different periods of event-time).} Both our main differences-in-differences and event study estimates are similar using robust estimators introduced in \citet{dechaisemartin2020}, \citet{borusyak2022revisiting}, \citet{callaway_2021}, and \citet{sun_estimating_2020}, as well as using traditional two-way fixed effects OLS. In fact, Appendix Figure \ref{fig_aiym} shows similar treatment effects across adoption cohorts (e.g. those that received AI access earlier or later, and were thus subject to potentially different versions of the model).  We also show our results our similar clustering at different levels of granularity (Appendix Table \ref{tab:robustness_clustering}) and instrumenting agent adoption with the date the first worker within the agent's team received AI access (Appendix Table \ref{tab:ivdd_companylocteam}). 

\section{Main Results}

\subsection{Overall Impacts}
Table \ref{tab:dd_main} examines the impact of the deployment of the AI model on our primary measure of productivity, resolutions per hour, using a standard two-way fixed effects model.  In Column 1, we show that, controlling for time and location fixed effects, access to AI recommendations increases resolutions per hour by 0.47 chats, up 23.9\% from their pre-treatment mean of 1.97.  In Column 2, we include fixed effects for individual agents to account for potential differences between treated and untreated agents.  In Column 3, we include additional fixed effects for agent tenure in months to account for time-varying experience levels.  As we add controls, our effects fall slightly, so that, with agent and tenure fixed effects, we find that the deployment of AI increases RPH by 0.30 chats or 15.2\%. 

\begin{figure}[ht!]
\begin{center}
\captionsetup{justification=centering}
\caption{\textsc{Figure \ref{fig:es_main_AS}: Event Studies, Productivity}}

\makebox[\linewidth]{
\begin{tabular}{c}
\includegraphics[scale=0.8]{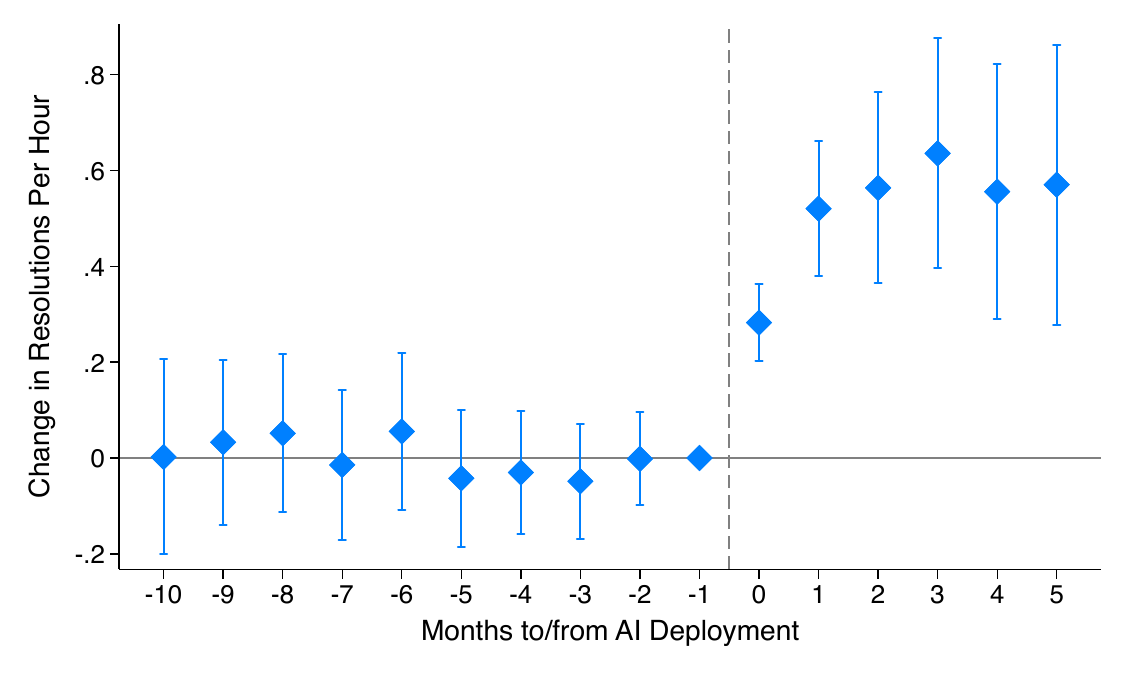} 
\end{tabular}
}
\label{fig:es_main_AS}
\end{center}
\end{figure}

Figure \ref{fig:es_main_AS} shows the accompanying IW event study estimates of \citet{sun_estimating_2020} for the impact of AI assistance on RPH.  We find a substantial and immediate increase in productivity in the first month of deployment.  This effect grows slightly in the second month and remains stable and persistent up to the end of our sample. 

In Table \ref{tab:dd_oth}, we report additional results using our preferred specification with fixed effects for year-month, agent, location and agent tenure.  Column 1 documents a 3.7 minute decrease in the average duration of customer chats, an 8.5\% decline from the baseline mean of 43 minutes (shorter handle times are considered better).  Next, Column 2 indicates a 0.37 unit increase in the number of chats that an agent can handle per hour.  Relative to a baseline mean of 2.4, this represents an increase of roughly 15\%.  Unlike average handle time, chats per hour account for the possibility that agents may handle multiple chats simultaneously.  The fact that we find a stronger effect on this outcome suggests that AI enables agents to both speed up chats and multitask more effectively.

Column 3 of Table \ref{tab:dd_oth} indicates a small 1.3 percentage point increase in chat resolution rates.  This effect is economically modest and insignificant, given a high baseline resolution rate of 82\%; we interpret this as evidence that improvements in chat handling do not come at the expense of problem solving on average.  Finally, Column 4 finds no economically significant change in customer satisfaction, as measured by net promoter scores: the coefficient is -0.12 percentage points and the mean is 80\%. 

Appendix Figure \ref{afig:es_oth} presents the accompanying event studies for additional outcomes.  We see immediate impacts on average handle time (Panel A) and chats per hour (Panel B), and relatively flat patterns for resolution rate (Panel C) and customer satisfaction (Panel D).  We therefore interpret these findings as saying that, on average, AI assistance increases productivity without negatively impacting resolution rates and surveyed customer satisfaction. 

\subsubsection{RCT Analysis} \label{sec:rct}

In August 2020, our data firm conducted a pilot analysis involving approximately 50 workers, about half of whom were randomized into treatment.  Unfortunately, we do not have information on the control group workers that were part of the experiment, but we are able to identify the treatment group because these are the only workers who were given access to the AI tool in August of 2020.  In this section, we compare the productivity of these treated workers with the productivity of their then-untreated colleagues.

Appendix Table \ref{rctdd.tex} presents the overall difference-in-difference estimates.  Focusing on the treatment group of 22 workers, we still see a significant increase in resolutions per hour, as well as a significant decrease in average handle time.  The magnitudes of these effects are similar to those in our main sample. As with our main results, we see no average impacts on resolution rate and customer satisfaction.  Appendix Figure \ref{fig_rct} reports the accompanying event studies for our various outcomes. 

\subsubsection{Instrumenting Individual AI adoption} \label{sec:instrumentingadoption}

During the rollout process, managers were in charge of deciding which agents to onboard on to the AI system, and scheduling when their training would occur.  One may therefore be concerned that the timing of an individual agent's AI adoption may be related to their performance if, for instance, managers prioritized onboarding particularly weak or strong workers.  This concern is less likely to apply to the timing of an entire team's onboarding.  In Appendix Table \ref{tab:ivdd_companylocteam}, we present results in which we instrument an individual agent's AI adoption date with the first adoption date of the worker's company (e.g. the main data firm or one of its subcontractors), office location, and team.\footnote{Information on team is missing for some workers.  Workers without this information are grouped into a single ``missing'' team.} Although AI-access was in theory introduced across the entire firm, variation in training schedules and time zones made it so that some teams began the process sooner than others.  Using this approach, we estimate that individual AI adoption increases resolutions per hour by 0.55 chats per hour, compared to 0.30 under our main specification.  Our effects for average handle time and chats per hour are essentially identical to our main effects: the larger magnitude for resolutions per hour comes from the fact that this IV approach estimates a significant and larger impact on resolution rates.

\subsubsection{Further Robustness}

We report a variety of robustness tests for these findings.  

Appendix Table \ref{atab:dd_main_robust} finds similar results using alternative difference-in-difference estimators introduced in \citet{callaway_2021}, \citet{borusyak2022revisiting}, \citet{dechaisemartin2020}, and \citet{sun_estimating_2020}.  Unlike traditional OLS, these estimators avoid comparing between newly treated and already treated units.  In most cases, we find larger effects of AI assistance using these alternatives.  Similarly, Appendix Figure \ref{afig:es_main_alt} reports that our results are similar under alternative event study estimators: \citet{callaway_2021}, \citet{borusyak2022revisiting}, \citet{dechaisemartin2020}, and traditional two-way fixed effects.  

In Appendix Table \ref{tab:robustness_clustering}, we show that our standard errors are similar whether clustering at the individual level, team level, or geographic location level.  We choose to cluster at the individual level to reflect the individual variation in the AI roll out.  While our standard errors are somewhat smaller at this level, our results are still highly significant when clustering at the most conservative location level.\footnote{Geographic location is usually available at the city level (e.g. Reno, Nevada) and one geographic location may contain firms employed by different subcontractors, although a small share workers work remotely.}  

Finally, we explore robustness to alternative weighting.  Our main results treat each worker-month observation equally in order to focus on the impact of generative AI on worker-level productivity.  In Appendix Table \ref{tab:robustdd_nchat_weight}, we reweight these estimates by the number of customer chats a worker conducts and show that our results are similar.  

\subsection{Heterogeneity by Agent Skill and Tenure}

There is substantial interest in the distributional consequences of AI-based technologies. An extensive literature suggests that earlier waves of information technology (e.g., the Internet and computers) complemented high-skilled workers, increasing their productivity and labor demand and widening wage differentials. Together with important changes in relative supply and demand for skilled labor, these changes shaped patterns of wage inequality in the labor market \citep{katzgoldinorigins1998, katzgoldineducationtech2008}.  Unlike earlier waves of IT, however, generative AI does not simply execute routine tasks. Instead, as outlined in Section \ref{sec:aibackground}, AI models identify patterns in data that replicate the behaviors of many types of workers, including those engaged in non-routine, creative, or knowledge-based tasks.  These fundamental technical differences suggest that generative AI may impact workers in different ways. 

In this section, we explore two components that are important for understanding the distributional consequences of AI adoption: its impact on the productivity of less skilled and less experienced contact center workers.  

\subsubsection{Pre-treatment Worker Skill}
We measure an agent's ``skill'' using an index incorporating three key performance indicators: call handling speed, issue resolution rates, and customer satisfaction.  To construct this index, we compute an agent's ranking within its employer company-month for each component productivity measure and then average these rankings into a single index.  Next, we calculate the average index value over the three months for each agent, to smooth out month-to-month shocks in agent performance.  An agent in the top quintile of this productivity index demonstrates excellence across all three metrics: efficient call handling, high issue resolution rates, and superior customer satisfaction scores. 

Panel A of Figure \ref{fig:dd_workerhet} shows how our productivity effects vary based across workers in each quintiles our skill index, measured in the month prior to AI-access.  To isolate the impact of worker skill, our regression specification includes a set of fixed effects for months of worker tenure. Appendix Section \ref{asec:specifications} includes the exact specification. 
We find that the productivity impact of AI assistance is most pronounced for workers in the lowest skill quintile (leftmost side), who see a 0.5 increase or 36\% increase in resolutions per hour. In contrast, AI assistance does not lead to any productivity increase for the most skilled workers (rightmost side).

In Figure \ref{fig:dd_byskill} we show that less-skilled agents consistently see the largest gains across our other outcomes.  For the highest-skilled workers, we find mixed results: a zero effect on average handle time (Panel A), a small, but positive effect for chats per hour (Panel B), and, interestingly, small but statistically significant \textit{decreases} in resolution rates and customer satisfaction (Panels C and D).  These results are consistent with the idea that generative AI tools may function by exposing lower-skill workers to the best practices of higher-skill workers.  Lower-skill workers benefit because AI assistance provides them with new solutions, whereas the best performers may see little benefit from being exposed to their own best practices.  Indeed, the fact that we find negative effects along measures of chat quality---resolution rate and customer satisfaction---suggests that AI recommendations may distract top performers, or lead them to choose the faster or less cognitively taxing option (following suggestions) rather than taking the time to come up with their own responses.  This is potentially important to address because the conversations of top agents are also used for ongoing training: failing to incentivize these agents to continue to produce responses could reduce the quality of the AI model in the future.

A potential concern is that our results could be driven by mean reversion: agents that performed well just prior to AI-adoption may see a natural decline in their productivity afterward, while lower-performing agents may rebound.  To address this concern, we plot raw resolutions per hour in event-time, graphed by skill tercile at AI treatment in Appendix Figure \ref{afig:meanrev}. If mean reversion were driving our observed effects, we would expect to see a convergence of productivity levels after treatment, with top-tercile agents showing decreased performance and the least-skilled agents demonstrating improved output. However, our analysis reveals a consistent linear increase in productivity across all skill levels after AI implementation, with no strong evidence of mean reversion, suggesting that productivity gains are attributable to AI assistance.

\subsubsection{Pre-treatment Worker Experience}
Next, we repeat our previous analysis for agent tenure to understand how the treatment effects of AI access vary by worker experience. To do so, we divide agents into five groups based on their months of tenure at the time the AI model is introduced.  Some agents have less than a month of tenure when they receive AI access, while others have more than a year of experience.  To isolate the impact of worker tenure, this analysis controls for the worker skill quintile at AI adoption, with the regression specification located in Appendix Section \ref{asec:specifications}. 

\begin{figure}[ht!]
\begin{center}
\captionsetup{justification=centering}
\caption{\textsc{Figure \ref{fig:dd_workerhet}: Heterogeneity of AI Impact, by Skill and Tenure}}
\makebox[\linewidth]{
\begin{tabular}{c}
\textsc{A. Impact of AI on Resolutions Per Hour, by Skill at Deployment} \\
\includegraphics[scale=0.65]{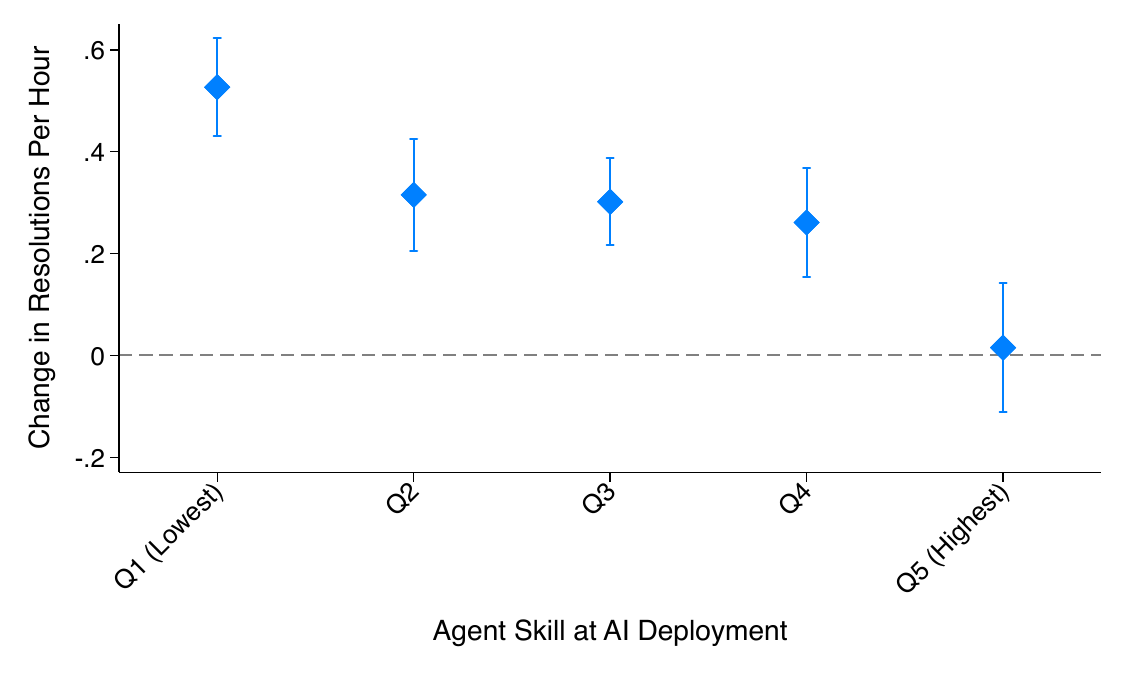} \\
\\
\textsc{B. Impact of AI on Resolutions Per Hour, by Tenure at Deployment} \\
\includegraphics[scale=0.65]{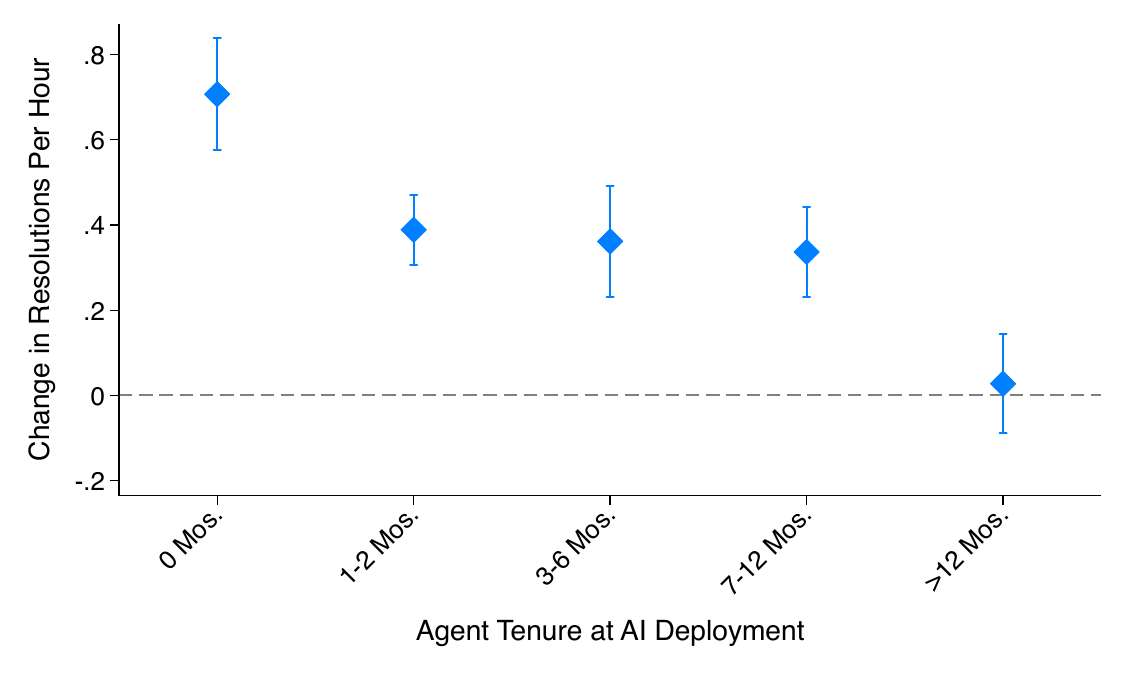} \\
\end{tabular}
}
\label{fig:dd_workerhet}
\end{center}
\end{figure}

In Panel B of Figure \ref{fig:dd_workerhet}, we see a clear monotonic pattern in which the least experienced agents see the greatest gains in resolutions per hour.  Agents with less than 1 month of tenure improve their resolutions per hour by 0.7 resolutions per hour, with larger effects for less-experienced workers.  In contrast, we see no effect for agents with more than a year of tenure. 

In Figure \ref{fig:dd_bytenure}, we report results for other outcomes.  In Panels A and B, we see that AI assistance generates large gains in call handling efficiency, measured by average handle times and chats per hour, respectively, among the newest workers.  In Panels C and D, we find positive impacts of AI assistance on chat quality, as measured by resolution rates and customer satisfaction, respectively.  For the most experienced workers, we see modest positive effects for average handle time (Panel A), positive but statistically insignificant effects on chats per hour (Panel B), and small but statistically significant negative effects for measures of call quality and customer satisfaction (Panels C and D). 

Overall, these patterns are very similar to our findings for agent skill, even though our regressions are designed to isolate the distinct roles of skill and experience (e.g. our skill regressions control for experience and vice versa).  This suggests that even within the same task, access to AI systems disproportionately improves the performance of both novice and less skilled workers.  

\subsubsection{Moving Down the Experience Curve}
\begin{figure}[ht!]
\begin{center}
\captionsetup{justification=centering}
\caption{\textsc{Figure \ref{fig:learning_bycohort}: Experience Curves by Deployment Cohort}}
\makebox[\linewidth]{
\begin{tabular}{c}
\textsc{Resolutions Per Hour, by Agent Tenure} \\
\includegraphics[scale=0.8]{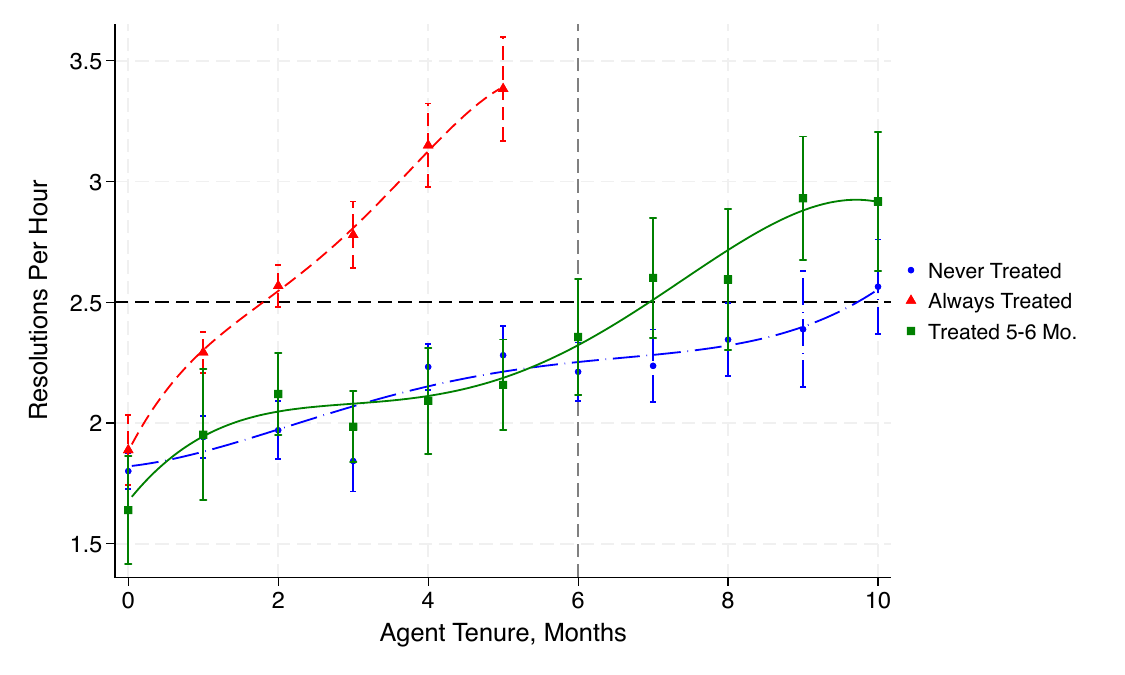} \\
\end{tabular}
}
\label{fig:learning_bycohort}
\end{center}
\end{figure}

To further explore how AI assistance impacts newer workers, we examine how worker productivity evolves on the job.\footnote{We avoid the term ``learning curve'' because we cannot distinguish if workers are learning or merely following recommendations.}  In Figure \ref{fig:learning_bycohort}, we plot productivity variables by agent tenure for three distinct groups: agents who never receive access to the AI model (``never treated''), those who have access from the time they join the firm (``always treated''), and those who receive access in their fifth month with the firm (``treated 5 mo.'').  

We see that all agents begin around 1.8 resolutions per hour.  Never-treated workers (long dashed blue line) slowly improve their productivity with experience, reaching approximately 2.5 resolutions per hour 8 to 10 months later.  In contrast, workers who always have access to AI assistance (short dashed red line) increase their productivity to 2.5 resolutions per hour after only two months and continue to improve until they are resolving more than 3 chats per hour after five months of tenure.\footnote{Our sample ends here because we have very few observations more than five months after treatment.} Comparing just these two groups suggests that access to AI recommendations helps workers move more quickly down the experience curve and reduces ramp-up time. 

The final group in Panel A tracks workers who begin their tenure without access to AI assistance but who receive access after five months on the job (solid green line).  These workers initially improve at the same rate as never-treated workers, but after gaining AI access in month 5, their productivity begins to more rapidly increase, following the same trajectory as the always-treated agents. These findings demonstrate that AI assistance not only accelerates ramp-up for new workers, but also improves the rate at which even experienced workers improve in their roles.

In Appendix Figure \ref{afig:learning_bycohort}, we plot these curves for other outcomes. We see clear evidence that the experience curve for always-treated agents is steeper for handle time, chats per hour, and resolution rates (Panels A through C). Panel D follows a similar but noisier pattern for customer satisfaction. Across many of the outcomes we examine, agents with two months of tenure and access to AI assistance perform as well as or better than agents with more than six months of tenure who do not have access. AI assistance alters the relationship between on-the-job productivity and time, with potential implications for how firms might value prior experience, or approach training and worker development.

\section{Adherence, Learning, Topic Handling, and Conversational Change}

In this section, we conduct a variety of analyses to better understand the mechanisms driving our main results.  

First, we examine how workers engage with AI recommendations.  We show that workers are selective about the recommendations they adopt, following the recommendations 35\% on average.  We find that the returns on AI assistance are highest for workers who choose to follow the recommendations.  Consistent with a story in which workers find AI recommendations helpful, we show that adherence rates increase over time for all workers, especially among more experienced workers: by the end of our sample, we see similar adherence rates across worker tenure and skill. 

Second, we explore whether AI-assistance helps workers learn.  Using information on software outages in which AI assistance is temporarily unavailable, we provide evidence that exposure to AI leads to durable changes in worker skills. We find that workers exposed to AI recommendations continue to perform better during outages, and this effect is greater after more exposure and for agents who follow AI recommendations more closely when the software is working.

Next, we provide evidence on two ways in which AI assistance may improve workers' conversations: providing them with the information to address specific customer problems and refining their overall conversational fluency.  Our findings indicate that AI access is particularly beneficial when workers encounter less routine issues, where they might otherwise struggle to find solutions. In addition, we also show that AI assistance improves workers' English language skills, with the most substantial gains seen among those with the lowest initial proficiency. 

Lastly, we analyze the text of conversations to assess how AI assistance influences the communication patterns of higher- and lower-skill workers.  In particular, we track how an agent's words change before and after AI adoption and find evidence of larger textual shifts for lower-skill workers.  We also show that the similarity of conversations conducted by high- and low-skill agents increases following AI-adoption.  These results suggest that AI assistance may help lower-skill agents converge towards their higher-skill peers.

Taken together, our results suggest that examining and following AI recommendations helps workers---particularly lower-skilled workers---learn to adopt best practices gathered from higher-skill and more experienced agents.

\subsection{Adherence to AI recommendations} \label{sec:receptivity}

The AI tool we study makes suggestions, but agents are ultimately responsible for what they say to the customer. Thus far, our analysis evaluates the effect of AI assistance, irrespective of the frequency with which users adhere to its suggestions.
Here, we examine how closely agents adhere to AI recommendations, and document the association between adherence and returns to adoption.  
We define ``adherence'' as the proportion of AI suggestions an agent typically adopts. The AI company considers an agent to have adhered when they either directly copy the AI's proposed text or manually enter highly similar content. To gauge initial adherence, we classify each treated agent into a quintile based on their level of adherence during their first month using the AI tool.

\begin{figure}[ht!]
\begin{center}
\captionsetup{justification=centering}
\caption{\textsc{Figure \ref{fig:dd_byreceptivity}: Heterogeneity of AI Impact, by AI Adherence}}
\makebox[\linewidth]{
\begin{tabular}{c}
\textsc{A. Distribution of AI Adherence} \\
\includegraphics[scale=0.75]{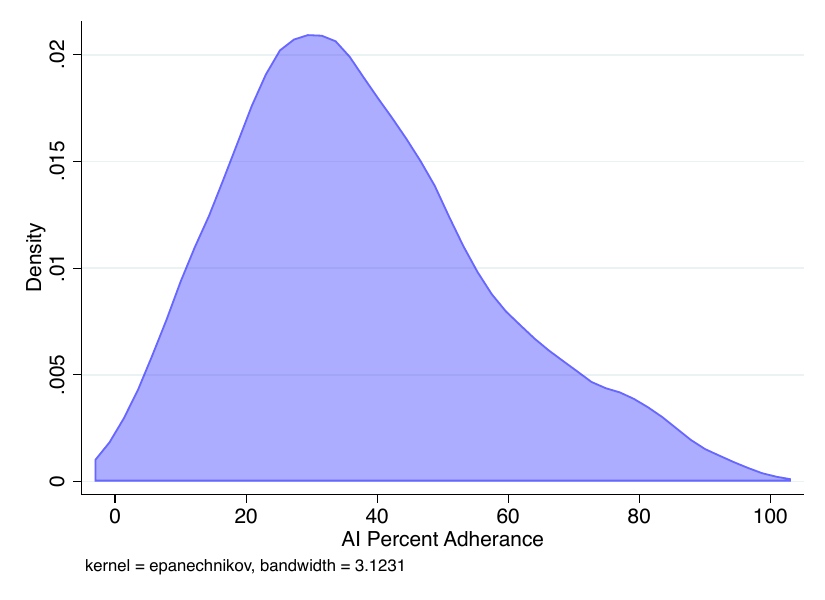} \\
\\

\textsc{B. Impact of AI on Resolutions Per Hour, by Initial Adherence} \\
\includegraphics[scale=0.75]{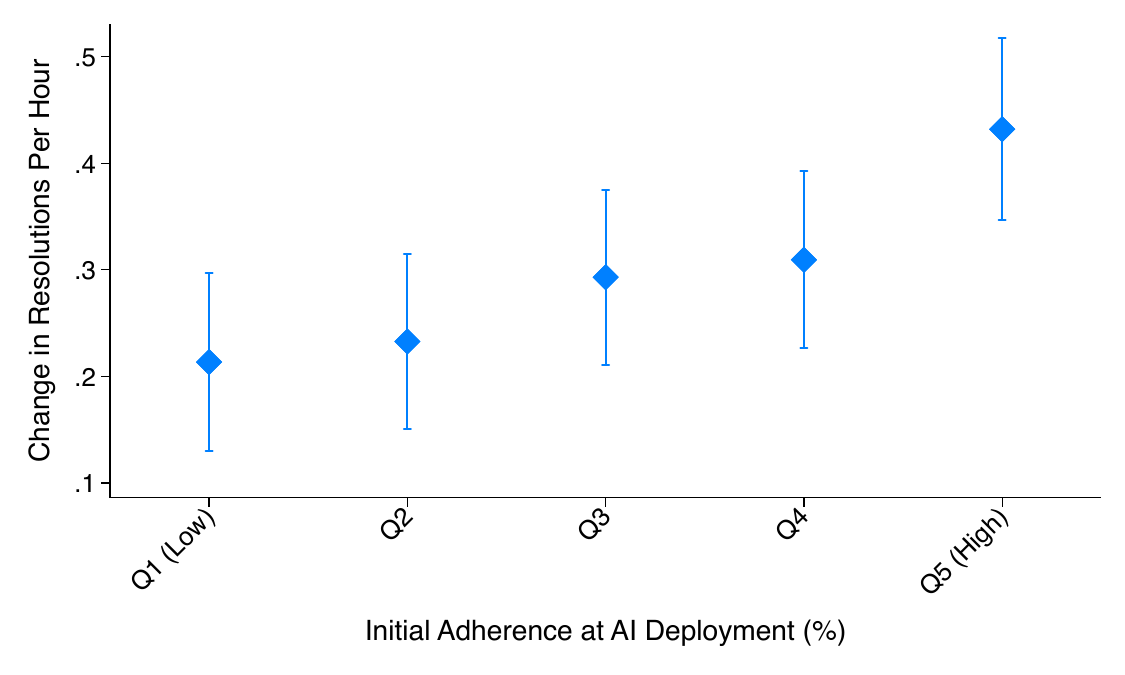} \\
\end{tabular}
}

\label{fig:dd_byreceptivity}
\end{center}
\end{figure}

Panel A of Figure \ref{fig:dd_byreceptivity} shows the distribution of average agent-month-level adherence for our post-AI sample, weighted by the log number of AI recommendations provided to that agent in that month.  The average adherence rate is 38\%, with an interquartile range of 23\% to 50\%: agents frequently ignore recommendations.  In fact, the share of recommendations followed is similar to the share of other publicly reported numbers for generative AI tools; a study of GitHub Copilot reports that individual developers use 27\% to 46\% of code recommendations \citep{zhao2023}. Such behavior may be appropriate, given that AI models may make incorrect or irrelevant suggestions.  In Appendix Figure \ref{fig:adherence_decomposed}, we further show that the variation in adherence is similar within locations and teams, indicating that this variation is not driven by some organizational segments being systematically more supportive than others.

Panel B of Figure \ref{fig:dd_byreceptivity} shows that \textit{returns} to AI model deployment are higher when agents actually follow recommendations.  To show this, we divide agents into quintiles based on the share of AI recommendations they follow in the first month of AI access. Following Equation \ref{aeq:dd_adherence}, we separately estimate the impact of AI assistance for each group, including year-month, agent and agent tenure fixed effects.

We find a steady and monotonic increase in returns by agent adherence: among agents in the lowest quintile, we still see a 10\% gain in productivity, but for agents in the highest quintile, the estimated impact is over twice as high, close to 25\%.  Appendix Figure \ref{afig:dd_byreceptivity} shows the results for our other four outcome measures. The positive correlation between adherence and returns holds most strongly for average handle time (Panel A) and chats per hour (Panel B), and more noisily for resolution rate (Panel C) and customer satisfaction (Panel D). 

Our results are consistent with the idea that there is a treatment effect of following AI recommendations on productivity.  We note, however, that this relationship could also be driven by other factors: selection (agents who choose to adhere are more productive for other reasons); or selection on gains (agents who follow recommendations are those with the greatest returns).  To further explore this, we consider the worker's revealed preference: do they continue to follow AI recommendations over time?  If our results were driven purely by selection, we would expect workers with low adherence to continue having low adherence, since it was optimal for them to do so.

Figure \ref{fig:receptivity_overtime} plots the evolution of AI adherence over time, for various categories of agents.  Panel A begins by considering agents who differ in their initial AI compliance, which we categorize based on terciles of AI adherence in the first month of model deployment (``initial adherence'').  Here, we see that compliance either stays stable or grows over time.  The most initially compliant agents continue to comply at the same rates (just above 50\%).  Less initially compliant agents increase their compliance over time: those in the bottom tercile initially follow recommendations less than 20\% of the time but, by month five, their compliance rates have increased by over 50\%, to just over half of the time.  Next, Panel B divides workers up by tenure at the time of AI deployment.  More senior workers are initially less likely to follow AI recommendations: 30\% for those with more than a year of tenure compared to 37\% for those with less than three months of tenure.  However, over time, all workers increase adherence, with more senior workers doing so faster so that the groups converge five months after deployment.  In Panel C, we show the same analysis by worker skill at AI deployment.  Here, we see that compliance rates are similar across skill groups, and all groups increase their compliance over time.  In Appendix Figure \ref{fig:receptivity_overtime_agentfe} we show that these patterns are robust to examining within-agent changes in adherence (that is, adherence rates residualized by agent fixed effects).  This suggests that increases in adherence over time are not driven exclusively by selection.   

The results in Figures \ref{fig:receptivity_overtime} and \ref{fig:receptivity_overtime_agentfe} are consistent with agents, particularly those who are initially more skeptical, coming to value AI recommendations over time.  We note, however, that high skill agents increase their adherence as quickly as their lower skill peers, even though their productivity gains are smaller and---in the case of some quality measures---even negative.  This suggests an alternate possibility: agents may be over-relying on AI recommendations beyond what is optimal in the long run. Top agents, in particular, may see little additional value in taking the time to provide the highest quality inputs when an adequate AI suggestion is readily available.  If this is indeed the case, high AI adherence in the present may reduce the quality or diversity of solutions used for AI training in the future.  While this may be a longer run concern, in the short run, our analysis finds no evidence that the model is declining in quality over our sample period: in Appendix Figure \ref{fig_aiym}, we show that workers who received later access to the AI system---and therefore to a more recently updated version---had similar first month treatment effects as those who were onboarded to an earlier version of the model.\footnote{When the AI model is updated, changes are made available to all onboarded workers.  In order to compare workers using different versions of the model, we restrict our comparisons to productivity impacts from the first month after adoption, before a version update would have been pushed.}

\subsection{Worker Learning} \label{sec:learning}

A key question raised by our findings so far is whether these improvements in productivity and changes in communication patterns reflect durable changes in the human capital of workers or simply their growing reliance on AI assistance.  In the latter case, the introduction of AI assistance could actually lead to an erosion in human capital, and we would expect treated workers to be less able to address customer questions if they are no longer able to access AI assistance.\footnote{For example, research in cognitive science has shown that individuals learn less about spatial navigation when they follow GPS directions, relative to using a map \citep{Brügger2019}.}

To study this, we examine how workers perform during periods in which they are not able to access AI-recommendations due to technical issues at the AI firm.  Outages occur occasionally in our data and can last anywhere from a few minutes to a few hours.  During an outage, the system fails to provide recommendations to some, but not necessarily all, workers.  For example, outages may affect agents who log into their computers after the system crashes, but not agents working at the same time who had signed in earlier.  They may also affect workers using one physical server but not another. Our AI firm tracks the most significant outages in order to perform technical reviews of what went wrong.  We compile these system reports to identify periods in which a significant fraction of chats are affected by outages.

Appendix Figure \ref{afig:outage_example} shows an example of such an outage, which occurred on September 10, 2020.  The $y$-axis plots the share of post-treatment chats (e.g. those occurring after the AI system has been deployed for a given agent) for which the AI software does not provide any suggestions, aggregated to the hour level.  The $x$-axis tracks hours in days leading up to and following the outage event (hours with fewer than 15 post-treatment chats are plotted as zeros for figure clarity).  During non-outage periods, the share of chats without AI recommendations is typically 30-40\%.  This reflects the fact that the AI system does not generate recommendations in response to all messages, even when it is functioning properly.  
On the morning of September 10th, however, we see a notable spike in the number of chats without recommendations, increasing to almost 100\%. Records from our AI firm indicate that this outage was caused by a software engineer running a load test that crashed the system.

Figure \ref{fig:ES_outage} examines the impact of access to the AI system for chats that occur during and outside these outage periods.  These regressions are estimated at the individual chat level in order to precisely compare conversations that occurred during outage periods, versus those that did not. Because we do not have information on chat resolution at this level of granularity, our main outcome measure is chat duration. Panel A considers the impact of AI assistance using only post-adoption periods in which no outages are reported.  Consistent with our main results, we see an immediate decline in the duration of individual chats by approximately 10\% to 15\%.  

In Panel B, we use the same pre-treatment observations, but now restrict to post-adoption periods that are impacted by large outages. We first note that our estimates are noisy and their magnitude appears larger than for non-outage periods (equivalent to 15\% to 25\% declines in chat duration). Because AI outages are rare and not necessarily random, this may reflect differences in the types of chats that are seen during outage periods than during non-outage periods.  However, focusing on the size of estimated effects over time, an interesting pattern emerges.  Rather than declining immediately post-adoption and staying largely stable as we see in Panel A for non-outage periods, Panel B shows that the benefit of exposure to AI assistance increases with time during outage periods.  That is, if an outage occurs one month after AI adoption, workers do not handle the chat much more quickly than their pre-adoption baseline.  Yet, if an outage occurs after three months of exposure to AI recommendations, workers handle the chat faster---even though they are not receiving direct AI assistance in either case.  

Panel B highlights the potential scope for improving existing employee training practices. Prior to AI assistance, training was limited to brief weekly coaching sessions where managers reviewed select conversations and provided feedback. However, by necessity, managers can only provide feedback on a small fraction of the conversations an agent conducts.  Moreover, because managers are often short on time and may lack training, they often simply point out weak metrics (``you need to reduce your handling time'') rather than identifying strategies for how an agent could better approach a problem (``you need to ask more questions at the beginning to diagnose the issue better.'') Such coaching can be ineffective and can be counterproductive to employee engagement \citep{berg2018}. In contrast, AI assistance offers workers specific, real-time, actionable suggestions, potentially addressing a limitation of traditional coaching methods.

To better understand how learning might occur, in Panels C and D of Figure \ref{fig:ES_outage}, we divide our main study of outage events by the initial adherence of the worker to AI, as described in Section \ref{sec:receptivity}. When a worker chooses not to follow a particular AI recommendation, they miss the opportunity to observe how the customer might respond.  AI suggestions may prompt workers to communicate in ways that differ from their natural style, such as by expressing more enthusiasm or empathy, or by frequently pausing to recap the conversation.  Workers who do not ``try out'' these recommendations may never realize that customers could react positively to them.

Panel C reveals that workers with high initial adherence to AI recommendations experience significant and rapid declines in chat processing times, even during outages, relative to their pre-adoption baseline. In contrast, Panel D shows no such improvement for workers who frequently deviate from AI suggestions; they see no reduction in chat times during outage periods, even after prolonged AI access.  These findings suggest that workers learn more by actively engaging with AI suggestions and observing first-hand how customers respond. These findings are consistent with other evidence from education that having students attempt math problems on their own before seeing LLM-generated responses positively impacted learning \citep{kuman2023}. Together, these results suggest that AI assistance can play a useful role in supplementing existing on-the-job training programs. 

\subsection{Handling Routine and Non-Routine Topics}

So far, our analyses have focused on how the impact of AI assistance varies by characteristics of the worker.  The impact of AI of course, can also depend on the types of problems it is asked to resolve.  Agents in our data encounter a diverse set of customer questions, ranging from common requests for help in onboarding an employee or changing a password, to less common questions such as setting up wage garnishments in child support cases or ensuring compliance with international tax treaties.  When an AI model is trained on these conversations, it will naturally have more examples of the most common problems, raising the question of whether it can effectively provide suggestions for topics it has encountered less frequently.

In this section, we examine the impact of AI-assistance for more and less routine customer problems.  To categorize the diverse conversations in our data, we employed Gemini, a large language model developed by Google DeepMind, to classify the interactions into topic categories. The details of this process, along with our human validation of the LLM classification process, are described in Appendix Section \ref{asec:key_vars} \citep{geminipaper}. 

Appendix Figure \ref{hist_topic} reports the distribution of conversation topics in our dataset.  Unsurprisingly for a customer service setting, we observe a small number of frequent issues, accompanied by a long tail of less common problems. Specifically, the two most prevalent topics---payroll and taxes, and account access and management---comprise half of all conversations and the top 16 topics represent over 90\% of all chats.  

To evaluate the impact of AI assistance based on the frequency of customer inquiries, we categorize conversations into four distinct groups. The ``Payroll/Account'' category, comprising 50\% of all chats and including the two most common topics, includes inquiries related to payroll, taxes, and account access and management. The next 25\% of chats cover 5 additional topic categories, for instance, those dealing with bank transfers or managing subscriptions.  The following 15\% of chats encompass nine additional topics while the final 10\% of chats includes the remaining topics. Our regression, in Appendix Section \ref{asec:specifications}, is conducted at the chat level, with a focus on chat duration.

\begin{figure}[ht!]
\begin{center}
\captionsetup{justification=centering}
\caption{\textsc{Figure \ref{fig:dd_bytopic}: AI Impact by Chat Topic}}
\makebox[\linewidth]{
\begin{tabular}{c}
\textsc{A. Impact of AI on Chat Duration, by Overall Topic Frequency} \\
\includegraphics[scale=0.7]{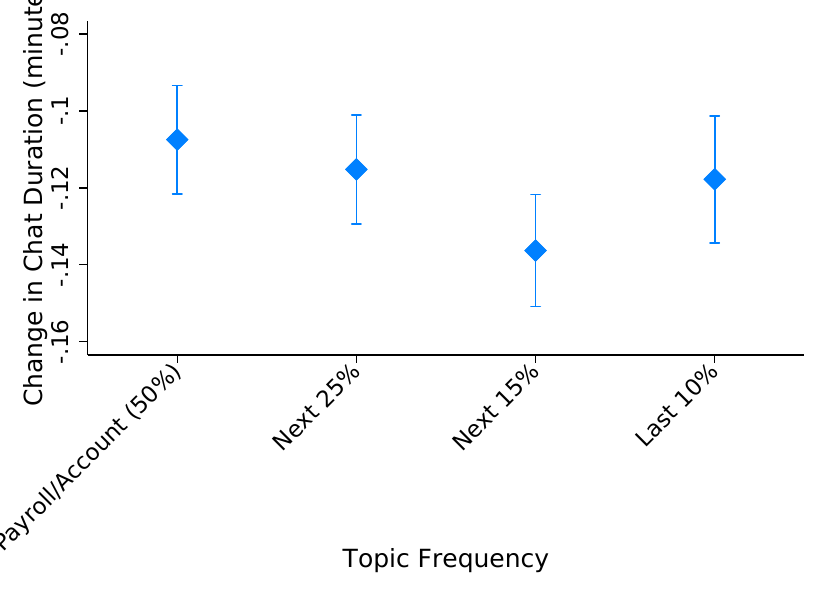} \\

\textsc{B. Impact of AI on Chat Duration, by Agent Topic Frequency} \\
\includegraphics[scale=0.7]{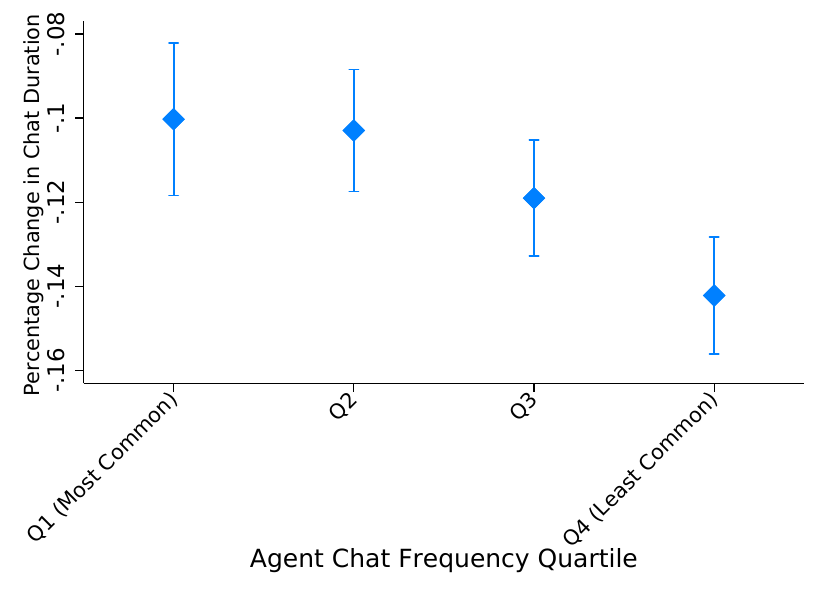} \\
\\

\end{tabular}
}
\label{fig:dd_bytopic}
\end{center}
\end{figure}

Panel A of Figure \ref{fig:dd_bytopic} shows the average treatment effect of AI assistance according to the routine nature of the customer support inquiry.  This pattern is non-monotonic: our results suggest that AI-assistance has the greatest impact on workers' ability to handle problems that are moderately rare.  Workers with access to AI assistance handle the most routine problems---payroll and account management---about 4 to 5 minutes faster, which corresponds to an approximately 10\% decrease off the pre-treatment mean duration for these topics.   We see the largest decline, 5 to 6 minutes, for issues that are in the 75th to 90th percentiles of topic rarity, corresponding to a 14\% reduction given the pre-treatment means for those topics.  Finally, we see a smaller 4 minute or 11\% decrease for the most rare problems.

These results highlight the difference between the technical quality of an AI system and its potential productivity impacts in real-world settings. AI models generally perform better when trained on large datasets, as these provide diverse examples and richer contextual information, enabling the model to learn more robust and generalizable patterns while reducing the risk of overfitting \citep{halevy2009}.  Consequently, we might expect an AI system to function best when dealing with routine problems, where training data are abundant.  

However, the value of AI systems, when used to complement human workers is less straightforward.  Customer service agents, especially those dealing with common issues, are specifically trained to address these routine problems efficiently and get the most experience answering common questions. For example, even novice workers are likely to know how to reset a customer's password.  In such cases, access to even high quality AI assistance may not have a large complementary impact on the most common problems. Rather, as our findings suggest, the impact of an AI system on workplace productivity depends critically on its capabilities relative to workers' baseline skills. The greatest productivity gains may occur not where the AI system is most capable in absolute terms, but where its capabilities most effectively complement or exceed those of human workers.

In our setting, the heterogeneous impact of AI access appears to reflect both factors: AI access has the smallest reduction in handle time for problems where human agents are already well trained (very routine problems) or where its training data may be sparse (very rare problems).  We see the largest improvements in the handle time for somewhat uncommon problems.  Here, the AI system is likely to have enough training data to assess these problems, while individual agents are less likely to have had much first-hand experience.  For example, the AI system we study provides workers with links to potentially relevant technical documentation.  This may be particularly valuable for the types of cases where agents may not be aware of the solution and would need to put the customer on hold while they search for an answer.

To examine the role of agent-specific experience, Panel B of Figure \ref{fig:dd_bytopic} plots the impact of AI assistance on chat duration by quartiles of topic frequency with respect to an individual agent, controlling for the overall frequency of a problem.  Here we show that AI assistance reduces conversation times most for problems that an agent has encountered least often: 15\% for the least common problems compared with 10\% for the most common.  Once we control for a topic's overall frequency, we now find a monotonic relationship between agent-specific exposure to a problem and the impact of AI. That is, holding constant the AI model's exposure to a problem, the impact of AI assistance is greatest for problems that a specific agent is least exposed to.  This suggests that while AI in isolation may be most effective where training data is most plentiful, the marginal value of AI assistance is highest where humans have a greater need for AI input.

\subsection{Conversational Style}

Our previous analysis examined how the effectiveness of AI assistance varies based on the technical nature of the problems being addressed. However, top customer service agents not only resolve technical issues but also communicate in a way that ensures customer satisfaction.  For our US-based customers, this involves clearly writing in fluent English and displaying subtle interpersonal skills.  In this section, we first demonstrate that AI assistance improves overall language fluency.  Then, to capture less quantifiable changes in communication style, we compare the evolution of conversational text produced by different workers.  Our analysis provides suggestive evidence that AI adoption helps lower-skill workers communicate more like their higher-skill peers.

\subsubsection{English Fluency} \label{sec:englishproficiency}
For contact center workers serving US customers, the ability to communicate in clear, idiomatic English is crucial to their job performance and customer satisfaction. We begin by assessing how AI assistance influences workers' ability to communicate clearly. In our data, 80\% of the agents are based in the Philippines, where many residents are fluent English speakers.\footnote{English is an official language of the Philippines and  Filipinos, especially those who are college educated, often have near-native fluency due to early exposure in schools.  Further, the Philippines is a former American colony and American English tends to be the dominant version of English used in the media.} However, cultural differences and language nuances can occasionally lead to misunderstandings or a sense of disconnect, even when an agent's technical language skills are strong.

We measure the proficiency of text in two ways: its \textit{comprehensiblity} and its \textit{native fluency}. The comprehensibility score assesses whether the agent produces text that is cogent and easy to understand, using a scale of 1 to 5, where 1 indicates ``very difficult to comprehend'' and 5 signifies ``very fluent and easily understandable, with no significant errors.''  In contrast, our native fluency focuses on whether the text was likely to have been produced by a native speaker of American English.  Our criteria for native-like fluency are based on the Interagency Language Roundtable ``functionally native'' proficiency standard.  This is also a 5 point scale where 1 indicates that a writer is ``Definitely not a native American English speaker'' and 5 indicates that they definitely are.  For instance, ``I could care less'' is grammatically incorrect, but a common English-language expression. On the other hand, Filipino agents often use the greeting ``to have a blessed day,'' which is grammatically correct, but not a common greeting in the United States.  We use Gemini, an LLM, to score agents' text in each conversation. For more information on our specific approach, prompts, and validation tests, see Appendix \ref{asec:language_scoring}. 

\begin{figure}[ht!]
\begin{center}
\captionsetup{justification=centering}
\caption{\textsc{Figure \ref{fig:englishfluency}: Impact of AI on Language Skills}}
\makebox[\linewidth]{
    \begin{tabular}{cc}
    \textsc{A.  Comprehensibility Score} &  \textsc{B.  Native Fluency Score}\\
    \includegraphics[scale=0.5]{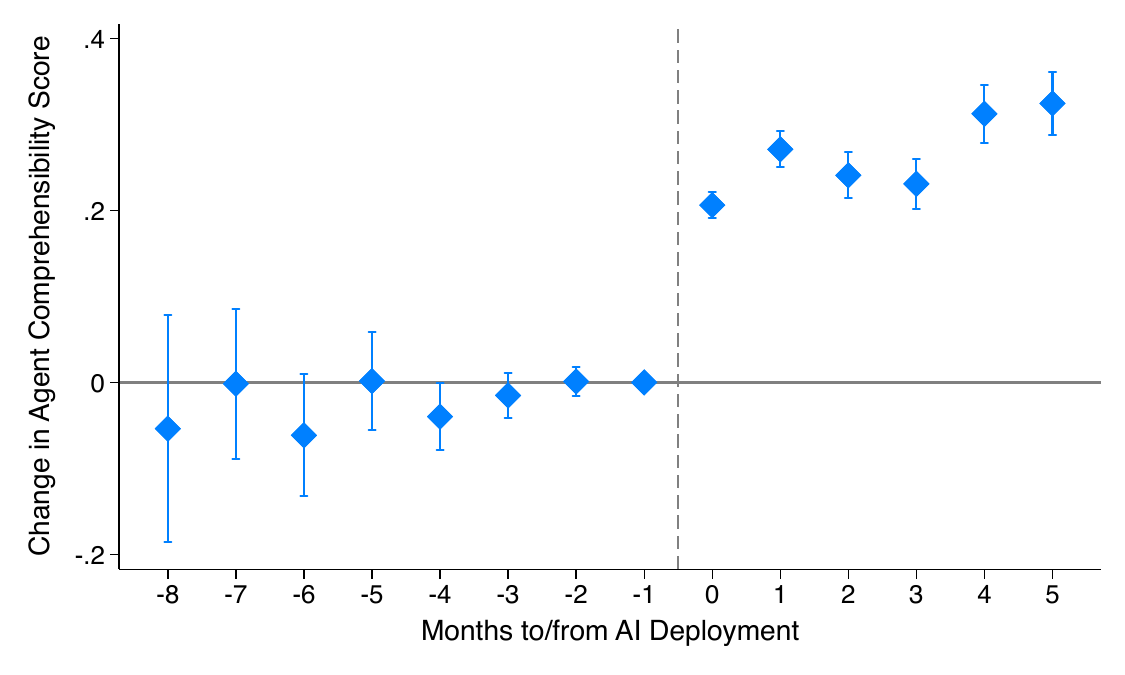} & 
  \includegraphics[scale=0.5]{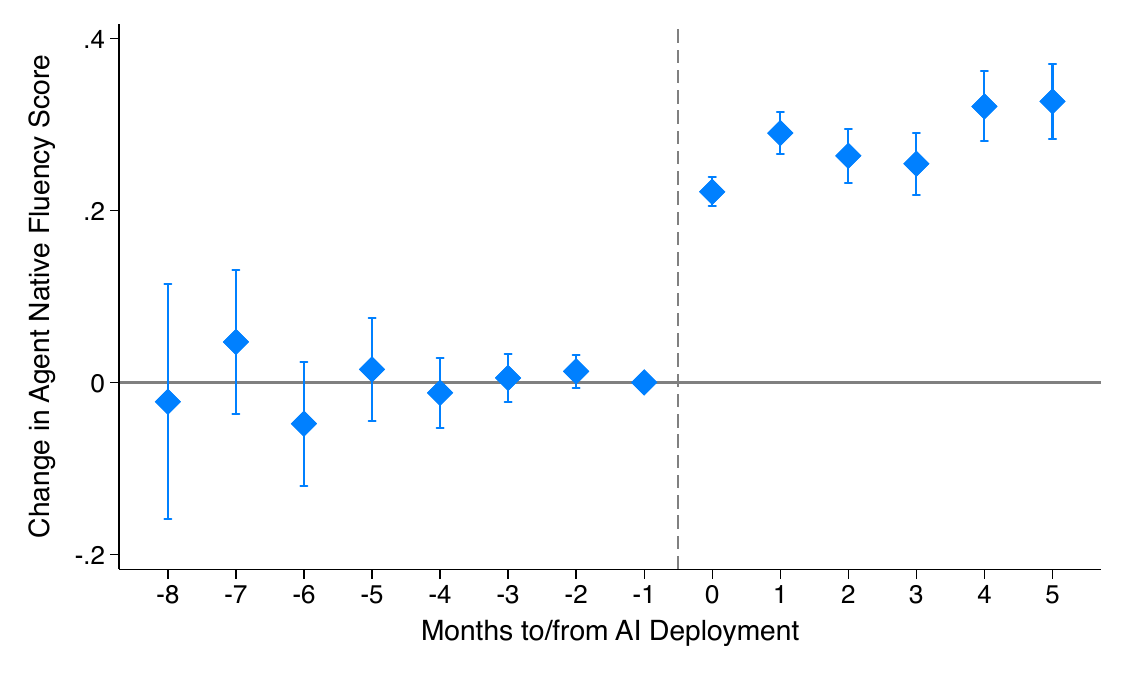}\\
\textsc{C. Comprehensibility Score, by Agent Country} & \textsc{D.  Native Fluency Score, by Agent Country}\\
\includegraphics[scale=0.5]{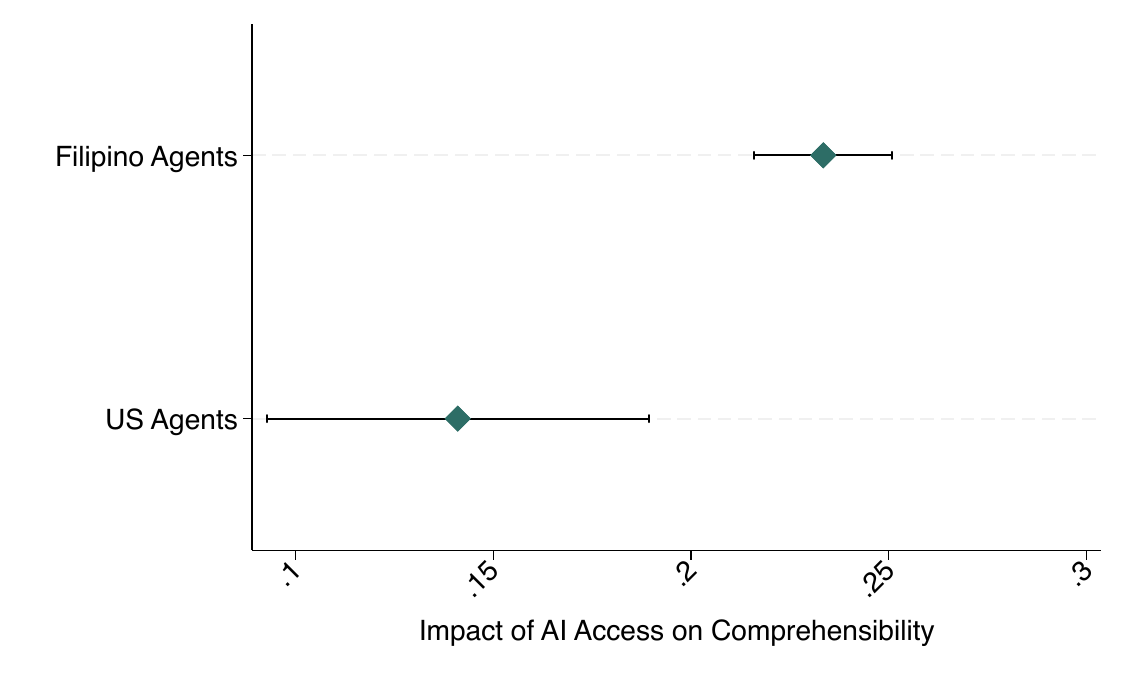} & \includegraphics[scale=0.5]{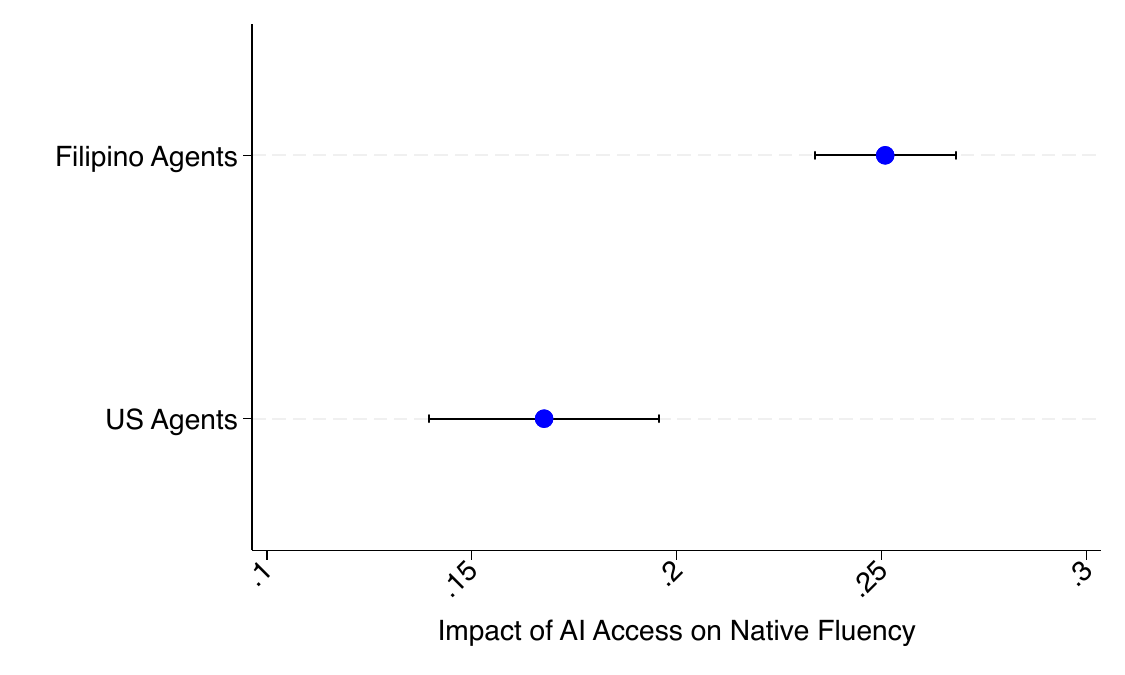} 
\end{tabular}
}
\label{fig:englishfluency} 
\end{center}
\end{figure}

First, we note that the general level of both comprehensibility and native fluency is high: prior to having AI access, the interquartile range of comprehensibility scores was 4.28 to 4.67; for native fluency it was 4.26 to 4.65.  Despite this high baseline level, we find clear evidence that access to AI assistance increases proficiency scores in our data.  This can be seen in Appendix Figure \ref{afig:englishfluency1} which simply presents the raw pre- and post-AI distribution of comprehensibility and native fluency scores for never-treated workers, pre-treatment workers, and post-treatment workers.  The never treated and pre-treatment workers have identical distributions but that we see markedly higher scores for post-treatment workers.  In Panels A and B of Figure \ref{fig:englishfluency}, we report the accompanying event studies: we see a marked and significant improvement in both comprehensiblity and native fluency.  Finally, in Panels C and D of Figure \ref{fig:englishfluency}, we report separate coefficients for US- and Philippines-based workers.  We see a positive impact for all workers, but a larger improvement for workers based in the Philippines.

\subsubsection{Textual Convergence}

The analysis above focuses on an important, but narrow, aspect of how workers communicate. To gain a broader understanding of AI's influence on communication patterns, we examine how the text produced by workers evolves over time: do they change how they write relative to their pre-AI baseline, and does AI access impact the relative communication patterns of high and low skill workers?
Because tacit knowledge is, by definition, not something that can be codified as a set of rules, we examine the overall textual similarity of conversations using textual embeddings, rather than looking for the presence of specific formulaic phrases \citep{sentence2023}.\footnote{For more details on this process, see Appendix Section \ref{asec:key_vars}.} 


Panel A of Appendix Figure \ref{afig:text_change} plots the evolution of agents' communication over time, before and after access to AI assistance. We compute the cosine similarity of agents' text in each given event-time week to their own chats from the month before AI deployment (week -4 to week -1). Cosine similarity runs from 0 to 1, with 0 meaning that two pieces of text are orthogonal (when represented as semantic vectors), and 1 indicating exact semantic similarity.  Prior to the deployment of AI, the similarity between a worker's own text from month to month is stable, at 0.67, which reflects consistency in an individual agent's language use, while also capturing differences in the topics and customers that she faces.  However, following AI deployment, the similarity of agents' text drops. The drop is equivalent to about half of a standard deviation of within-agent cosine similarity across the pre-period. This is consistent with the idea that AI assistance changes the content of agents' messages, rather than merely leading workers to type the same content but faster.  Panel B of Figure \ref{afig:text_change} plots the average change in textual content separately by pre-AI worker skill. Lower-skill agents experience greater textual change after AI adoption, relative to top performers. 


Next, we examine across-worker changes in communication, focusing on the gap between high and low skill workers.  Panel C of Appendix Figure \ref{afig:text_change} plots the cosine similarity between high- and low-skill agents at specific moments in calendar time, separately, for workers without (blue dots) and with (red diamonds) access to AI assistance.\footnote{For non-AI users, we define skill levels based on monthly quintiles of our skill index. For AI users, we use skill quintiles at the time of AI deployment.}  Without AI, high- and low-productivity workers show a moderate level of similarity in their language use, with an average cosine similarity between high and low workers of 0.55.  This similarity remains stable over time, suggesting that there are no divergent trends between skill groups that do not have access to AI assistance.  Post-AI adoption, however, text similarity between high and slow skill workers begins increasing, from 0.55 to 0.61. While this change may seem modest, it represents a substantial narrowing of language gaps, given that the average similarity of a high-skill worker's own pre and post AI text is only 0.67. The change is equivalent to half of a standard deviation of the average high and low worker textual similarity.  

Taken together, the patterns in Appendix Figure \ref{afig:text_change} are consistent with the idea that AI assistance leads lower-skill workers to change how they communicate more, and that these changes ultimately lead them to communicate more like high skill workers. We caution, however, that changes in agent text can reflect many factors that are not directly related to a worker's style or tacit skills, such as changes in conversation topics driven by customers.  As a result, we stress that this analysis is only suggestive.

\section{Effects on the Experience of Work}

Qualitative studies suggest that working conditions for contact center agents can be unpleasant.  They are regularly exposed to challenging, emotionally charged and anonymized conversations, and called upon to absorb customer frustrations while limiting one's own emotional reaction \citep{hochschild2019managed}.  The stress associated with this type of emotional labor is often cited as a key cause of burnout and attrition among customer service workers \citep{leePhilippinesHasBecome2015}.  Additionally, contact center work for US-based businesses is frequently outsourced to countries such as India and the Philippines, meaning that agents often work difficult hours and may face cultural barriers or judgements when speaking with customers.  jMoreover, increases in productivity may not improve the work environment, especially if workers feel pressure to work faster. 

In this section, we examine the impact of generative AI on two particularly salient aspects of the workplace experience: how agents are treated by customers and how often they face requests to speak with a manager.  We also examine the impact of AI assistance on worker turnover as an overall indicator of worker satisfaction. 

\subsection{Customer Sentiment}

Customers often vent their frustrations to anonymous service agents and, in our data, we see regular instances of swearing, verbal abuse, and ``yelling'' (typing in all caps).

Access to AI-assistance may impact how customers treat agents, but the direction and magnitude of these impacts are ambiguous.  AI assistance may improve the tenor of conversations by helping agents set customer expectations or resolve their problems more quickly.  Alternatively, customers may become more frustrated if AI-suggested language feels ``corporate'' or insincere.  

To assess this, we capture the affective nature of both agent and customer text, using sentiment analysis \citep{mejovaSentimentAnalysisOverview}. For this analysis, we use SiEBERT, an LLM that is fine-tuned for sentiment analysis using a variety of datasets, including product reviews and tweets \citep{hartmann2023}.  Sentiment is measured on a scale from $-1$ to $1$, where $-1$ indicates negative sentiment and 1 indicates positive.  In a given conversation, we compute separate sentiment scores for both agent and customer text.  We then aggregate these chat-level variables into a measure of average agent sentiment and average customer sentiment for each agent-year-month.

Panel A and B of Appendix Figure \ref{afig:expwork} shows the distribution of customer sentiment scores.  On average, customer sentiments in our data are mildly positive and normally distributed around a mean of 0.14, except for a mass of very positive and very negative scores.  Panel B shows the distribution of sentiments associated with agents: agents are unfailingly positive, with a mean sentiment score of 0.89.  This reflects the fact that agents are trained to be extremely polite and friendly, even prior to AI access.

Panels A and B of Figure \ref{fig:experienceofwork} consider how sentiment scores respond following the rollout of AI assistance.  In Panel A, we see an immediate and persistent improvement in customer sentiment.  This effect is economically large: according to Column 1 of Table \ref{tab:sentiment}, access to AI improves the mean customer sentiments (averaged over an agent-month) by 0.18 points, equivalent to half of a standard deviation.  In Panel B, we see no detectable effect for agent sentiment, which is already very high at baseline.  Column 2 of Table \ref{tab:sentiment} indicates that agent sentiments increase by only 0.02 points or about 1\% of a standard deviation.

Focusing on customer sentiment, Panels C and D of Appendix Figure \ref{afig:expwork} examine whether access to AI has different impacts across agents.  We find that access to AI assistance significantly improves how customers treat agents of all skill and experience levels, with the largest effects for agents in the lower to lower-middle range of both the skill and tenure distributions.  Consistent with our productivity results, the highest-performing and most-experienced agents see the smallest benefits of AI access.  These results suggest that AI recommendations, which were explicitly designed to prioritize more empathetic responses, may improve agents' demonstrated social skills and have a positive emotional impact on customers.  

\subsection{Customer Confidence and Managerial Escalation}
Changes in individual productivity may have broader implications for organizational workflows \citep{garicano_hierarchies_2000, athey_allocation_1994, atheystern_1998}.  
In most customer service settings, front-line agents attempt to resolve customer problems but can seek the help of supervisors when they are unsure how to proceed.  Customers, knowing this, will sometimes attempt to escalate a conversation by asking to speak to a manager.  This type of request generally occurs when the customer becomes frustrated and feels that the current agent is not equipped to address their problem.  

\begin{figure}[ht!]
	\begin{center}
		\captionsetup{justification=centering}
		\caption{\textsc{Figure \ref{fig:experienceofwork}: Experience of Work}}		
		\makebox[\linewidth]{
			\begin{tabular}{cc}
				\textsc{A.  Customer Sentiment, Event Study}  \\
    \includegraphics[scale=0.45]{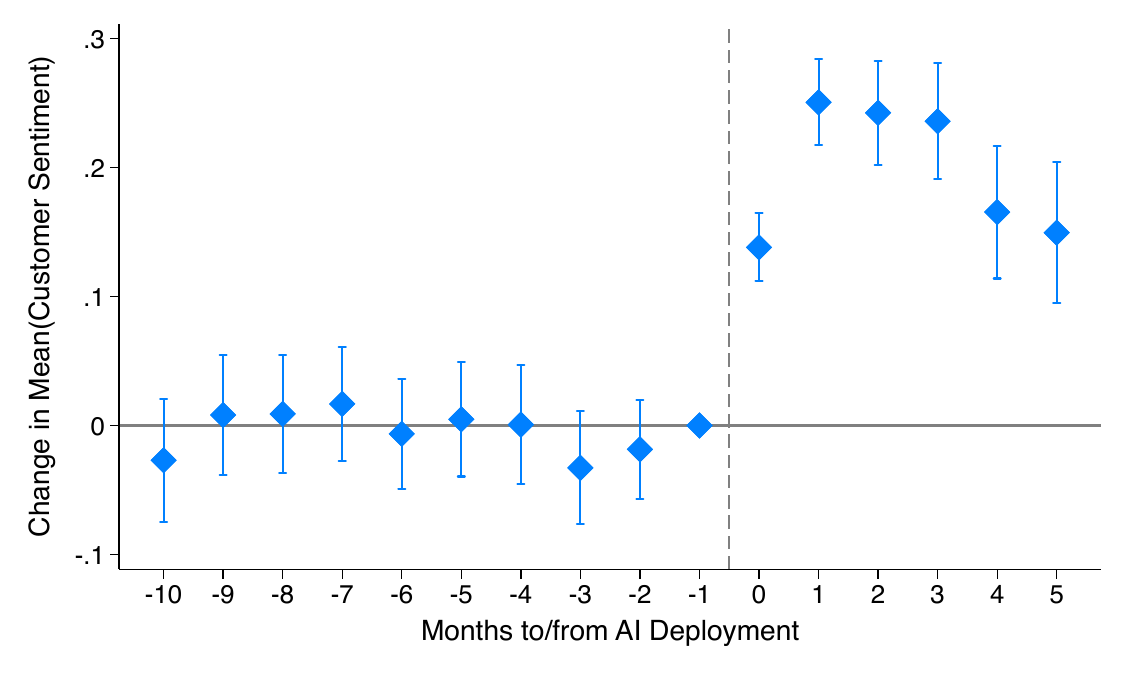} \\
     	\textsc{B.  Agent Sentiment, Event Study}\\
				\includegraphics[scale=0.45]{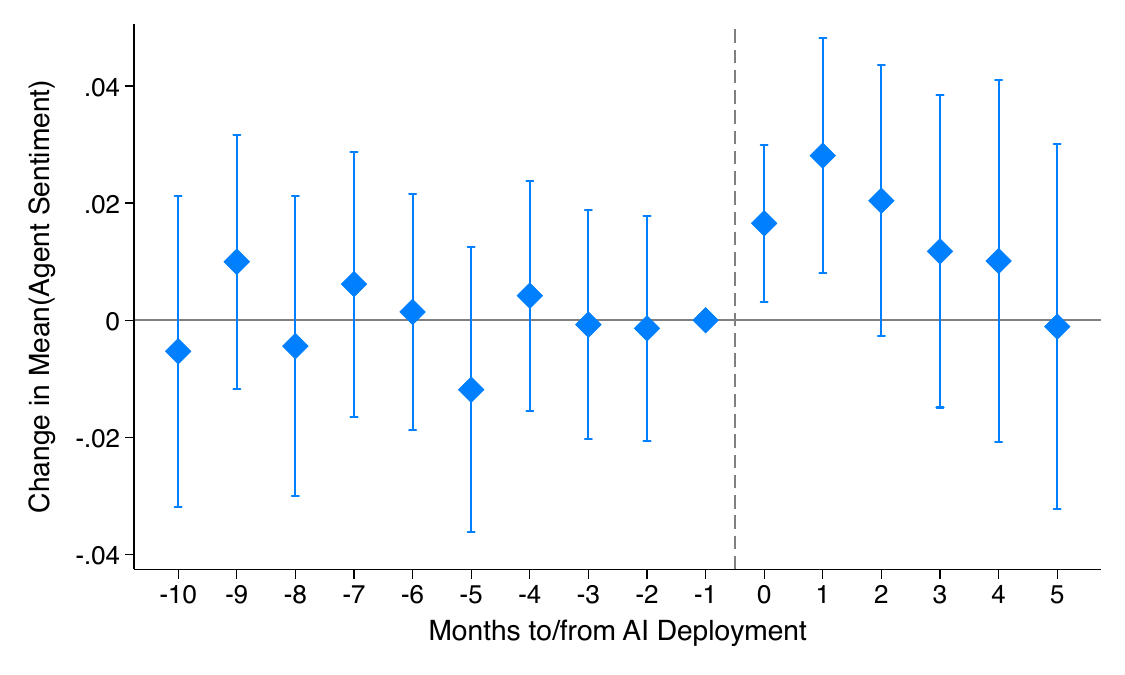} \\
    \textsc{C.  Manager Escalation, Event Study}\\
	\includegraphics[scale=0.45]{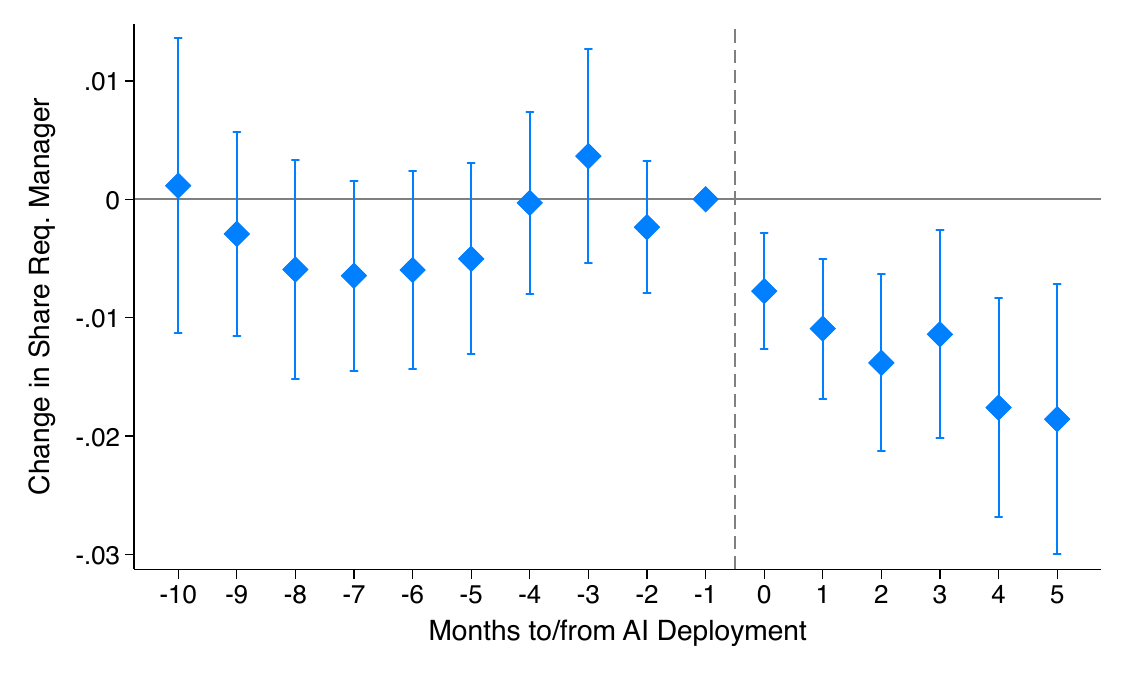}		
			\end{tabular}
		}
		
		\label{fig:experienceofwork}
	\end{center}
\end{figure}

In Panel C of Figure \ref{fig:experienceofwork}, we consider the impact of AI-assistance on the frequency of chat escalation.  The outcome variable we focus on is the share of an agent's chats in which a customer requests to speak to a manager or supervisor, aggregated to the year-month level.  We focus on requests for escalation rather than actual escalations both because we lack data on actual escalations and because requests are a better measure of customer confidence in an agent's competence or authority.  Following the introduction of AI assistance, we see a gradual decline in requests for escalation.  Relative to a baseline rate of approximately 6 percentage points, these coefficients suggest that AI assistance generates an almost 25\% decline in customer requests to speak to a manager.  In Panels E and F of Appendix Figure \ref{afig:expwork}, we consider how these patterns change depending on the skill and experience of the worker.  While these results are relatively noisy, our point estimates suggest that requests for escalation are disproportionately reduced for agents who were less skilled or less experienced at the time of AI adoption.  

\subsection{Attrition}

The adoption of generative AI tools can affect workers in various ways, including their productivity, the level of stress they experience, how customers perceive them, and their overall job satisfaction. While we cannot directly observe all these factors, we can analyze turnover patterns as a broad measure of how workers respond to AI implementation.

In this analysis, we compare attrition rates between AI-assisted agents and untreated agents with equal tenure. We drop observations for treated agents before treatment because they do not experience attrition by construction (they must survive to be treated in the future), and control for location and time fixed effects.\footnote{See Appendix Section \ref{asec:specifications} for the exact specification.}
 
\begin{figure}[ht!]
\begin{center}
\captionsetup{justification=centering}
\caption{\textsc{Figure \ref{fig:attrition}: Impact of AI Model Deployment on Worker Attrition}}

\makebox[\linewidth]{
\begin{tabular}{c}
\textsc{A.  By Productivity at AI Model Deployment}\\
\includegraphics[scale=0.65]{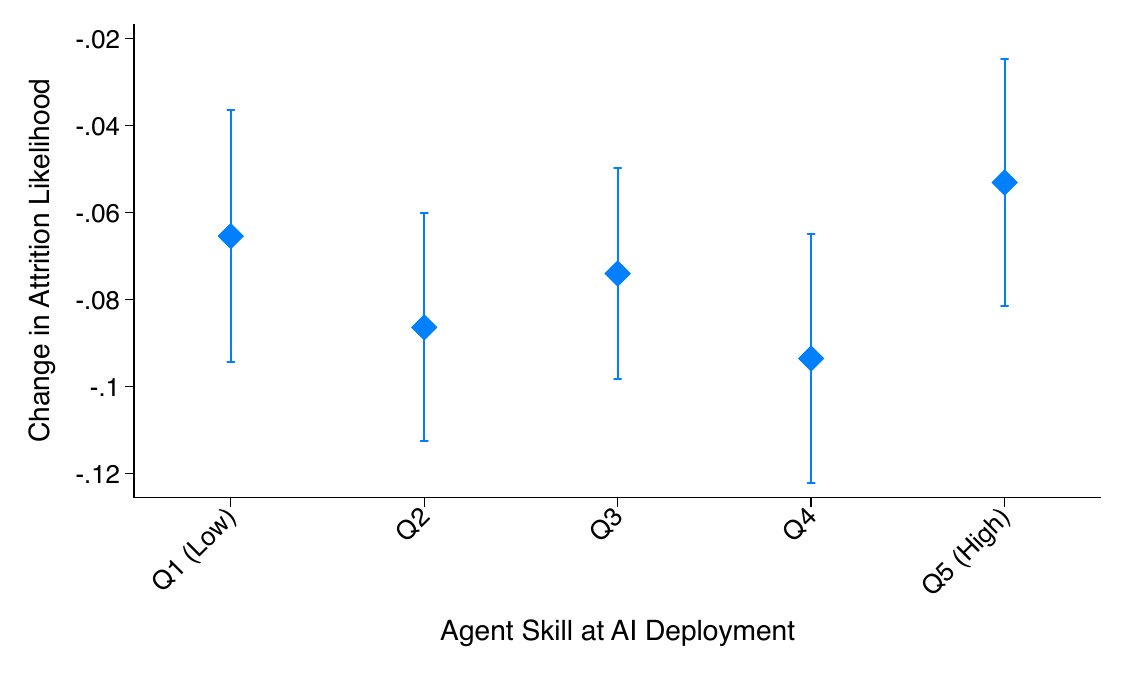} \\
\textsc{B.  By Tenure at AI Model Deployment}\\
\includegraphics[scale=0.65]{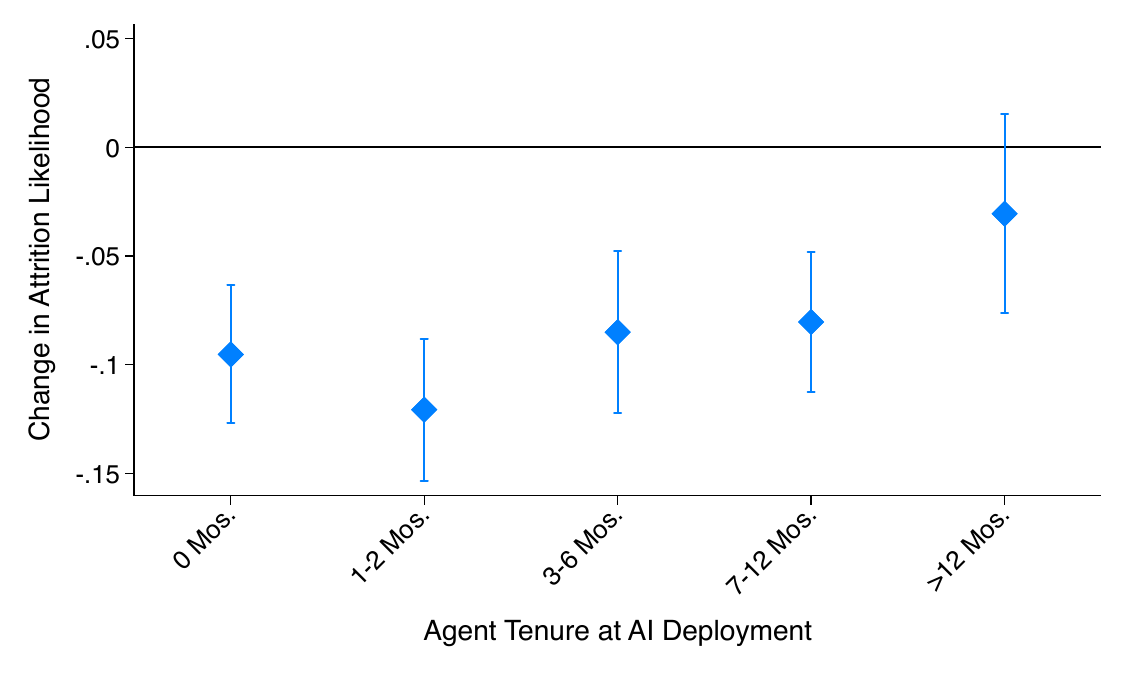} \\
\end{tabular}
}

\label{fig:attrition}
\end{center}
\end{figure}

Consistent with our findings so far, Panel A of Figure \ref{fig:attrition} shows that access to AI assistance is associated with the strongest reductions in attrition among newer agents, those with less than 6 months of experience.  The magnitude of this coefficient, around 10 percentage points, translates into a 40\% decrease relative to a baseline attrition rate in this group of 25\%.   In Panel B, we examine attrition by worker skill.  Here, we find a significant decrease in attrition for all skill groups, although without a clear gradient.  

These results should be taken with more caution relative to our main results because we are unable to include agent fixed effects because attrition only occurs once per worker. Our results may overstate the impact of AI access on attrition if, for example, the firm is more likely to give AI access to agents deemed more likely to stay.

\section{Conclusion}




Advancements in AI technologies open up a broad set of economic possibilities.  Our paper provides the first empirical evidence on the effects of a generative AI tool in a real-world workplace. In our setting, we find that access to AI-generated recommendations increases overall worker productivity by 15\%, with even larger impacts for lower-skill and novice agents.  These productivity gains in part reflect durable worker learning rather than rote reliance on AI suggestions.  Furthermore, AI assistance appears to improve worker's on the job experiences, such as by improving customer sentiment and confidence, and is associated with reductions in turnover.

Our analysis is subject to some caveats and raises many unanswered questions.

First, we note again that our findings apply for a particular AI tool, used in a single firm, within a single occupation, and should not be generalized across all occupations and AI systems.  For example, our setting has a relatively stable product and set of technical support questions. In areas where the product or environment is changing rapidly, the relative value of AI recommendations may be different: they may be better able to synthesize changing best practices, or they may actually impede learning by promoting outdated practices observed in historical training data.  Indeed recent work  by \citet{otisUnevenImpactGenerative2023} and \citet{Perry2022DoUW} have found instances in which AI adoption has limited or even negative effects.  

Second, we report partial equilibrium short to medium run impacts of AI deployment.  While we do not have access to pay data, the managers we spoke to believed that workers may have received higher performance pay as a result of AI assistance, since these bonuses were typically tied to targets related to average handle time and resolution rates. They caution, however, that potential gains in bonus pay may not be long lived because it is common practice to adjust performance targets if too many agents were hitting the goals.  As a result, workers may eventually be subject to a ratchet effect if AI assistance leads performance targets to be readjusted upwards.  

More generally, we are not able to observe longer run equilibrium responses in worker demand or job design.  If customer demand for assistance is inelastic, then the productivity gains we document will likely translate into less demand for human labor. A back-of-the-envelope calculation suggests the firm could field the same number of customer support issues with 12\% fewer worker-hours.  Conversely, individuals may currently avoid contacting customer service because of the long wait times and low-quality service.  AI assistance that improves this experience may boost consumer demand for product support, resulting in increased labor demand \citep{berg2018, korinekantonHowInnovationAffects2023}.  It is also possible that the use of AI could create new jobs for customer service agents, such as testing and training AI models \citep{autornewwork2022}.  One manager we spoke with reports that, in some contact centers, high-skill workers are already being tasked with reviewing AI-suggestions and providing better alternatives.  Other work shows that even low levels of AI adoption can impact market equilibrium prices and quantities, highlighting the need for more work on the equilibrium effects of AI on the labor market \citep{raymondjmp}.

Furthermore, the eventual impacts of AI tools on workers, wages, and aggregate productivity effects will depend on context-specific factors and the development of complementary organizational practices and educational policies \citep{brynjcurve}.  For example, \citet{humlumAdoptionChatGPT2024} shows that workers' willingness to use AI depends critically on access to training and on regulations such as those related to data confidentiality.  

Finally, our findings also raise questions about the nature of worker productivity.  Traditionally, a support agent's productivity refers to their ability to help the customers they come in contact with.  Yet, in a setting where customer service conversations are fed into training datasets, a worker's productivity also includes their ability to provide ML models with examples of successful behaviors that can be shared with others.  Top performers, in particular, contribute many of the examples used to train the AI system we study, but our results suggest that access to AI suggestions may lead them to put too little effort into coming up with new solutions.  Going forward, compensation polices that provide incentives for workers to contribute to model training could be important. Given the early stage of generative AI, these and other questions deserve further scrutiny.  

\clearpage
\begin{spacing}{0.8}
\bibliography{AugI_Bibliography.bib}
\end{spacing}




\clearpage
\begin{footnotesize} 
\begin{singlespace}
\noindent \textsc{Notes}:  This figure shows the distribution various outcome measures.  We split this sample into agent-month observations for agents who eventually receive access to the AI system before deployment (``Pre AI''), after deployment (``Post AI''), and for agent-months associated with agents who never receive access (``Never AI'').  Our primary productivity measure is ``resolutions per hour,'' the number of customer issues the agent is able to successfully resolve per hour.  We also provide descriptives for ``average handle time,'' the average length of time an agent takes to finish a chat; ``chats per hour,'' the number of chats completed per hour incorporating multitasking; ``resolution rate,''  the share of conversations that the agent is able to resolve successfully; and ``net promoter score'' (NPS), which are calculated by randomly surveying customers after a chat and calculating the percentage of customers who would recommend an agent minus the percentage who would not. All data comes from the firm's software systems. 
\end{singlespace}
\end{footnotesize}

\clearpage

\begin{footnotesize} 
\begin{singlespace}
\noindent \textsc{Notes}: This figure plot the coefficients and 95\% confidence intervals from event study regressions of AI model deployment on our measure of productivity, resolutions per hour, using the \citet{sun_estimating_2020} interaction weighted estimator. Our specification follows Equation \ref{eq:dd_main} and includes fixed effects for agent, chat year-month and agent tenure in months.  Observations are at the agent-month level, which is the most granular level at which resolutions per hour is available. Robust standard errors are clustered at the agent level. Section \ref{sec:rollout} describes the rollout and Appendix Section \ref{asec:specifications} outlines the regression specification.  

\end{singlespace}
\end{footnotesize}

\clearpage

\begin{footnotesize} 
\begin{singlespace}
\noindent \textsc{Notes}:  These figures plot the impacts of AI model deployment on resolutions per hour for agents grouped by pre-AI skill and experience. For Panel A, within each month and company (e.g. the employer of an agent), agents are grouped into quintiles of a trailing three month performance index incorporating handle time, call resolution, and customer satisfaction, with the most productive agents in quintile 5 and the least productive in quintile 1. In Panel B, pre-AI worker tenure is the number of months an agent has been employed when they receive access to AI recommendations. Our regression specifications, outlined in Appendix Section \ref{asec:specifications}, include fixed effects for agent, chat year-month and agent tenure in months, for Panel A, and agent skill at deployment in Panel B. The month of AI model deployment is the month that AI output is turned on for each agent. We estimate these regressions with OLS and cluster standard errors at the agent level. Section \ref{sec:rollout} describes the rollout, Appendix Section \ref{asec:key_vars} explains construction of key variables and includes detailed regression specifications. 
\end{singlespace}
\end{footnotesize}


\clearpage

\begin{footnotesize} 
\begin{singlespace}
\noindent \textsc{Notes}:  This figure plot the relationship between productivity and job tenure. The short dashed red line plots the performance of always-treated agents, those who have access to AI assistance from their first month on the job. The long dashed blue line plots agents who are never treated. The solid green line plots agents who spend their first four months of work without the AI assistance, and gain access to the AI model during their fifth month on the job.  95\% confidence intervals are shown. Observations are at the agent-month level. 
\end{singlespace}
\end{footnotesize}


\clearpage

\begin{footnotesize} 
\begin{singlespace}
\noindent \textsc{Notes}: Panel A plots the distribution of AI adherence, averaged at the agent-month level, weighted by the log of the number of AI recommendations for that agent-month. Panel B shows the impact of AI assistance on resolutions by hour, by agents grouped by their initial adherence, defined as the share of AI recommendations they followed in the first month of treatment. The regression, outlined in Appendix Section \ref{asec:specifications}, is run at the agent-month level and includes fixed effects for agent, chat year-month and agent tenure in months. Standard errors are clustered at the agent level. 
\end{singlespace}
\end{footnotesize}

\clearpage
\begin{figure}[ht!]
\begin{center}
\captionsetup{justification=centering}
\caption{\textsc{Figure \ref{fig:ES_outage}: Chat Duration during AI system Outages}}	
\makebox[\linewidth]{
\begin{tabular}{cc}

\textsc{A.  Post-treatment Non-Outage Periods} & \textsc{B.  Post-treatment Outage Periods}\\
\includegraphics[scale=0.45]{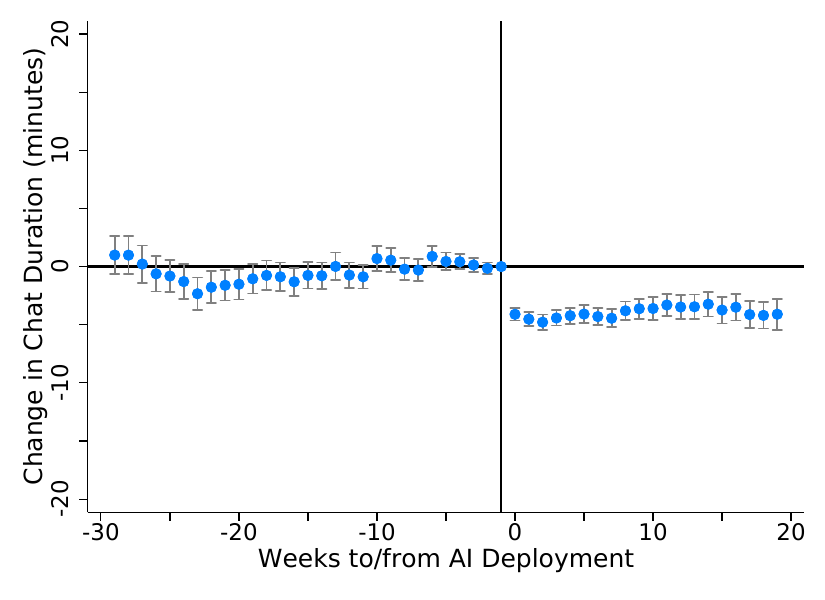} &
\includegraphics[scale=0.45]{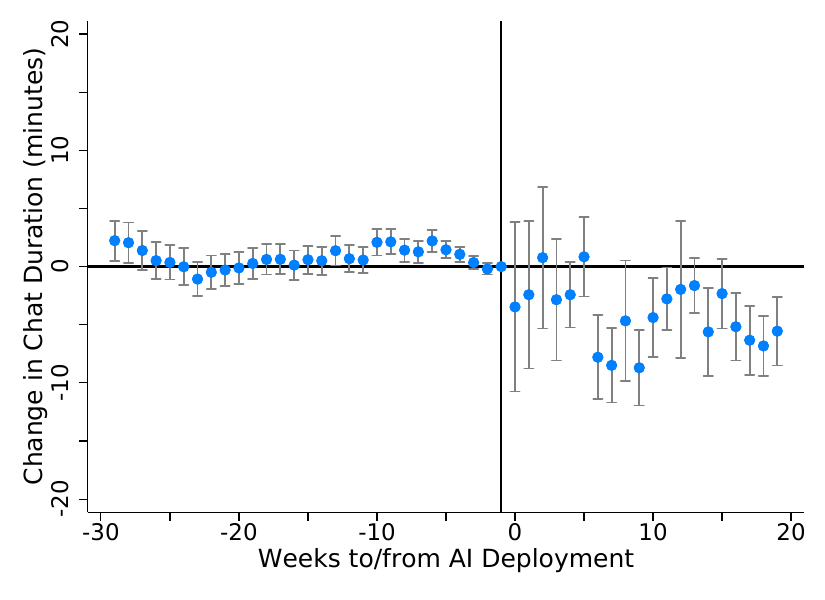} \\

\textsc{C.  Outage--High Initial Adherence} & \textsc{D.  Outage--Low Initial Adherence} \\
\includegraphics[scale=0.53]{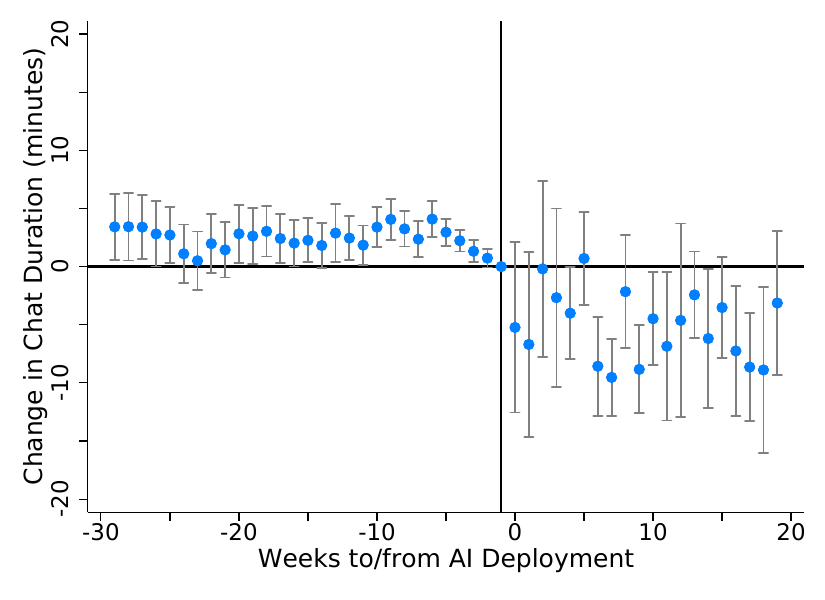} & \includegraphics[scale=0.53]{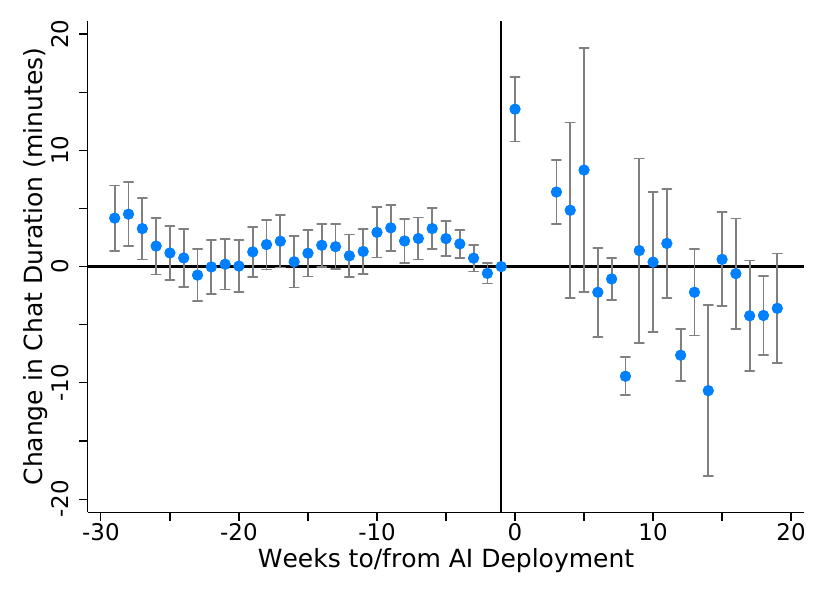} \\

\end{tabular}
}
\label{fig:ES_outage}
	\end{center}
\end{figure}

\begin{footnotesize} 
\begin{singlespace}
\noindent \textsc{Notes}: These figures plot event studies for the impact of AI system rollout of chat duration at the individual chat level.  Panel A restricts to post-treatment chats that do not occur during any period where there is a AI system outage.  Panel B restricts to post-treatment chats that only occur during a large system outage.  Panels C and D focus on outage only post-periods.  Panel C restricts to only chats generated by ever-treated agents who with high initial AI adherence (top tercile) while Panel D restricts to agents with low initial adherence (bottom tercile). Agents who are never treated are excluded from this analysis. The regressions are run at the chat level with agent, year-month and tenure fixed effects with standard errors clustered at the agent level.    
\end{singlespace}
\end{footnotesize}


\clearpage

\begin{footnotesize} 
\begin{singlespace}
\noindent \textsc{Notes}:  Panel A shows the impact of AI assistance on chat duration in minutes for each category of conversation topic frequency. Data is at the chat level and the regressions control for topic frequency, year-month fixed effects, agent fixed effects, and fixed effects in months of agent tenure. Panel B shows the impact of AI assistance grouped instead by topic frequency as encountered by the individual agent. Appendix Section \ref{asec:specifications} details the regression specification and topic category construction.
\end{singlespace}
\end{footnotesize}


\clearpage

\begin{footnotesize} 
\begin{singlespace}
\noindent \textsc{Notes}: These figures show the impact of AI access on scores of agent comprehensibility in Panel A and native fluency in Panel B. Observations for this regression are at the agent-chat level, aggregate to the agent-month level. Regressions follow Equation \ref{eq:dd_main} and include agent, chat year-month and months of agent tenure fixed effects. Robust standard errors are clustered at the agent level in Panels A and B and agent location in Panels C and D. For more details on construction of the comprehensibility and native fluency scores, refer to Appendix Section \ref{asec:key_vars}.  
\end{singlespace}
\end{footnotesize}


\clearpage
\begin{figure}[ht!]
	\begin{center}
		\captionsetup{justification=centering}
		\caption{\textsc{Figure \ref{afig:text_change}: Within Agent Textual Analysis}}		
\makebox[\linewidth]{
\begin{tabular}{c}
\textsc{A.  Within-Person Textual Similarity to Month Prior to AI} \\
\includegraphics[scale=0.45]{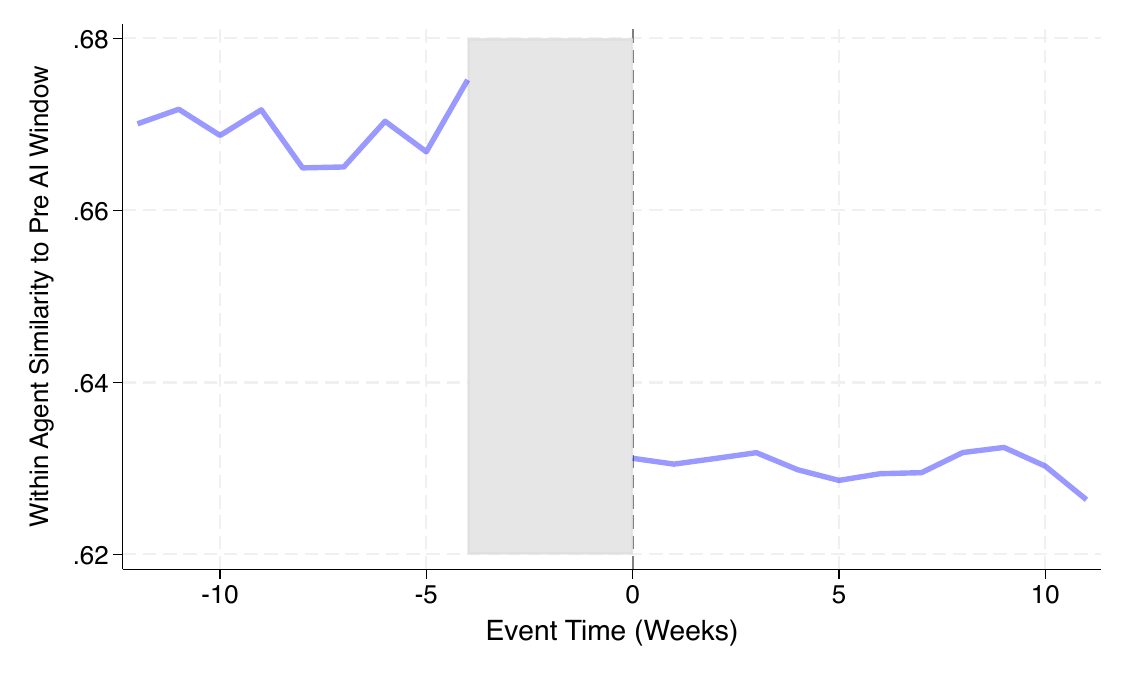} \\
\textsc{B.  Within-Person Textual Change, Low vs. High Skill} \\
\includegraphics[scale=0.55]{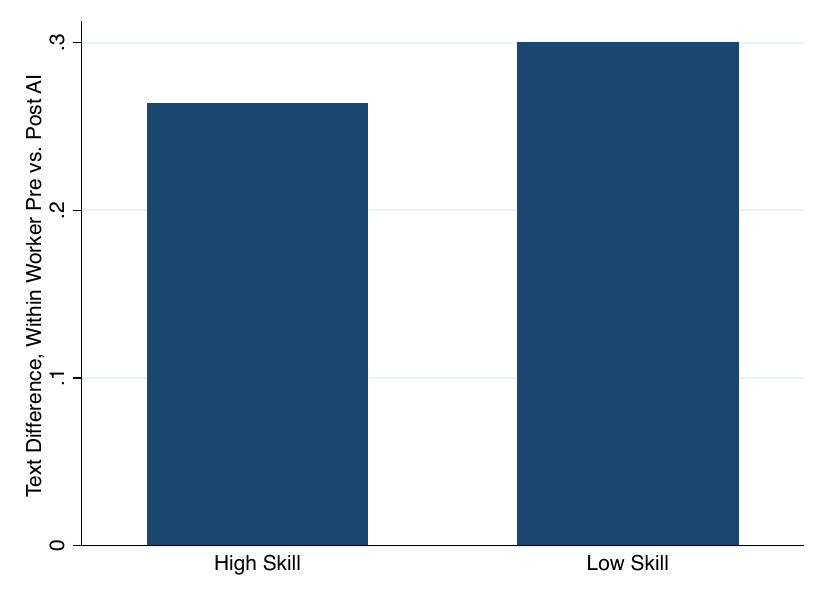} \\
\textsc{C.  Text Similarity Between Low-Skill and High-Skill Workers, Pre and Post AI} \\
\includegraphics[scale=.5]{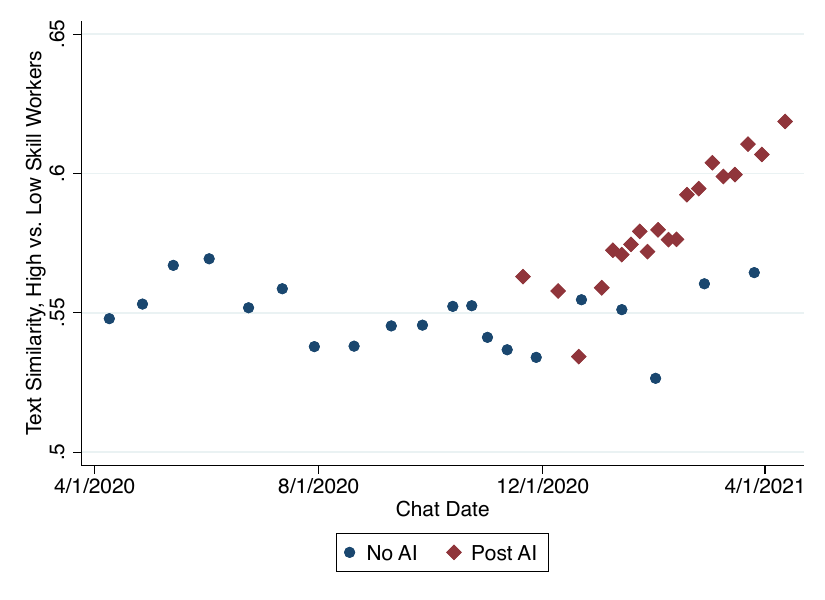}
\end{tabular}
}
		\label{afig:text_change}
	\end{center}
\end{figure}
\vspace{-10pt}
\begin{footnotesize} 
\begin{singlespace}
\noindent \textsc{Notes}: Panel A plots the average similarity between an agent's chats each week and a comparison group of their conversations in the month prior to AI deployment (the gray section). Panel B plots the average difference between an agent's pre-AI corpus of chat messages and that same agent's post-AI corpus, controlling for year-month and agent tenure.  The first bar represents the average pre-post text difference for agents in the highest quintile of pre-AI skill and the second bar represents those in the bottom quintile.  Panel C plots the average text similarity between the top and bottom quintile of agents. The blue line plots the similarity for never treated or pre-treatment agents, the red line plots the similarity for agents with access to the AI model. For agents in the treatment group, we define agent skill at AI model deployment.
\end{singlespace}
\end{footnotesize}


\clearpage

\begin{footnotesize} 
	\begin{singlespace}
		\noindent \textsc{Notes}: Each panel of this figure plots the impact of AI model deployment on the experience of work. Panel A plots the impact of AI model deployment on customer sentiment, Panel B plots the corresponding estimate for agent sentiment, and Panel C show the impacts of AI assistance on customer requests for manager assistance.  Sentiment is measured using SiEBERT, a fine-tuned checkpoint of a RoBERTA, an English language transformer model. Regressions follow Equation \ref{eq:dd_main} and include agent, chat year-month and months of agent tenure fixed effects. Observations are at the chat-level, aggregated to the agent-month and robust standard errors are clustered at the agent level.   
	\end{singlespace}
\end{footnotesize}


\clearpage

\begin{footnotesize} 
\begin{singlespace}
\noindent \textsc{Notes}:  This figure presents the results of the impact of AI model deployment on workers' likelihood of attrition.  Panel A plots the same impact by agent skill index at AI model deployment.  Panel B graphs the effects of AI assistance on attrition by agent tenure at AI model deployment.   All specifications include chat year and month fixed effects, as well as agent location, company and agent tenure.  Observations for these regressions, detailed in \ref{asec:specifications}, are at the agent-month and all robust standard errors are clustered at the agent level.  
\end{singlespace}
\end{footnotesize}


\clearpage

\begin{table}[ht!]
\begin{center}
                \caption{\textsc{Table \ref{sumstat_agents}: Summary Statistics for the Sample of Customer-Service Agents}}
                  \vspace{20pt}        
\scalebox{1}{\makebox[\linewidth]{\begin{tabular}{lcccc} \toprule \toprule
Variable & All & Never Treated & Treated, Pre & Treated, Post \\
\addlinespace \hline \addlinespace

\hspace{3mm} Chats   &    3,006,395 &      944,848  &      881,101 &    1,180,446  \\
\hspace{3mm} Agents   &        5,172 &        3,517  &        1,340&        1,636   \\
\hspace{3mm} Number of Teams   &          133 &          111  &           80 &           81  \\
\hspace{3mm} Share US Agents  &          .11 &          .15  &         .081  &         .072  \\
\hspace{3mm} Distinct Locations   &           25 &           25  &           18 &           17  \\
\hspace{3mm} Average Chats per Month   &          128 &           83  &          147 &          188  \\
\hspace{3mm} Average Handle Time (Min)   &           41 &           43  &           43  &           35 \\
\hspace{3mm} St. Average Handle Time (Min)   &           23 &           24  &           24  &           22 \\
\hspace{3mm} Resolution Rate   &          .82 &          .78  &          .82 &          .84   \\
\hspace{3mm} Resolutions Per Hour  &          2.1 &          1.7  &            2 &          2.5   \\
\hspace{3mm} Customer Satisfaction (NPS)  &           79 &           78  &           80   &           80 \\
\addlinespace \bottomrule \bottomrule
\addlinespace \bottomrule \bottomrule
\end{tabular}
}}
  \label{sumstat_agents}
\end{center}
\end{table}

\begin{singlespace}
\footnotesize
\noindent \textsc{Notes}: This table shows summary statistics of conversations, agent characteristics and issue resolution rates, customer satisfaction and average call duration.  Column 1 consists of all agents in our sample, Column 2 includes control agents who were never receive AI access. Column 3 presents statistics for treated agents before they receive AI access and Column 4 includes treated agents after AI model deployment. 
\end{singlespace}
\normalsize

 %

\clearpage
\begin{table}[ht!]
\begin{center}
                \caption{\textsc{Table \ref{tab:dd_main}: Main Effects: Productivity (Resolutions per Hour)}}
                  \vspace{20pt}        
\scalebox{1}{\makebox[\linewidth]{\begin{tabular}{lccc} \hline
 & (1) & (2) & (3) \\
VARIABLES & Resolutions/Hr & Resolutions/Hr & Resolutions/Hr \\ \hline
 &  &  &  \\
Post AI X Ever Treated & 0.469*** & 0.371*** & 0.301*** \\
 & (0.0325) & (0.0318) & (0.0329) \\
Ever Treated & 0.110** &  &  \\
 & (0.0440) &  &  \\
 &  &  &  \\
Observations & 13,192 & 12,295 & 12,295 \\
R-squared & 0.249 & 0.562 & 0.575 \\
Year Month FE & Yes & Yes & Yes \\
Location FE & Yes & Yes & Yes \\
Agent FE & - & Yes & Yes \\
Agent Tenure FE & - & - & Yes \\
 DV Mean & 2.123 & 2.176 & 2.176 \\ \hline
\multicolumn{4}{c}{ Robust standard errors in parentheses} \\
\multicolumn{4}{c}{ *** p$<$0.01, ** p$<$0.05, * p$<$0.10} \\
\end{tabular}
}}
  \label{tab:dd_main}
\end{center}
\end{table}

\begin{singlespace}
\footnotesize
\noindent \textsc{Notes}:  This table presents the results of difference-in-difference regressions estimating the impact of AI model deployment on our main measure of productivity, resolutions per hour, the number of technical support problems resolved by an agent per hour (resolutions/hour).  Post AI X Ever Treated captures the impact of AI model deployment on resolutions per hour. Column 1 includes agent geographic location and year-by-month fixed effects.  Columns 2 and 3 include agent-level fixed effects, and Column 3, our preferred specification described by Equation \ref{eq:dd_main}, also includes fixed effects that control for months of agent tenure.  Observations for this regression are at the agent-month level and all standard errors are clustered at the agent level. Section \ref{sec:rollout} describes the AI rollout procedure.
\end{singlespace}
\normalsize

\clearpage
\begin{table}[ht!]
\begin{center}
                \caption{\textsc{Table \ref{tab:dd_oth}: Main Effects: Additional Outcomes}}
                  \vspace{20pt}        
\scalebox{1}{\makebox[\linewidth]{\begin{tabular}{lcccc} \hline
 & (1) & (2) & (3) & (4) \\
VARIABLES & AHT & Chats/Hr & Res. Rate & NPS \\ \hline
 &  &  &  &  \\
Post AI X Ever Treated & -3.746*** & 0.365*** & 0.0132 & -0.119 \\
 & (0.369) & (0.0345) & (0.00882) & (0.524) \\
 &  &  &  &  \\
Observations & 21,839 & 21,839 & 12,295 & 12,541 \\
R-squared & 0.591 & 0.563 & 0.371 & 0.526 \\
Year Month FE & Yes & Yes & Yes & Yes \\
Agent FE & Yes & Yes & Yes & Yes \\
Agent Tenure FE & Yes & Yes & Yes & Yes \\
 DV Mean & 40.64 & 2.559 & 0.822 & 79.59 \\ \hline
\multicolumn{5}{c}{ Robust standard errors in parentheses} \\
\multicolumn{5}{c}{ *** p$<$0.01, ** p$<$0.05, * p$<$0.10} \\
\end{tabular}
}}
  \label{tab:dd_oth}
\end{center}
\end{table}

\begin{singlespace}
\footnotesize
\footnotesize
\noindent \textsc{Notes}:  This table presents the results of difference-in-difference regressions estimating the impact of AI model deployment on additional measures of productivity and agent performance.  Post AI X Treated measures the impact of AI model deployment after deployment on treated agents for average handle time (AHT) in Column 1, chats per hour (Chats/Hr), the number of chats an agent handles per hour in Column 2, resolution rate (Res. Rate), the share of technical support problems they can resolve in Column 3 and net promoter score (NPS), an estimate of customer satisfaction in Column 4. Our regression specification, Equation \ref{eq:dd_main}, includes fixed effects for each agent, chat year-month and agent months of tenure. Observations for this regression are at the agent-month level and all standard errors are clustered at the agent level. Section \ref{sec:rollout} describes the AI rollout procedure.
\end{singlespace}
\normalsize

\clearpage
\begin{table}[ht!]
\begin{center}
                \caption{\textsc{Table \ref{tab:sentiment}: Experience of Work}}
                  \vspace{20pt}        
\scalebox{1}{\makebox[\linewidth]{\begin{tabular}{lccc} \hline
 & (1) & (2) & (3) \\
VARIABLES & Mean(Customer Sentiment) & Mean(Agent Sentiment) & Share Req. Manager \\ \hline
 &  &  &  \\
Post AI X Ever Treated & 0.177*** & 0.0198*** & -0.00875*** \\
 & (0.0116) & (0.00599) & (0.00201) \\
 &  &  &  \\
Observations & 21,218 & 21,218 & 21,839 \\
R-squared & 0.485 & 0.596 & 0.482 \\
Year Month FE & Yes & Yes & Yes \\
Agent FE & Yes & Yes & Yes \\
Agent Tenure FE & Yes & Yes & Yes \\
 DV Mean & 0.141 & 0.896 & 0.0377 \\ \hline
\multicolumn{4}{c}{ Robust standard errors in parentheses} \\
\multicolumn{4}{c}{ *** p$<$0.01, ** p$<$0.05, * p$<$0.10} \\
\end{tabular}
}}
  \label{tab:sentiment}
\end{center}
\end{table}

\begin{singlespace}
\footnotesize
\footnotesize
\noindent \textsc{Notes}:  This table presents the results of difference-in-difference regressions estimating the impact of AI model deployment on measures of conversation sentiment and requests to speak to a manager (``Share Req. Manager''). Our regression specification, Equation \ref{eq:dd_main}, includes fixed effects for each agent, chat year-month and months of agent tenure. Observations for these regressions are at the agent-month level and all standard errors are clustered at the agent level. Measures of customer sentiment are created from conversation transcripts using SiBERT and aggregated to the agent-month level. Appendix Section \ref{asec:key_vars} elaborates on sentiment construction and Section \ref{sec:rollout} describes the AI rollout procedure. 
\end{singlespace}
\normalsize

\newpage

\noindent \textbf{\Large{Appendix Materials}}


\newpage

\begin{appendix}
\renewcommand{\thefigure}{A.\arabic{figure}}
\setcounter{figure}{0}
\renewcommand{\thetable}{A.\arabic{table}}
\setcounter{table}{0}

\newpage
\clearpage

\section{Data Appendix}\label{asec:datamodelsappendix}
\subsection{Sample Construction}\label{asec:sample_details}
We begin with 3,006,395 chat conversations conducted by agents employed at our data firm over the period between September 2019 and June 2021. Each chat includes the number of messages in the conversation, the start and end times, and identifiers associated with the conversation and agent. We drop chats with only one agent or one customer message to avoid capturing interactions without meaningful content.

We merge the chat data with a set of internal company datasets and datasets from our AI firm that allow us to track agent information, conversation outcomes, and AI model output. To do this, we use the system-generated chat identifier to merge chat-level information across database systems. We also merge our conversation-level data into a message-level dataset that includes the text of each message and message-level AI output. At the conversation level, the difference between the chat start and end time gives us the chat duration, which is available for each call. We drop the small number of chats that are missing start or end times, not associated with an agent identifier or missing a chat identifier. There are also a small number of chats that remain ``open'' for days because the customer and agents forget to close the chat, so we winsorize call duration at the 99th percentile. 

We aggregate our chat-level dataset to the agent-month level and merge it with agent data. This information includes the firm they are employed by, their location, their manager/team, their tenure at the firm, and the date treated agents are onboarded onto the AI assistant (defined as the date the agent's account on the AI system is created). Some employees work flexible schedules, including part-time or seasonal roles. As a result, we measure performance metrics only for active periods, and an agent's tenure increases solely during months when they are actively handling customer chats. 


\subsection{Construction of Key Variables}\label{asec:key_vars}

\subsubsection{Call Duration, Resolution and Customer Satisfaction}
Our firm and associated subcontractors track call resolution, average call duration, and customer satisfaction at the agent-month level. Customer satisfaction is collected by randomly sending surveys to customers who have interacted with contact center agents. From the survey results, our firm calculates a monthly average agent customer satisfaction score. Not all customers complete these surveys, so customer satisfaction is calculated monthly rather than at the chat level. Call resolution is also calculated at the agent-month level. To calculate call resolution, our data firm also uses an algorithm that incorporates elements of chat text and future interactions between the customer and data firms to calculate a monthly average call resolution score. Although our firm calculates an agent-month average monthly call duration when tracking agent performance, our chat-level dataset allows us to calculate call duration data for each chat in our sample. However, because call resolution is only available on a monthly basis, we report our omnibus productivity measure at the agent-month level.

\subsubsection{Measuring Agent Skill, Firm and Tenure}
We measure the tenure of the agent in months of work experience. The turnover of the contact center is extremely high; Our firm considers agents with over 6 months of experience to be ``very experienced'' while agents during their first two months are considered to be ``in training''. In our regression specifications, we control for agent tenure using a set of fixed effects for each month of agent tenure. We do not count weeks when agents are not actively working due to vacation or management practices like staggered scheduling intended to reduce worker burnout. 

Our regressions also include controls for each agent's firm, which is the company or subcontractor employing each agent. Agents are employed directly by our data firm as well as by four other business process outsourcing firms. For instance, the firm name for an outsourced worker is ``Convergys'', one of our outsourcing firms, while the company of a directly employed call center agent is ``data firm.''

We also include controls for agent location---the physical location where the agent is employed. For example, several of our subcontractor have locations in Cebu City, so many agents are located in Cebu. While many of the US-based call center agents are employed directly by our data firm, the outsourcing firms also employ US-based workers. Contact center agents based in the US tend to be clustered in cities with a low cost of living, such as Hazard, Kentucky, and Reno, Nevada. Outside of the United States, most workers are based in the Philippines. The turnover of workers is high, so contact center firms frequently open and close new locations.

We construct an agent skill index that incorporates call handling speed, issue resolution rates, and customer satisfaction. For each firm, we rank agents based on three monthly performance indicators: average handle time (lower times receive higher ranks), call resolution rate, and customer satisfaction (higher rates receive higher ranks). The agent's rank is calculated within firm; data firm agents are ranked against their peers, while the Convergys agents are compared to other Convergys employees. These rankings are then averaged into a single skill index at the agent-month level. We then categorize agents into quintiles based on their average skill index from the previous quarter. The skill index incorporates call speed and quality, while comparing individuals within the same organization. 

\subsubsection{Productivity}
Our omnibus productivity measure is monthly resolutions per hour. To calculate this, we divide the share of calls resolved (the monthly call resolution score) by the number of calls handled per hour. Although call duration is available for each chat, resolution data is only available at the agent-month level. Consequently, we report our productivity results at the agent-month level. 

\subsubsection{Customer Sentiment}
The text-based nature of customer support sheds light into the on-the-job experience of contact center agents. Although a call may ultimately be resolved or receive a high customer satisfaction score, customers often start conversations feeling stressed, frustrated, or angry. Regularly dealing with complaints and dissatisfied customers is emotionally taxing and contributes to high worker turnover. Conversations also tend to be fairly long, close to 40 minutes, and often involve complex topics such as managing payroll for employees, issues connecting with banks, and calculating taxes. 

To capture the experience of customer-agent interactions, we use natural language processing to capture the sentiment by analyzing the language, context, and emotional content of chat transcript. We employ SiEBERT, an LLM fine-tuned for sentiment analysis on numerous datasets, including product reviews and tweets, with similar syntax and content to chat-based custoemr support. This model has demonstrated robustness to noisy, real-world data, such as social media posts or customer reviews. SiEBERT uses the text surrounding each word to capture word meaning, which as been shown to outperform other methods of sentiment analysis \citep{hartmann2023}.

For each piece of text, the model produces a sentiment score measured on a scale from $-1$ to $1$, where $-1$ indicates negative sentiment and $1$ indicates positive. We separately calculate sentiment scores for agents' text (agent sentiment) and customers' text (customer sentiment). We then aggregate these chat-level variables into measures of average agent sentiment and average customer sentiment for each agent-month.

\subsubsection{Language Comprehensibility and Fluency}\label{asec:language_scoring}
We use Gemini Pro, a large language model (LLM), to measure comprehensibility and native fluency \citep{geminipaper}. Our criteria for native-like fluency are based on the Interagency Language Roundtable (ILR) ``functionally native'' language proficiency standard, adapted for written text. The ILR is an organization comprising federal government agencies that coordinates and shares information on foreign language activities. Functionally Native Proficiency, the highest level on the ILR scale, describes language ability where an individual communicates with complete fluency and precision on all levels normally pertinent to professional needs, displays cultural understanding equivalent to a native speaker and demonstrates the ability to counsel, persuade, and negotiate in the language as effectively as a well-educated native speaker.

Using the prompt below, we ask the LLM to score each agent's text from a scale of 1 to 5 where: 1 = Definitely not a native American English speaker 2 = Probably not a native American English speaker 3 = Uncertain if native American English speaker 4 = Probably a native American English speaker 5 = Definitely a native American English speaker. The full prompt is:

\begin{displayquote} 
You are given a conversation transcript from a customer service agent helping a customer that only includes what the customer service agent writes to the customer. Based on the transcript of agents' conversations, provide a score that determines if the customer service agent is likely a native speaker of American English. For the native speaker assessment, look for traits such as: Correct grammar based on standard American English conventions, use of American English vocabulary, idioms, and phrasing rather than other dialects, natural, fluent-sounding language that does not appear stilted or translated, adherence to American cultural norms and reference points in word choice and examples, the writer can produce written material with the proficiency of a highly educated native speaker, demonstrating a superior command of the language. For each excerpt, provide a score from 1-5 for the native speaker assessment, where: 1 = Definitely not a native American English speaker 2 = Probably not a native American English speaker 3 = Uncertain if native American English speaker 4 = Probably a native American English speaker 5 = Definitely a native American English speaker. Only return your score of native speaker assessment. Do not include an explanation. 
\end{displayquote}

To validate the LLM's scores, we had two independent human reviewers evaluate 100 randomly selected agent conversations. The reviewers did not have access to each other’s scores or the LLM results. The mean score given by the LLM was 4.29, while the average score from the human evaluators was 4.22. The difference between these scores was not statistically significant, with a p-value of 0.22.

The comprehensibility score captures the general fluency and ease of understanding in the responses of the customer service agents, ranging from very difficult to comprehend (1) to very fluent and easily understandable (5). The score assesses the clarity of communication, frequency and impact of errors, and general language proficiency demonstrated in the writing, regardless of whether the writer is a native speaker. A Philippines-based agent may have a very high level of fluency, but may not speak like a native speaker. For instance, a common sign-off is to tell customers to ``have a blessed day.'' While that phrase is not a common colloquial English phrase, it is grammatically correct and fully comprehensible. The full prompt is:

\begin{displayquote} 
You are given a conversation transcript from a customer service agent helping a customer that only includes what the customer service agent writes to the customer. Based on the transcript of agents' conversations,  score the overall fluency of each excerpt from 1-5, regardless of whether it seems to be written by a native speaker, where: 1 = Very difficult to comprehend, multiple errors impeding comprehension 2 = Somewhat difficult to understand, some errors impeding comprehension 3 = Mostly understandable but with some errors 4 = Fluent and understandable with only minor errors that do not impact meaning 5 = Very fluent and easily understandable with no significant errors. Only return your score of fluency. Do not include an explanation.
\end{displayquote}

We conducted a similar validation exercise with a random sample of 100 agent conversations, asking human evaluators to assess the comprehensibility of the agents' speech. The mean comprehensibility score assigned by the LLM was 4.31, while the average score given by human evaluators was 4.40. The difference between these scores was not statistically significant, with a p-value of 0.12.

\subsubsection{Conversation Topic}
Conversations between agents and customers are fairly complex interactions, often discussing details of tax filings, setting up sick leave, firing an employee, or correcting login issues. We use Gemini Pro to capture the topic of each conversation \citep{geminipaper}.

First, we select a random sample of 5,000 conversations. Using the prompt below, we first ask the LLM to define the topic of the conversation in one to three words, following a procedure similar to \citet{choi2024}. Using this list of 5,000 conversation topics, we ask the LLM to group these topics into no more than 50 categories that describe the subject matter of the conversation. Using the LLM and reviewing the conversations, we come up with a list of 50 topic categories and a one or two sentence description associated with each category. The total number of categories was chosen based on conversations with business management. 

\begin{figure}[ht!]
\begin{center}
\captionsetup{justification=centering}
\caption{\textsc{Figure \ref{hist_topic}: Distribution of Conversational Topics}}
\makebox[\linewidth]{
\begin{tabular}{c}
\textsc{A. Distribution of Conversational Topics} \\
\includegraphics[scale=0.4]{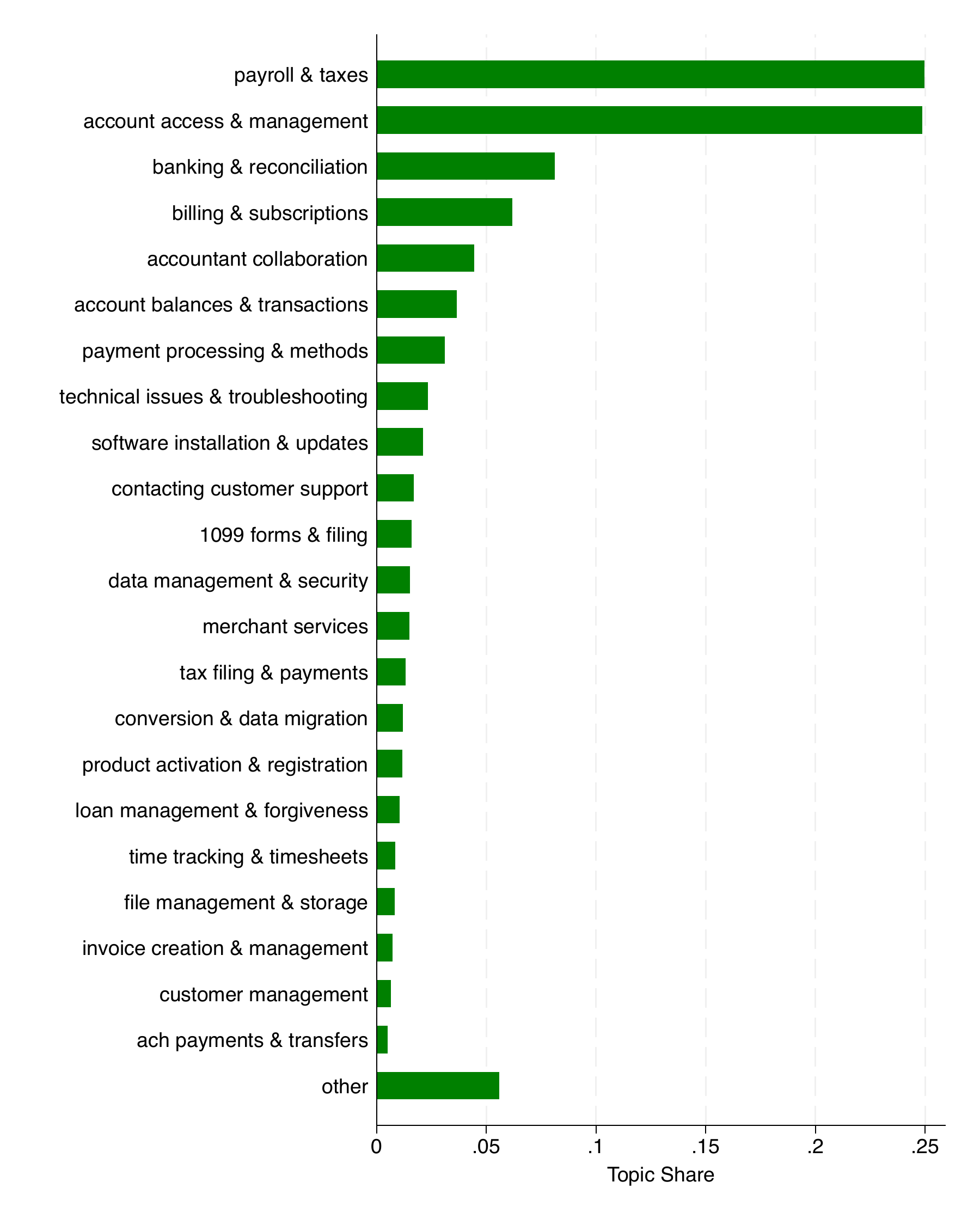} \\
\\
\end{tabular}
}
\label{hist_topic}
\end{center}
\end{figure}

We then use the LLM to cluster these topics into 50 distinct groups, each with a concise single-sentence definition, and validate these categories with contact center personnel. We use our LLM to classify each of our conversations into a topic category, ``other'' or ``unsure.'' Using this approach, we classify over 98\% of conversations into one of 53 main topics.  Appendix Figure \ref{hist_topic} displays a breakdown of the most common chat topics encountered by agents.  Conversation topics in our data are highly skewed, with the two most common topics---payroll and taxes and account access and management issues---making up 50\% of all chat topics.  The next five topics account for the next 25\% of conversations.  In total, the top 16 topics account for over 90\% of chats in the data. This pattern is consistent across customer support; generally a small number of fairly common issues make up the bulk of customer inquiries. The full prompt is:

\begin{displayquote}
    The following is the text of a chat between a customer service agent and a customer who has reached out to the Intuit customer support team. The customer is usually a small business owner based in the US. Based on the transcript, categorize the topic of the conversation into a 1 to 3 word topic. If the topic is about a customer not being able to log into their account, categorize the chat as ``Login.'' If the conversation is about paying their employees, the correct topic is ``Payroll.'' If the conversation is about paying suppliers, the correct topic is ``Supplier Payments”. If the topic is about tax related issues, the correct topic is ``Taxes.'' If the topic is about ensuring that two sets of records (usually the balances of two accounts) are in agreement, the right topic is ``Reconciliation.''  When analyzing the transcript, do not focus on details like the identity of the customer, filler greeting text, or metadata around the agent joining the conversation or variables that obscure sensitive details like \${SSN}, \${EMAIL},  \${NAME-1}, \${ADDRESS}, hyperlink metadata like <br>,  or case numbers. The output format should be: {topic, problem type, issue resolution, probability resolved, skill required, covid-19 related}. 
\end{displayquote}

Once we have a list of 50 categories and a clear definition, we then ask the LLM to classify the topic of each conversation into one of the 50 categories. The LLM could also classify a chat as ``other'' if the conversation does not fit any of the above categories, or ``unsure'' if the LLM is unable to classify the conversation into any specific category. 

All together, we are able to classify 85\% of all conversations into a topic category. The subjects follow a highly skewed distribution---almost 50\% of contact center questions fall into payroll \& taxes or account access \& management. 

We validate the LLM-generated topic categorizations by employing three independent human evaluators to classify a random sample of 100 conversations into our predefined topic categories. In 30\% of the sample, all three human evaluators unanimously agree on the same topic (``unanimous topic''). In 75\% of the sample (which includes the unanimous subset), at least two evaluators agree on a topic (``modal topic''). The remaining 25\% of the sample consists of discordant conversations where each evaluator selects a different topic. When comparing the LLM's topic selections to these human-evaluated subsets, we find that for unanimous-consensus conversations, the LLM selects the same topic as the human evaluators 87\% of the time. For modal-consensus conversations, the LLM selects the modal topic 66\% of the time. In the case of discordant conversations, where there is no agreement between human reviewers, the LLM's selected topic matches one of the three distinct human-selected topics 74\% of the time. 

We use these topic categories to rank topics by overall topic frequency across all conversations. We also categorize topics according to their overall frequency with respect to an individual agent.

\subsubsection{Conversation Sentiment}
Sentiment analysis involves determining the emotions or attitudes expressed in a piece of text. The valence of unstructured text data is widely used in consumer marketing, predicting stock market returns, measuring consumer sentiment, monitoring social media and understanding voting behavior. 

Machine learning-based methods, particularly those utilizing transfer learning models, generally outperform other sentiment classification approaches. \citet{hartmann2023} performs a meta-analysis of over 1,100 experimental results, demonstrating that transfer-learning models are superior for sentiment classification. They also provide an open-sourced, fine-tuned language model, SiEBERT, incorporating transfer-learning best practices, which we use for our sentiment analysis. The model generates a sentiment measure on a scale from -1 to 1, with -1 indicating negative sentiment and 1 indicating positive sentiment.

In our analysis, we classify sentiment separately for agents' speech and customers' text in each conversation. Customer sentiment reflects the emotional experience of the 40-minute or so chat-based interaction, while agent sentiment captures the tone of the agents' responses.

\subsubsection{Conversation Similarity}

We create textual embeddings of agent-customer conversations and compare similarity of these embeddings across workers and over time. Textual embeddings take a given body of text and transform it into a high-dimensional vector that represents its ``coordinates'' in linguistic space.  Two pieces of text will have more similar coordinates if they share a common meaning or style.  The specific embedding given to a body of text will depend on the embedding model used.  We form our text embeddings using all-MiniLM-L6-v2, an LLM that is specifically intended to capture and cluster semantic information to assess similarity across text \citep{sentence2023}. Once we create an embedding for each conversation, we can compare the similarity of conversations by looking at the cosine similarities of their associated vectors; this common approach yields a score of 0 if two pieces of text are semantically orthogonal and a score of 1 if they have the same meaning \citep{koroteev2021bert}. For context, the sentences ``Can you help me with logging in?'' and ``Why is my login not working?'' have a cosine similarity of $0.68$ in our model.

\subsection{Empirical Specifications}\label{asec:specifications}
\subsubsection{Pre-treatment Worker Skill Specification}

\begin{equation} \label{aeq:dd_skill}
y_{it}= \delta_{t} + \alpha_{i} + \sum_{q=1}^{5} \beta_q (AI_{it} \times Q_{iq}) + \gamma X_{it} + \epsilon_{it}
\end{equation}

Our worker skill specification allows us to estimate how the impact of assistance varies by agent skill when they receive access to AI assistance. The regression is conducted at the agent-month level, where:

\begin{itemize}
    \item $y_{it}$ is the resolutions per hour of agent $i$ in year-month $t$. 
    \item $\delta_{t}$ represents year-month fixed effects.
    \item $\alpha_{i}$ denotes agent-level fixed effects.
    \item $AI_{it}$ is an indicator equal to 1 if agent $i$ has access to AI assistance at time $t$, 0 otherwise.
    \item $Q_{iq}$ is an indicator function that equals 1 if agent $i$ belonged to skill quintile $q$ at the time of treatment, where $q$ ranges from 1 (lowest skill) to 5 (highest skill). 
    \item $X_{it}$ is a set of time-varying controls, specifically fixed effects for agent tenure in months.
    \item $\beta_q$ represents the average treatment effect of AI assistance for agents in skill quintile $q$
\end{itemize}

We estimate this equation separately for each of our outcome variables, which include our main measure of productivity for agent $i$ in year-month $t$ (resolutions per hour), as well as call resolution rate, customer satisfaction score, average call duration, calls handled per hour and requests to speak to the manager.  We cluster standard errors at the agent level to account for within-agent correlations in the error terms. In our main analysis, we find very similar results across estimators and similar main effects across adoption cohorts, so we estimate this regression specification using OLS.

\subsubsection{Pre-treatment Worker Tenure Specification}
\begin{equation} \label{aeq:dd_tenure}
y_{it}= \delta_{t} + \alpha_{i} + \sum_{e=1}^{5} \beta_e (AI_{it} \times Exp_{ie})  + \epsilon_{it}
\end{equation}

Our tenure specification allows us to estimate how the impact of assistance varies by agent experience when they receive access to AI assistance. The regression is conducted at the agent-month level, where:
\begin{itemize}
    \item $y_{it}$ is the resolutions per hour of agent $i$ in year-month $t$.
    \item $\delta_{t}$ represents year-month fixed effects.
    \item $\alpha_{i}$ denotes agent-level fixed effects.
    \item $AI_{it}$ is an indicator equal to 1 if agent $i$ has access to AI assistance at time $t$, 0 otherwise.
    \item $Exp_{ie}$ is an indicator that equals $1$ if agent $i$ has $e$ months of experience at the time of treatment, and 0 otherwise. We divide months of experience into five categories: agents in their first month on the job (``0 months''), 1-2 months of experience, 3-6 months of experience, 7-12 months, and over 12 months of experience.
    \item $X_{i}$ includes a set of fixed effects for agent $i$'s skill quintile at the time of treatment. This is only available for treated agents and is time-invariant. 
    \item $\beta_e$ is the average treatment effect of AI assistance for agents with $e$ months of experience when treated. 
\end{itemize}

We estimate this equation separately for each of our outcome variables, which include our main measure of productivity for agent $i$ in year-month $t$ (resolutions per hour), as well as call resolution rate, customer satisfaction score, average call duration, and calls handled per hour. The skill quintile at AI treatment is time invariant and is not defined for control group workers. Standard errors are clustered at the agent level, and we estimate this regression using OLS.

\subsubsection{Adherence to AI recommendations}
\begin{equation} \label{aeq:dd_adherence}
y_{it}= \delta_{t} + \alpha_{i} + \sum_{a=1}^{5} \beta_a (AI_{it} \times Adh_{ia}) + \gamma X_{it} + \epsilon_{it}
\end{equation}

This specification allows us to estimate how the impact of AI assistance varies by treated agents' adherence in their first month of AI access. The regression is conducted at the agent-month level, where:
\begin{itemize}
    \item $y_{it}$ is the resolutions per hour of agent $i$ in year-month $t$.
    \item $\delta_{t}$ represents year-month fixed effects.
    \item $\alpha_{i}$ denotes agent-level fixed effects.
    \item $AI_{it}$ is an indicator equal to 1 if agent $i$ has access to AI assistance at time $t$, 0 otherwise.
    \item $Adh_{ia}$ is an indicator equal to 1 if agent $i$ is in the $a$th quintile of adherence in their first month of access, 0 otherwise.
    \item $X_{it}$ includes time-varying controls, specifically fixed effects for agent tenure.
    \item $\beta_a$ is the average impact of AI assistance for agents in the $a$th quintile of initial adherence. 
\end{itemize}

Standard errors are clustered at the agent level. We estimate this regression with OLS.

\subsubsection{Heterogeneity by Chat Topic}

\begin{equation} \label{aeq:dd_topic}
y_{itc} = \delta_{t} + \alpha_{i} + \sum_{r=1}^{4} \beta_r (AI_{it} \times Topic_{cr}) + \gamma X_{itc} + \epsilon_{itc}
\end{equation}

This topic-based specification allows us to estimate how the impact of AI assistance varies by the routine nature of customers' problems. The regression is conducted at the chat level, where:
\begin{itemize}
    \item $y_{itc}$ is the duration of chat $c$ assigned to agent $i$ in year-month $t$.
    \item $\delta_{t}$ represents year-month fixed effects.
    \item $\alpha_{i}$ denotes agent-level fixed effects.
    \item $AI_{it}$ is an indicator equal to 1 if agent $i$ has access to AI assistance at time $t$, 0 otherwise.
    \item $Topic_{cr}$ is an indicator equal to 1 if chat $c$ belongs to topic frequency category $r$ (where $r$ ranges from 1 to 4), 0 otherwise. The frequency of the topic is defined using frequency across all chats and agents.  
    \item $X_{it}$ includes fixed effects for agent tenure and the overall category of topic frequency ($Topic_{cr}$).
    \item $\beta_r$ estimates the impact of AI assistance on chat duration for chats in the topic frequency category $r$.
\end{itemize}

Standard errors are clustered at the agent level and we estimate this regression with OLS. We drop the small number of topics that are classified as unsure or ``other.'' We also estimate a separate specification on the agent-specific frequency of the technical support issue. 

\begin{equation} \label{aeq:dd_topic_agent}
y_{itc} = \delta_{t} + \alpha_{i} + \sum_{r=1}^{4} \beta_r (AI_{it} \times AgentTopic_{icr}) + \gamma X_{cit} + \epsilon_{itc}
\end{equation}

\begin{itemize}
    \item $AgentTopic_{icr}$ is an indicator equal to 1 if chat $c$ belongs to agent-specific topic frequency category $r$ (where $r$ ranges from 1 to 4), 0 otherwise. Topic frequency is defined relative to conversations conducted by agent $i$.
    \item $X_{cit}$ includes time-varying controls, specifically fixed effects for agent tenure, aggregate topic category defined over all conversations ($Topic_{cr}$), and agent-specific topic frequency category ($ AgentTopic_{icr})$.
    \item $\beta_r$ estimates the impact of AI assistance on chat duration for chats in agent-specific topic frequency category $r$.
\end{itemize}

\subsubsection{Attrition}
\begin{equation} \label{aeq:dd_attrition}
attrit_{it}= \delta_{t} + \beta AI_{it} + \gamma X_{it} + \epsilon_{it}
\end{equation}

This specification allows us to look at how AI assistance impacts agent attrition. The regression is conducted at the agent-month level, where:
\begin{itemize}
    \item $attrit_{it}$ is equal to 1 if agent $i$ leaves in year-month $t$.
    \item $\delta_{t}$ represents year-month fixed effects.
    \item $AI_{it}$ is an indicator equal to 1 if agent $i$ has access to AI assistance at time $t$, 0 otherwise.
    \item $X_{it}$ includes time-varying controls, specifically fixed effects for agent tenure, agent location, country and company employing the agent (data firm or subcontractor).
    \item $\beta_a$ is the average impact of AI assistance on attrition. 
\end{itemize}

Attrition captures both voluntary or involuntary separations, which are undistinguishable in our data.  Standard errors are clustered at the agent level and we estimate this regression with OLS. We cannot include for agent-fixed effects, because agents only leave once, so agent-fixed effects are colinear with attrition.  We drop all observations for treated agents before treatment because, by construction, agents must survive through treatment to receive AI assistance. We also estimate specifications where we estimate the attrition effects by agent tenure and skill at AI deployment.

\clearpage

\begin{figure}[ht!]
\begin{center}
\captionsetup{justification=centering}
\caption{\textsc{Figure \ref{afig:AI-sample}: Sample AI Output}}
\makebox[\linewidth]{
\begin{tabular}{c}
\\
\textsc{A.  Sample Customer Issue}\\
  \includegraphics[scale=0.45]{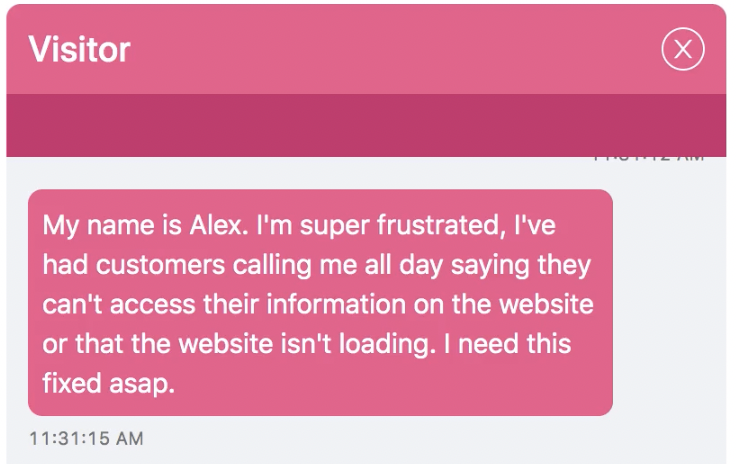} \\
\\
  
\textsc{B.  Sample AI-generated Suggested Response}\\
  \includegraphics[scale=0.35]{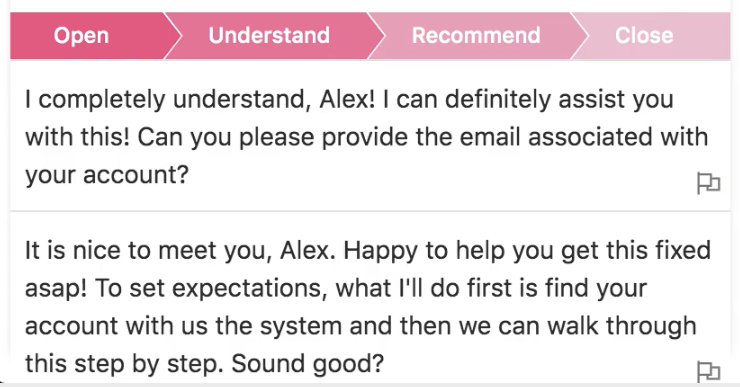} \\

    \textsc{C.  Sample AI-generated Technical Link}\\
   \includegraphics[scale=0.3, angle=270]{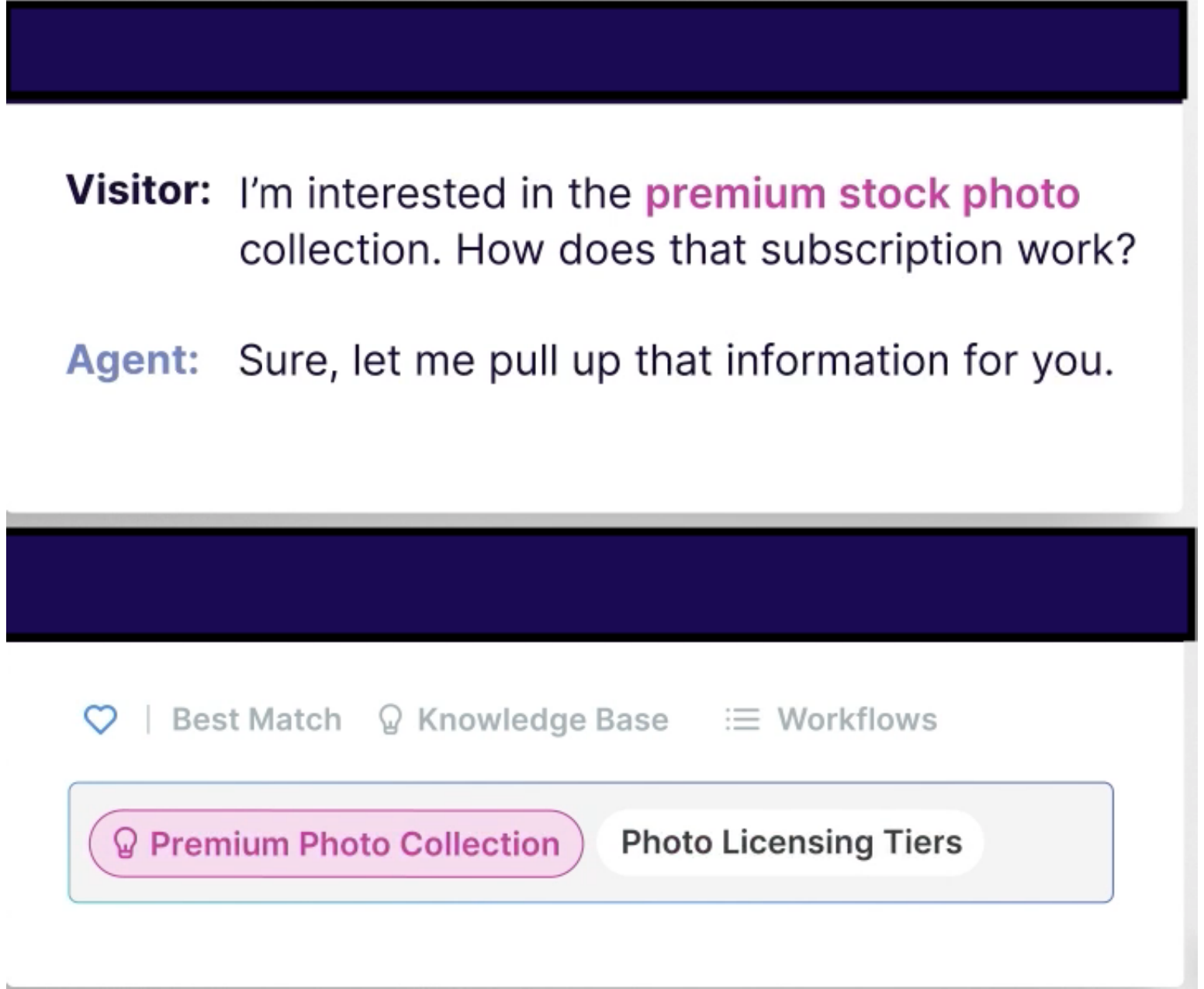} 
\end{tabular}
}
\label{afig:AI-sample}
\end{center}
\end{figure}
\begin{footnotesize} 
\begin{singlespace}
\noindent \textsc{Notes}: This figure illustrates AI-generated suggestions for customer service agents. Panel A shows a sample customer issue. Panel B displays AI-suggested responses for greeting and setting call expectations. Panel C presents a AI-recommended technical documentation excerpt from the company's internal knowledge base. Suggestions are only visible to agents, not customers, and agents can choose to use, modify, or ignore these suggestions when responding. 
\end{singlespace}
\end{footnotesize}


\clearpage
\begin{figure}[ht!]
\begin{center}
\captionsetup{justification=centering}
\caption{\textsc{Figure \ref{afig:deploymenttimeline}: Deployment Timeline}}
\makebox[\linewidth]{
\begin{tabular}{c}
  \includegraphics[scale=1.2]{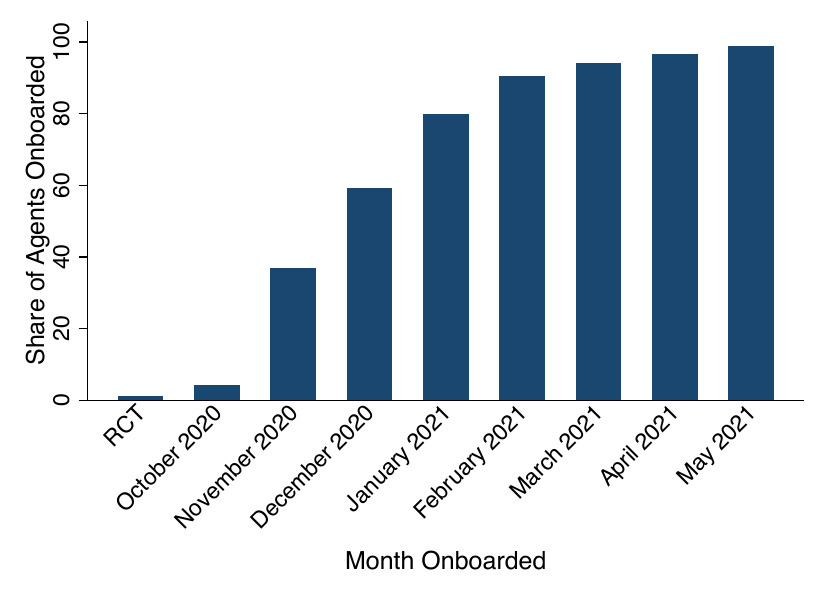} \\
\end{tabular}
}
\label{afig:deploymenttimeline}
\end{center}
\end{figure}

\begin{footnotesize} 
\begin{singlespace}
\noindent \textsc{Notes}:  This figure shows the share of agents deployed onto the AI system over the study period. Agents are deployed onto the AI system after a training session as described in Section \ref{sec:rollout}. The small randomized control trial in August 2020 is analyzed in Section \ref{sec:rct}. All data are from the firm's internal software systems.    
\end{singlespace}
\end{footnotesize}

\clearpage
\section{Average Productivity Effects}\label{asec:productivityresults}

\clearpage
\begin{figure}[ht!]
\begin{center}
\captionsetup{justification=centering}\caption{\textsc{Figure \ref{afig:es_oth}: Event Studies, Additional Outcomes}}

\makebox[\linewidth]{
\begin{tabular}{cc}
\textsc{A. Average Handle Time} & \textsc{B.  Chats Per Hour} \\
\includegraphics[scale=0.45]{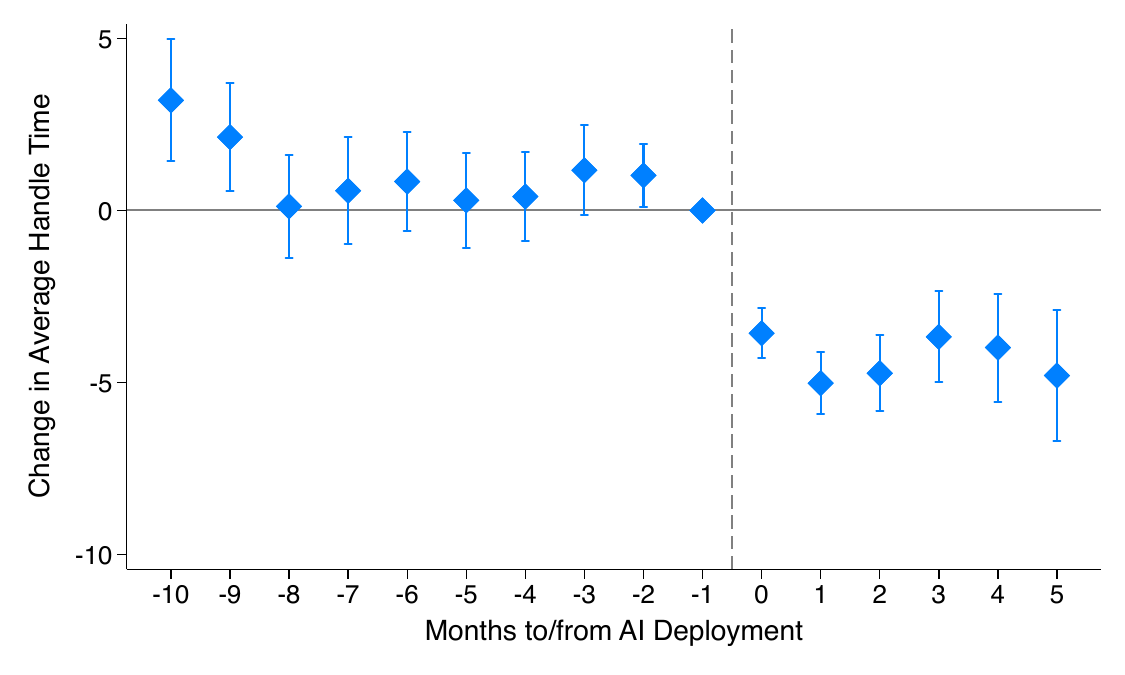} & \includegraphics[scale=0.45]{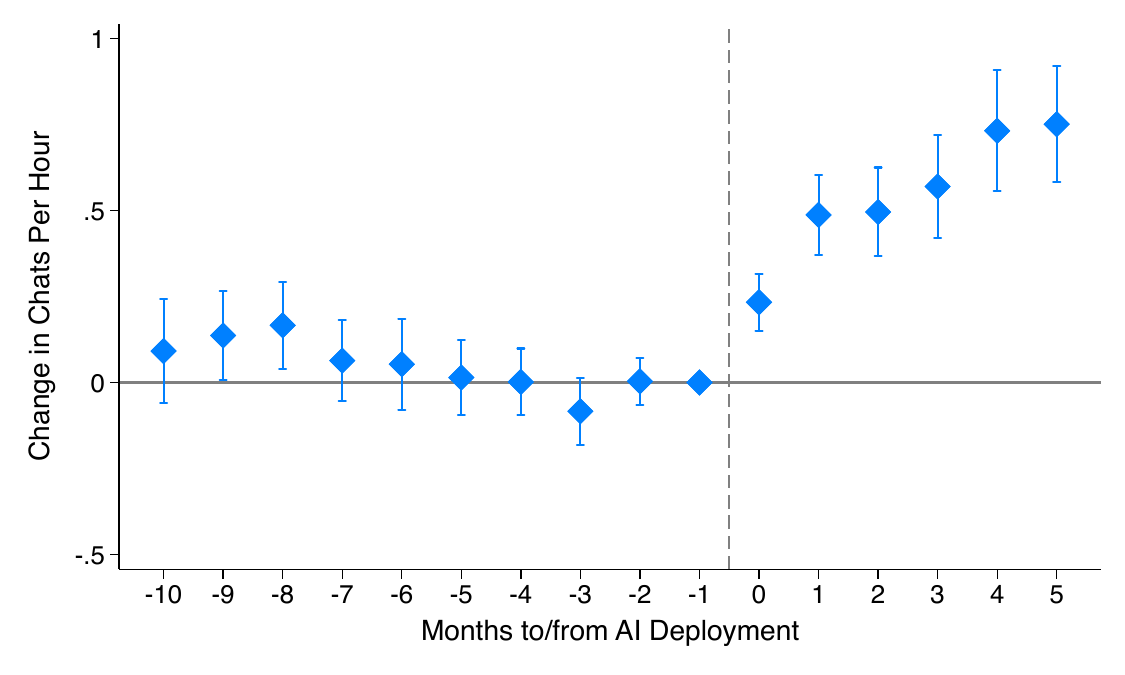}  \\

\textsc{C. Resolution Rate} & \textsc{D. Customer Satisfaction (NPS)} \\
\includegraphics[scale=0.45]{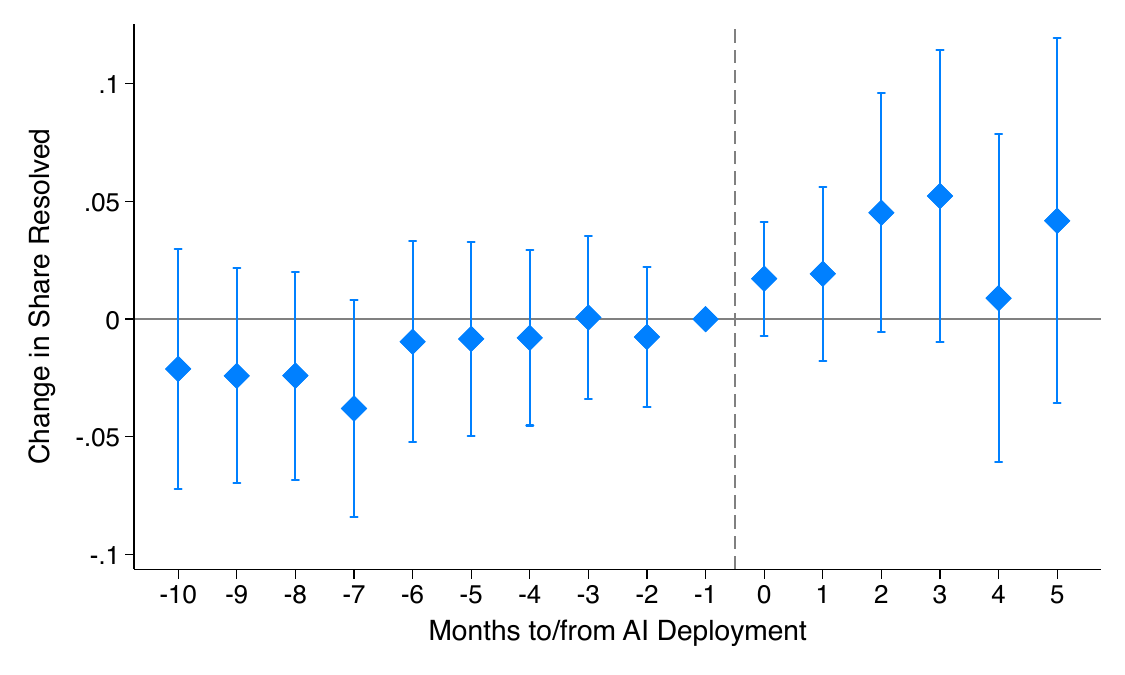} & \includegraphics[scale=0.45]{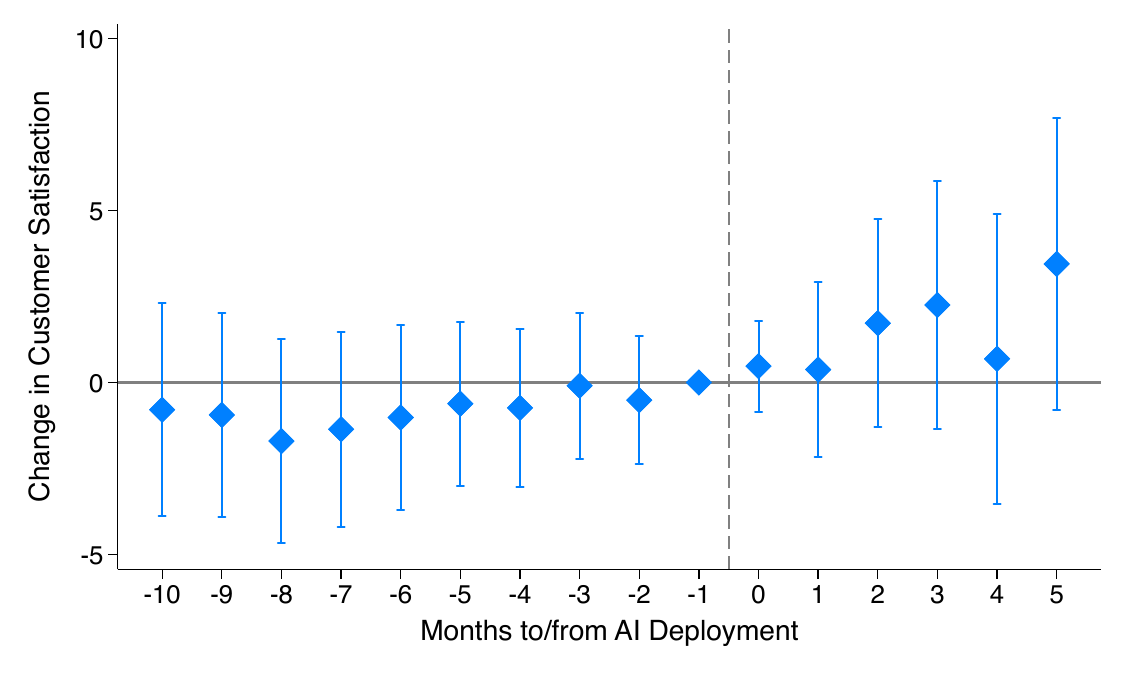} 
\\
\end{tabular}
}

\label{afig:es_oth}
\end{center}
\end{figure}

\begin{footnotesize} 
\begin{singlespace}
\noindent \textsc{Notes}:  These figures plot the coefficients and 95\% confidence intervals from event study regressions of AI model deployment using the \citet{sun_estimating_2020} interaction weighted estimator.  Panel A plots the average handle time or the average duration of each technical support chat.  Panel B plots the number of chats an agent completes per hour, incorporating multitasking. Panel C plots the resolution rate, the share of chats successfully resolved, and Panel D plots net promoter score, which is an average of surveyed customer satisfaction. All specifications follow Equation \ref{eq:dd_main} and include agent and chat year-month and months of agent tenure fixed effects. Data is at the agent-month level and robust standard errors are clustered at the agent level. 
\end{singlespace}
\end{footnotesize}

\clearpage
\begin{figure}[ht!]
\begin{center}
\captionsetup{justification=centering}
\caption{\textsc{Figure \ref{afig:es_main_alt}: Event Studies, Resolutions Per Hour}}
 \makebox[\linewidth]{
 \begin{tabular}{c}
 \textsc{A.  Resolutions Per Hour}\\
 \includegraphics[scale=0.8]{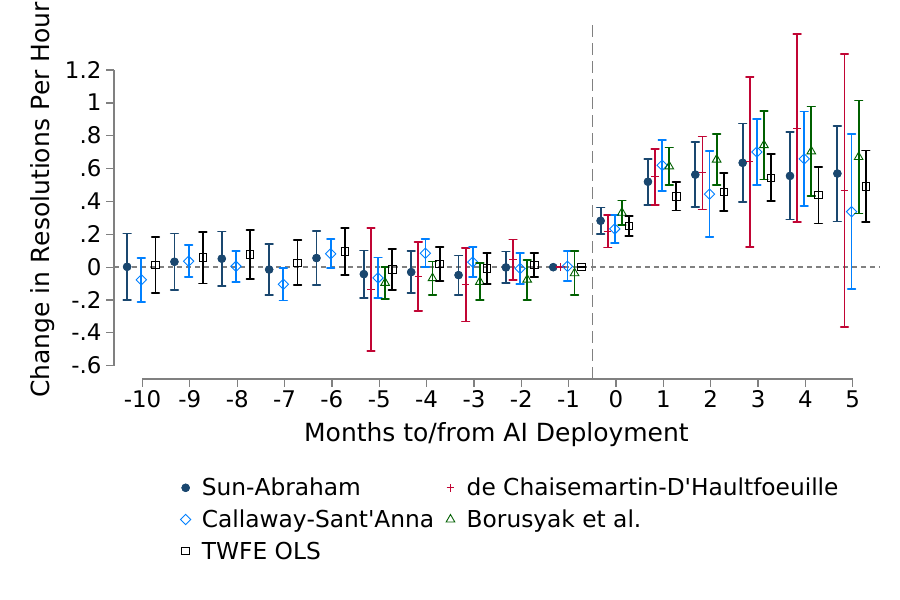} \\
 \end{tabular} }

\label{afig:es_main_alt}
\end{center}
\end{figure}

\begin{footnotesize} 
\begin{singlespace}
\noindent \textsc{Notes}: This figure presents the effect of AI model deployment on our main productivity outcome, resolutions per hour, using a variety of robust dynamic difference-in-differences estimators introduced in \citet{borusyak2022revisiting}, \citet{callaway_2021}, \citet{dechaisemartin2020} and \citet{sun_estimating_2020} and a standard two-way fixed effects regression model. Regressions follow Equation \ref{eq:dd_main} and include agent level, chat-year, and months of agent tenure fixed effects. Data is available at the agent-month level and robust standard errors are clustered at the agent level. Because of the number of post-treatment periods and high turnover of agents in our sample, we can only estimate five months of pre-period data using \citet{borusyak2022revisiting} and \citet{dechaisemartin2020}. 

\end{singlespace}
\end{footnotesize}

\clearpage
\begin{figure}[ht!]
\begin{center}
\captionsetup{justification=centering}
\caption{\textsc{Figure \ref{fig_rct}: RCT Analysis}}

\vspace{15pt}
\makebox[\linewidth]{
\begin{tabular}{c}
\textsc{A. Resolutions Per Hour}\\
\includegraphics[scale=0.45]{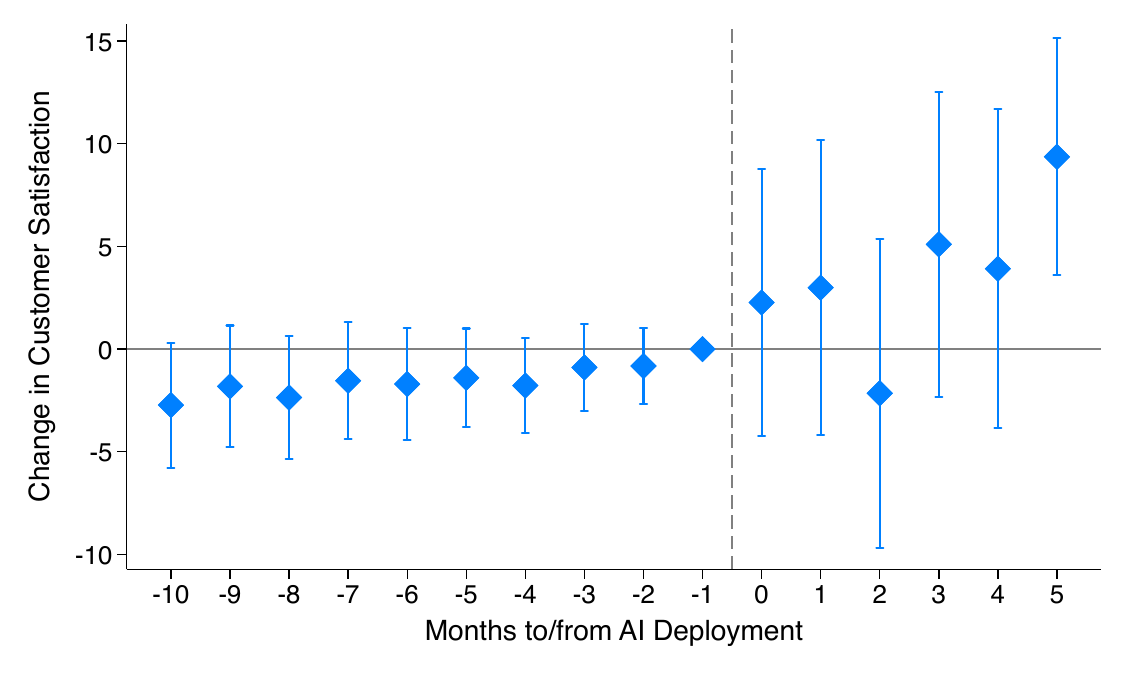} \\
\end{tabular}
}
\vspace{10pt}\\
\makebox[\linewidth]{
\begin{tabular}{cc}
\textsc{B. Average Handle Time} & \textsc{C.  Chats Per Hour}\\
\includegraphics[scale=0.4]{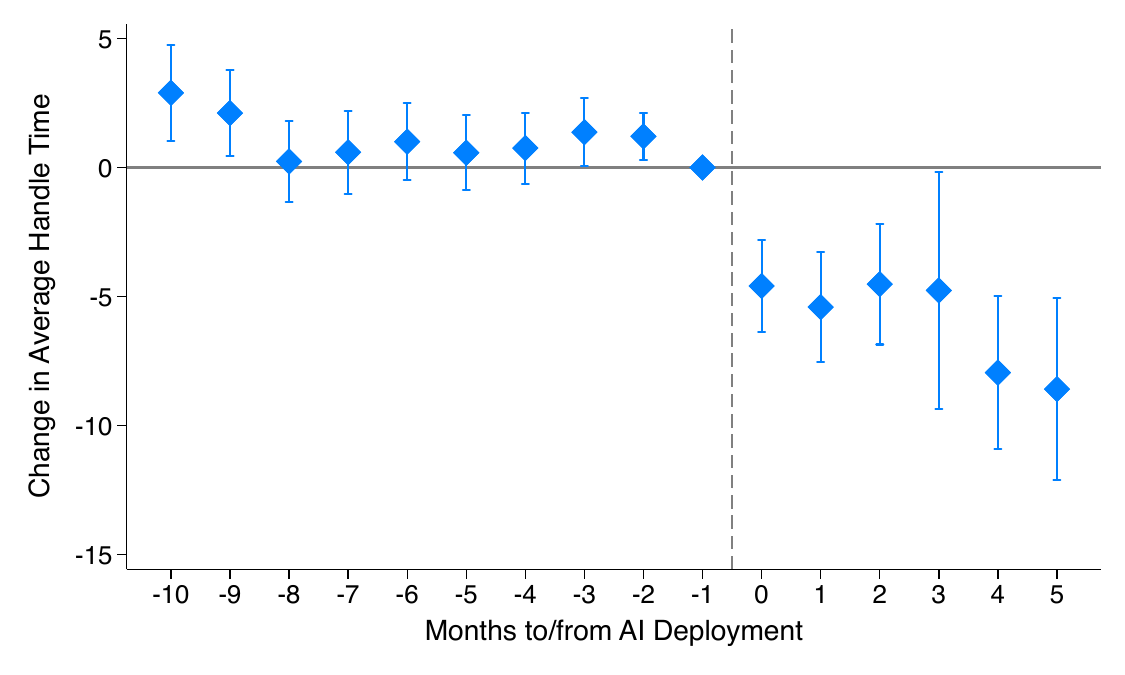} &\includegraphics[scale=0.4]{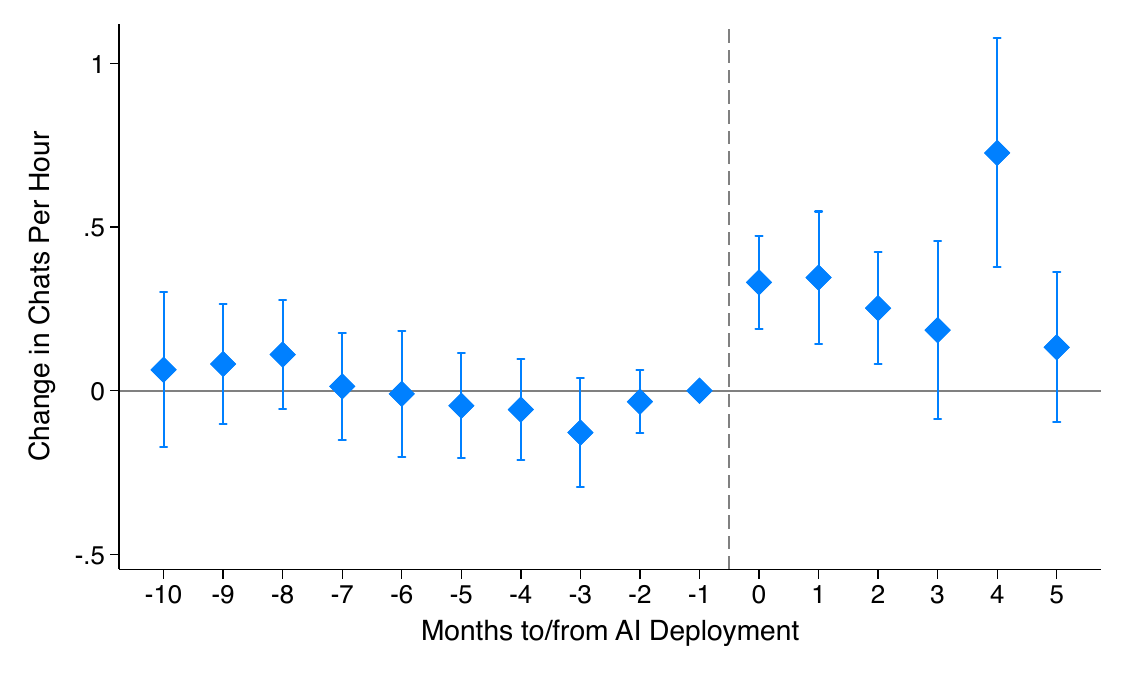} 
\\
\textsc{D. Resolution Rate} & \textsc{E. Customer Satisfaction (NPS)}\\
\includegraphics[scale=0.4]{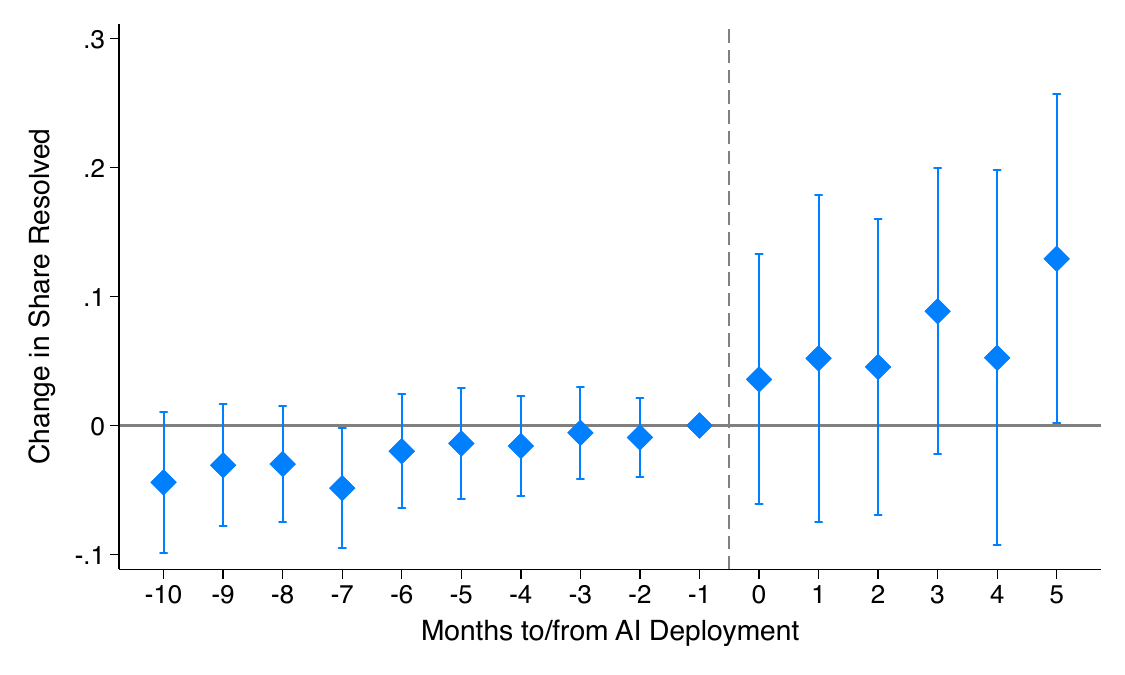} &\includegraphics[scale=0.4]{figures/15Aug2024/ES_tnps_cresta_AS_rct.pdf} 
\end{tabular}
}

\label{fig_rct}
\end{center}
\end{figure}

\begin{footnotesize} 
\begin{singlespace}
\noindent \textsc{Notes}:  This figures plots event studies focusing on the 22 agents who were onboarded as part of the pilot RCT.  Because we do not have information on the specific agents (there were around 25) selected to be part of the control group, we compare the 22 pilot-treated agents with observations for all pre-treatment agents, controlling for our usual agent, chat year-month, and months of agent tenure fixed effects in agent-month level regressions following Equation \ref{eq:dd_main}.  Robust standard errors are clustered at the agent level. 
\end{singlespace}
\end{footnotesize}

\clearpage
\begin{table}[ht!]
\begin{center}
\caption{\textsc{Table \ref{rctdd.tex}: RCT Analysis}}
\vspace{20pt}        
\scalebox{1}{\makebox[\linewidth]{\begin{tabular}{lccccc} \hline
 & (1) & (2) & (3) & (4) & (5) \\
VARIABLES & Res./Hr & AHT & Chats/Hr & Res. Rate & NPS \\ \hline
 &  &  &  &  &  \\
Post AI X Ever Treated & 0.202** & -3.713*** & 0.105 & 0.0169 & -1.393 \\
 & (0.0850) & (1.045) & (0.0718) & (0.0246) & (1.529) \\
 &  &  &  &  &  \\
Observations & 6,998 & 15,100 & 15,100 & 6,998 & 7,176 \\
R-squared & 0.568 & 0.574 & 0.566 & 0.383 & 0.517 \\
Year Month FE & Yes & Yes & Yes & Yes & Yes \\
Agent FE & Yes & Yes & Yes & Yes & Yes \\
Agent Tenure FE & Yes & Yes & Yes & Yes & Yes \\
 DV Mean & 1.956 & 43 & 2.378 & 0.808 & 79.68 \\ \hline
\multicolumn{6}{c}{ Robust standard errors in parentheses} \\
\multicolumn{6}{c}{ *** p$<$0.01, ** p$<$0.05, * p$<$0.10} \\
\end{tabular}
}}
\label{rctdd.tex}
\end{center}
\end{table}

\begin{singlespace}
\footnotesize
\footnotesize
\noindent \textsc{Notes}:  This table focuses on the 22 agents who were onboarded as part of the pilot RCT.  Because we do not have information on the specific agents (there were around 25) selected to be part of the control group, we compare the 22 pilot-treated agents with observations for all pre-treatment agents, controlling for agent, chat year-month, and months of agent tenure fixed effects.  Data are at the agent-month level and the standard errors are clustered at the agent level.  
\end{singlespace}
\normalsize

\clearpage
\begin{landscape}
\begin{table}[ht!]
\begin{center}
                \caption{\textsc{Table \ref{tab:ivdd_companylocteam}: Main Effects: Company Team IV}}
                  \vspace{20pt}        
\scalebox{1}{\makebox[\linewidth]{\begin{tabular}{lcccccc} \hline
 & (1) & (2) & (3) & (4) & (5) & (6) \\
VARIABLES & Individual Treatment & Res./Hr & AHT & Chats/Hr & Res. Rate & NPS \\ \hline
 &  &  &  &  &  &  \\
Earliest Team Treatment & 0.311*** &  &  &  &  &  \\
 & (0.0163) &  &  &  &  &  \\
Post AI X Individual Treatment &  & 0.550*** & -3.622** & 0.412*** & 0.0740*** & -0.518 \\
 &  & (0.0943) & (1.444) & (0.0791) & (0.0184) & (0.946) \\
 &  &  &  &  &  &  \\
Observations & 21,839 & 12,295 & 21,839 & 21,839 & 12,295 & 12,541 \\
R-squared & 0.813 & 0.177 & 0.102 & 0.115 & 0.007 & 0.007 \\
Year Month FE & Yes & Yes & Yes & Yes & Yes & Yes \\
Agent FE & Yes & Yes & Yes & Yes & Yes & Yes \\
Agent Tenure FE & Yes & Yes & Yes & Yes & Yes & Yes \\
F-statistic & 365.7 &  &  &  &  &  \\
 Number of agent\_id &  & 2,163 & 3,633 & 3,633 & 2,163 & 2,214 \\ \hline
\multicolumn{7}{c}{ Robust standard errors in parentheses} \\
\multicolumn{7}{c}{ *** p$<$0.01, ** p$<$0.05, * p$<$0.10} \\
\end{tabular}
}}
  \label{tab:ivdd_companylocteam}
\end{center}
\end{table}

\begin{singlespace}
\footnotesize
\footnotesize
\noindent \textsc{Notes}:  In this table we instrument agent level date of AI adoption with the minimum date of AI deployment on that agent's team. Column 1 shows the first stage regression of earliest team adoption on individual adoption and Columns 2 through 6 shows the 2SLS estimates on measures of productivity and call quality and efficiency. Regressions follow Equation \ref{eq:dd_main} and include agent level, chat-year, and months of agent tenure fixed effects. Data is available at the agent-month level and robust standard errors are clustered at the agent level.  
\end{singlespace}
\normalsize
\end{landscape}

\clearpage
\begin{table}[ht!]
\begin{center}
                \caption{\textsc{Table \ref{tab:robustness_clustering}: Main Effects: Robustness to alternative clustering}}
                  \vspace{20pt}        
\scalebox{1}{\makebox[\linewidth]{\begin{tabular}{lccc} \hline
 & (1) & (2) & (3) \\
VARIABLES & Res./Hr & Res./Hr & Res./Hr \\ \hline
 &  &  &  \\
Post AI X Ever Treated & 0.301*** & 0.301*** & 0.301*** \\
 & (0.0329) & (0.0455) & (0.0498) \\
 &  &  &  \\
Observations & 12,295 & 12,295 & 12,295 \\
R-squared & 0.575 & 0.575 & 0.575 \\
Year Month FE & Yes & Yes & Yes \\
Agent FE & Yes & Yes & Yes \\
Agent Tenure FE & Yes & Yes & Yes \\
 DV Mean & 2.176 & 2.176 & 2.176 \\ \hline
\multicolumn{4}{c}{ Robust standard errors in parentheses} \\
\multicolumn{4}{c}{ *** p$<$0.01, ** p$<$0.05, * p$<$0.10} \\
\end{tabular}
}}
  \label{tab:robustness_clustering}
\end{center}
\end{table}

\begin{singlespace}
\footnotesize
\footnotesize
\noindent \textsc{Notes}:  This table shows robustness of our Column 3 in Table \ref{tab:dd_main} to different levels of clustering. Column 1 clusters our robust standard errors at the agent level, Column 2 clusters at the company/team level and Column 3 clusters at the location (usually the agent's city of work). Regressions follow Equation \ref{eq:dd_main} and include agent level, chat-year, and months of agent tenure fixed effects. Data is available at the agent-month level and robust standard errors are clustered at the agent level.
\end{singlespace}
\normalsize

\clearpage
\begin{table}[ht!]
\begin{center}
                \caption{\textsc{Table \ref{tab:robustdd_nchat_weight}: Main Effects: Chat-weighted}}
                  \vspace{20pt}        
\scalebox{1}{\makebox[\linewidth]{\begin{tabular}{lccccc} \hline
 & (1) & (2) & (3) & (4) & (5) \\
VARIABLES & Res./Hour & AHT & Chats/Hr & Res. Rate & NPS \\ \hline
 &  &  &  &  &  \\
Post AI X Ever Treated & 0.252*** & -3.100*** & 0.295*** & 0.00325 & -0.119 \\
 & (0.0318) & (0.315) & (0.0263) & (0.00725) & (0.524) \\
 &  &  &  &  &  \\
Observations & 12,295 & 21,839 & 21,839 & 12,295 & 12,541 \\
R-squared & 0.650 & 0.756 & 0.721 & 0.465 & 0.526 \\
Year Month FE & Yes & Yes & Yes & Yes & Yes \\
Agent FE & Yes & Yes & Yes & Yes & Yes \\
Agent Tenure FE & Yes & Yes & Yes & Yes & Yes \\
 DV Mean & 2.457 & 38.66 & 2.886 & 0.839 & 79.59 \\ \hline
\multicolumn{6}{c}{ Robust standard errors in parentheses} \\
\multicolumn{6}{c}{ *** p$<$0.01, ** p$<$0.05, * p$<$0.10} \\
\end{tabular}
}}
  \label{tab:robustdd_nchat_weight}
\end{center}
\end{table}

\begin{singlespace}
\footnotesize
\footnotesize
\noindent \textsc{Notes}:  This table presents the results of difference-in-difference regressions estimating the impact of AI model deployment on measures of productivity and agent performance.  Observations are at the agent-month level, weighted by the number of chats associated with that agent-month. Post AI X Treated measures the impact of AI model deployment after deployment on treated agents for average handle time or average call duration, chats per hour, the number of chats an agent handles per hour, resolution rate, the share of technical support problems they can resolve and net promoter score (NPS), an estimate of customer satisfaction.  Regressions follow Equation \ref{eq:dd_main} and include agent level, chat-year, and months of agent tenure fixed effects. Data is available at the agent-month level and robust standard errors are clustered at the agent level.  
\end{singlespace}
\normalsize

\clearpage
\begin{table}[ht!]
\begin{center}
\caption{\textsc{Table \ref{atab:dd_main_robust}: Main Effects: Productivity (Resolutions per Hour), Alternative Difference-in-Difference Estimators}}
\vspace{20pt}        
 \scalebox{1}{\makebox[\linewidth]{{
\def\sym#1{\ifmmode^{#1}\else\(^{#1}\)\fi}
\begin{tabular}{l*{1}{cccc}}
\hline\hline
                &\shortstack{Point\\Estimate}&\shortstack{Standard\\Error}&\shortstack{Lower Bound\\ 95\% Confidence\\Interval}&\shortstack{Upper Bound\\ 95\% Confidence\\Interval}\\
\hline
TWFE-OLS        &    0.296&    0.032&    0.233&    0.360\\
Borusyak-Jaravel-Spiess&    0.576&    0.070&    0.438&    0.714\\
Callaway-Sant'Anna&    0.489&    0.059&    0.374&    0.605\\
DeChaisemartin-D'Haultfeuille&    0.219&    0.042&    0.137&    0.302\\
Sun-Abraham     &    0.521&    0.094&    0.337&    0.705\\
\hline\hline
\end{tabular}
}
}}
   \label{atab:dd_main_robust}
\end{center}
\end{table}

\begin{singlespace}
\footnotesize
\noindent \textsc{Notes}: This table shows the impact of AI model deployment on our main productivity outcome, resolutions per hour, using robust difference-in-differences estimators introduced in \citet{borusyak2022revisiting}, \citet{callaway_2021}, \citet{dechaisemartin2020} and \citet{sun_estimating_2020}. Regressions follow Equation \ref{eq:dd_main} and include agent level, chat-year, and months of agent tenure fixed effects. Data is available at the agent-month level and standard errors are clustered at the agent level. Because of the number of post-treatment periods and high turnover of agents in our sample, we can only estimate five months of pre-period data using \citet{borusyak2022revisiting} and \citet{dechaisemartin2020}.  
\end{singlespace}
\normalsize

\clearpage
\section{Impacts by Agent Skill and Tenure}\label{asec:heteroresults}

\clearpage
\begin{figure}[ht!]
\begin{center}
\captionsetup{justification=centering}
\caption{\textsc{Figure \ref{fig:dd_byskill}: Heterogeneity of AI Impact by pre-AI Worker Skill and Controlling for Tenure, Additional Outcomes}}

\makebox[\linewidth]{
\begin{tabular}{cc}
\textsc{A. Average Handle Time} & \textsc{B.  Chats Per Hour}\\
\includegraphics[scale=0.5]{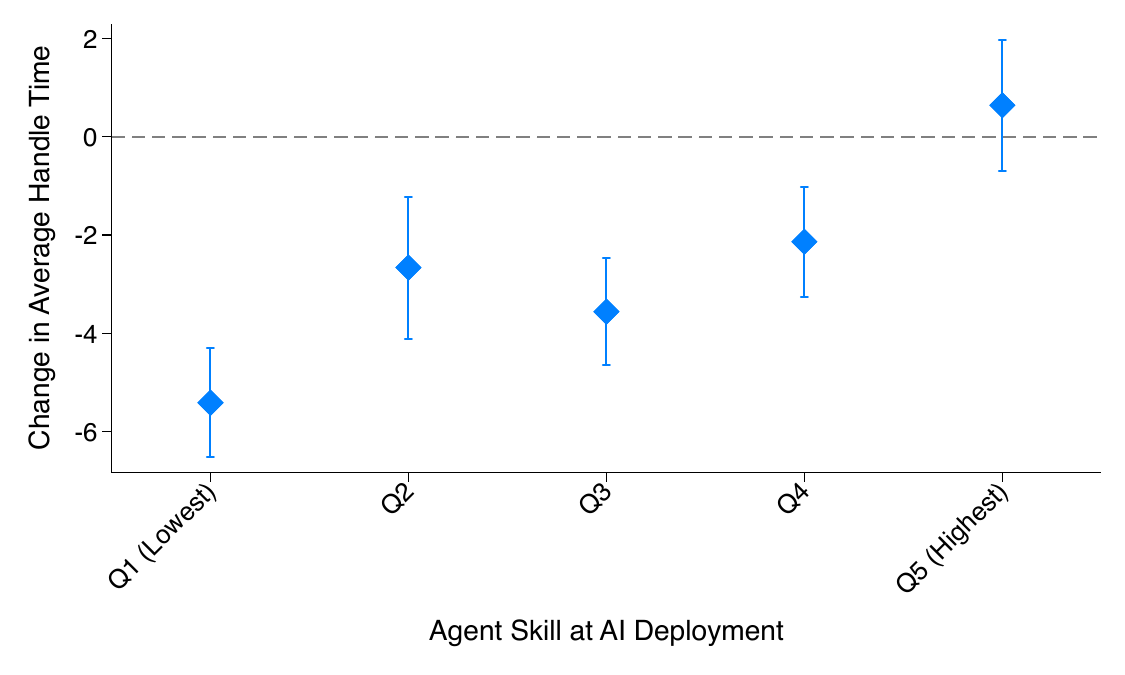} &\includegraphics[scale=0.5]{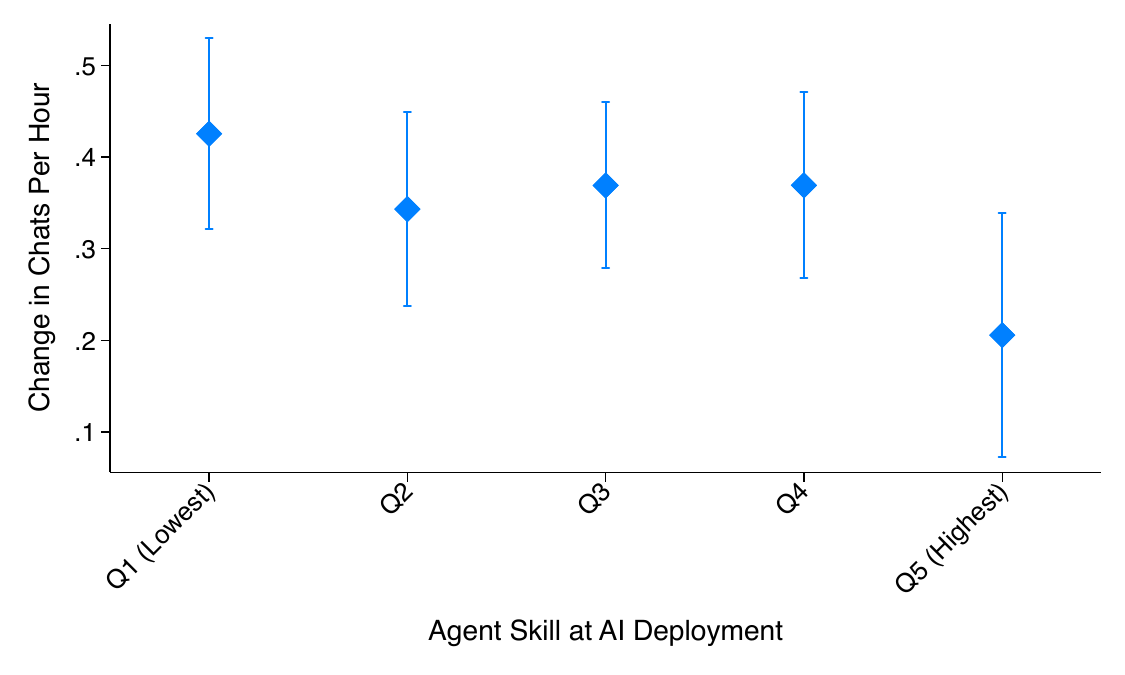} 
\\
\textsc{C. Resolution Rate} & \textsc{D.  Customer Satisfaction (NPS)}\\
\includegraphics[scale=0.5]{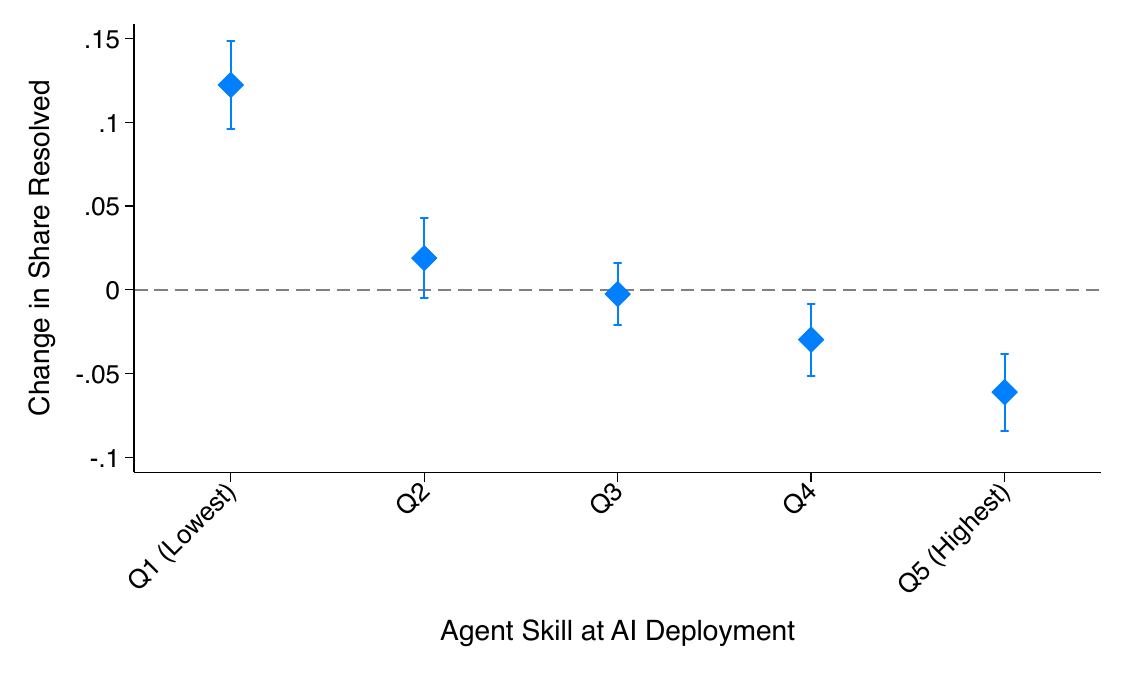} &\includegraphics[scale=0.5]{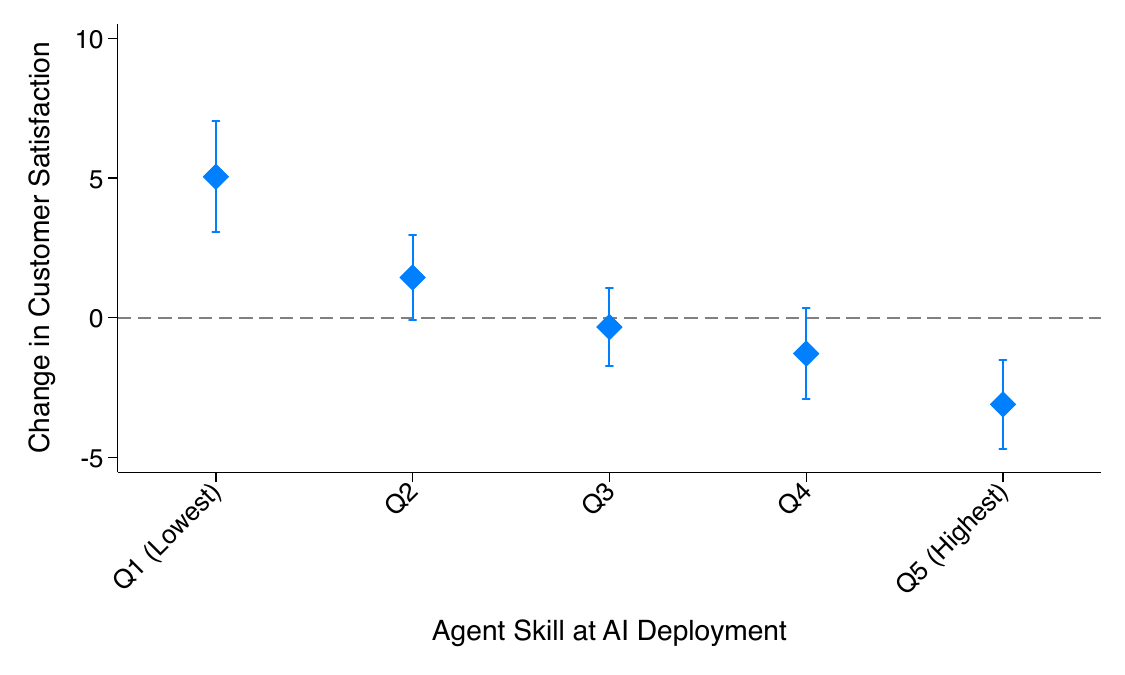} 
\\
\end{tabular}
}

\label{fig:dd_byskill}
\end{center}
\end{figure}

\begin{footnotesize} 
\begin{singlespace}
\noindent \textsc{Notes}:  These figures plot the impacts of AI model deployment on four measures of productivity and performance, by pre-deployment worker skill. Agent skill is calculated as the agent's trailing three month average of performance on average handle time, call resolution, and customer satisfaction, the three metrics our firm uses for agent performance. Within each month and company, agents are grouped into quintiles, with the most productive agents within each firm in quintile 5 and the least productive in quintile 1. Panel A plots the average handle time or the average duration of each technical support chat. Panel B graphs chats per hour, or the number of chats an agent can handle per hour. Panel C plots the resolution rate, and Panel D plots net promoter score, an average of surveyed customer satisfaction. All specifications include fixed effects for the agent, chat year-month and months of tenure. Robust standard errors are clustered at the agent level. The regression specifications are available in Appendix section \ref{asec:specifications}.
\end{singlespace}
\end{footnotesize}

\clearpage
\begin{figure}[ht!]
\begin{center}
\captionsetup{justification=centering}
\caption{\textsc{Figure \ref{fig:dd_bytenure}: Heterogeneity of AI Impact by pre-AI Worker Tenure Controlling for Skill, Additional Outcomes}}

\makebox[\linewidth]{
\begin{tabular}{cc}
\textsc{A. Average Handle Time} & \textsc{B.  Chats Per Hour}\\
\includegraphics[scale=0.5]{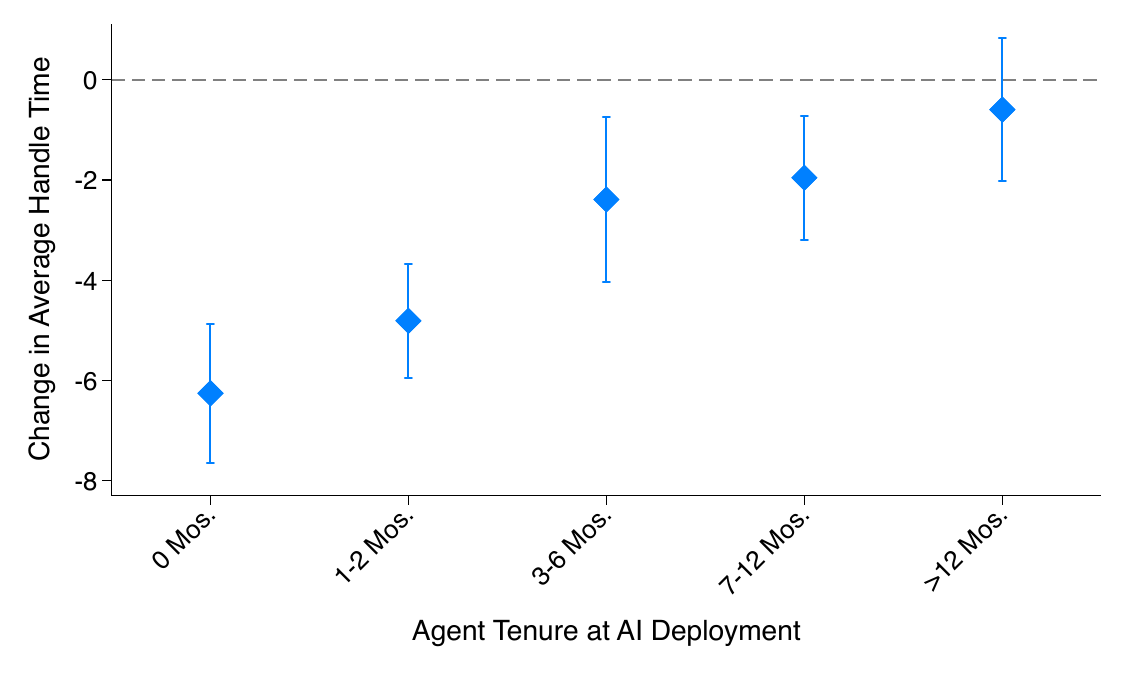} &\includegraphics[scale=0.5]{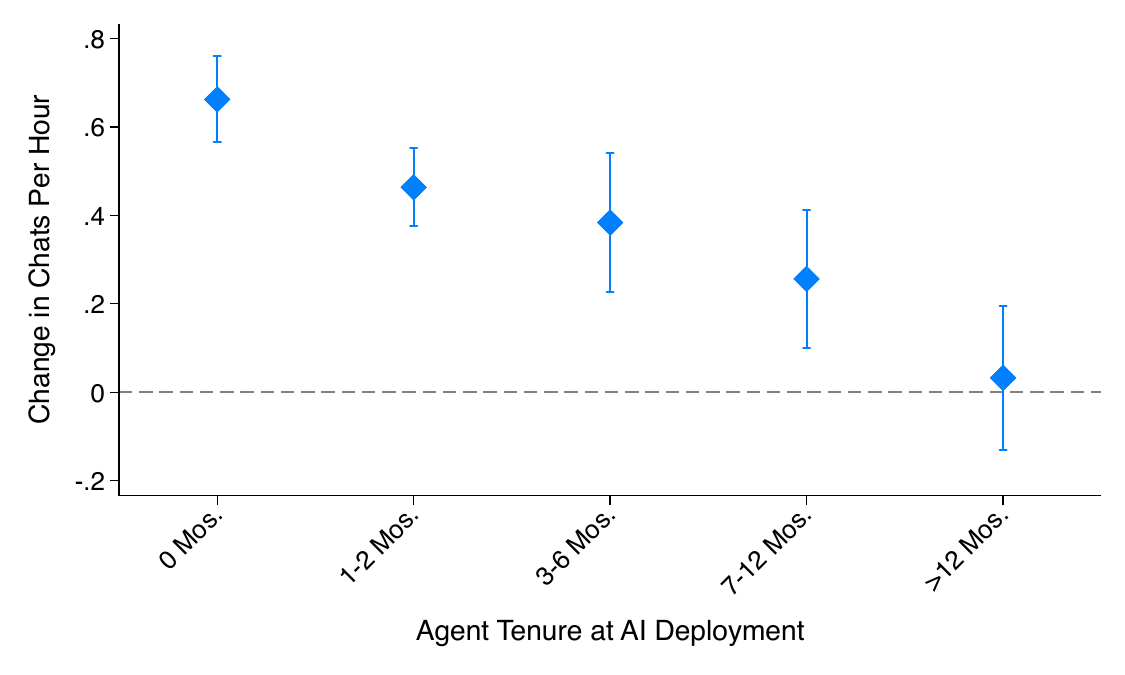} 
\\
\textsc{C. Resolution Rate} & \textsc{D.  Customer Satisfaction (NPS)}\\
\includegraphics[scale=0.5]{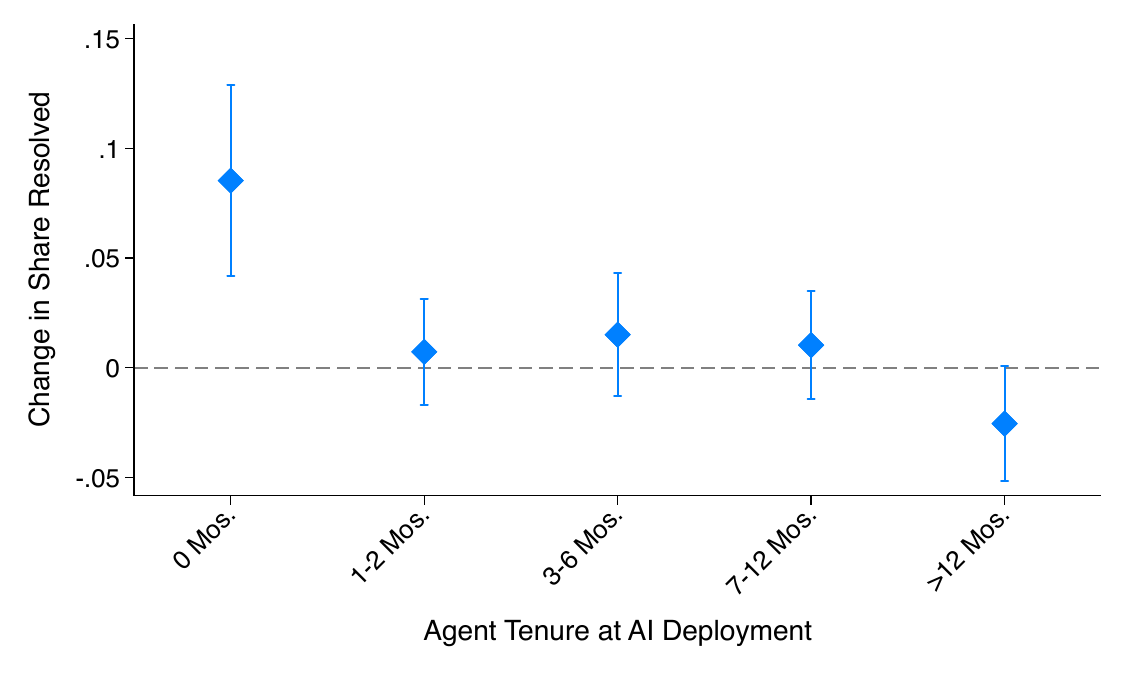} &\includegraphics[scale=0.5]{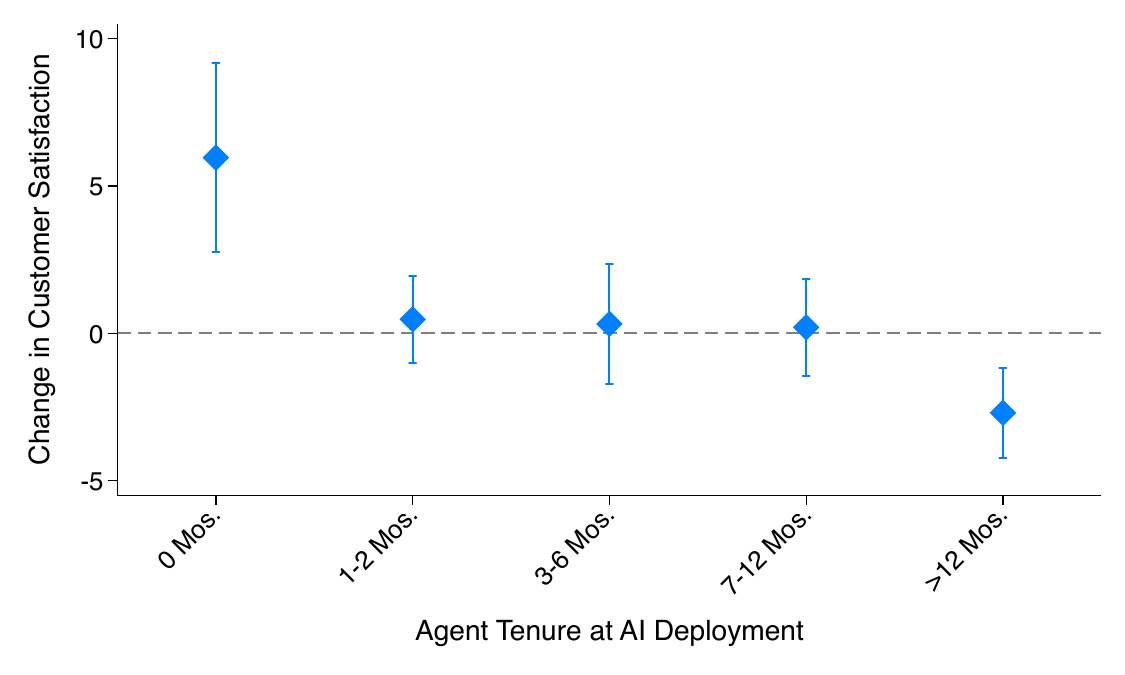} 
\\
\end{tabular}
}
\label{fig:dd_bytenure}
\end{center}
\end{figure}
\begin{footnotesize} 
\begin{singlespace}
\noindent \textsc{Notes}:  These figures plot the impacts of AI model deployment on measures of productivity and performance by pre-AI worker tenure, defined as the number of months an agent has been employed when they receive access to the AI model. Panel A plots the average handle time or the average duration of each technical support chat. Panel B graphs chats per hour, or the number of chats an agent can handle per hour. Panel C plots the resolution rate, and Panel D plots net promoter score, an average of surveyed customer satisfaction. All specifications include fixed effects for the agent, chat year-month and months of tenure. Robust standard errors are clustered at the agent level. The regression specifications are available in Appendix section \ref{asec:specifications}.
\end{singlespace}
\end{footnotesize}

\clearpage
\begin{table}[ht!]
\begin{center}
                \caption{\textsc{Table \ref{atab:heteroeffects}: Heterogeneity of AI Impact by Skill and Tenure, Resolutions per Hour}}
                  \vspace{20pt}        
\scalebox{1}{\makebox[\linewidth]{{
\def\sym#1{\ifmmode^{#1}\else\(^{#1}\)\fi}
\begin{tabular}{l*{2}{l}}
\toprule
                &\multicolumn{1}{c}{(1)}&\multicolumn{1}{c}{(2)}\\
                &\multicolumn{1}{c}{By Skill at AI}&\multicolumn{1}{c}{By Tenure at AI}\\
\midrule
Q1 (Lowest Skill)&0.527\sym{***}&            \\
                &(0.049)         &            \\
\addlinespace
Q2              &0.315\sym{***}&            \\
                &(0.056)         &            \\
\addlinespace
Q3              &0.302\sym{***}&            \\
                &(0.043)         &            \\
\addlinespace
Q4              &0.261\sym{***}&            \\
                &(0.055)         &            \\
\addlinespace
Q5 (Highest Skill)&0.015         &            \\
                &(0.065)         &            \\
\addlinespace
< 1 Mos.)       &            &0.707\sym{***}\\
                &            &(0.067)         \\
\addlinespace
1-2 Mos.        &            &0.388\sym{***}\\
                &            &(0.042)         \\
\addlinespace
3-6 Mos.        &            &0.361\sym{***}\\
                &            &(0.067)         \\
\addlinespace
7-12 Mos.       &            &0.337\sym{***}\\
                &            &(0.054)         \\
\addlinespace
> 12 Mos.)      &            &0.028         \\
                &            &(0.059)         \\
\midrule
Year Month FE   &Yes         &Yes         \\
Agent FE        &Yes         &Yes         \\
Other FE        &Tenure         &Skill at AI         \\
DV Mean         &2.176         &2.284         \\
Observations    &12,295         &8,148         \\
\bottomrule
\multicolumn{3}{l}{\footnotesize Standard errors in parentheses}\\
\multicolumn{3}{l}{\footnotesize \sym{*} \(p<0.10\), \sym{**} \(p<0.05\), \sym{***} \(p<0.01\)}\\
\end{tabular}
}
}}
  \label{atab:heteroeffects}
\end{center}
\end{table}

\begin{singlespace}
\footnotesize
\footnotesize
\noindent \textsc{Notes}:  This table presents the results of difference-in-difference regressions estimating the impact of AI model deployment on resolutions per hour.  Column 1 estimates the impact of AI access by worker skill at AI, including agent, year-month and months of agent tenure fixed effects. Column 2 estimates the effects by worker tenure at AI deployment, including agent, year-month and agent skill at AI deployment fixed effects. Observations are at the agent-month level and all standard errors are clustered at the agent level.  
\end{singlespace}
\normalsize

\clearpage
\begin{figure}[ht!]
\begin{center}
\captionsetup{justification=centering}
\caption{\textsc{Figure \ref{afig:learning_bycohort}: Experience Curves by Deployment Cohort, Additional Outcomes}}
\makebox[\linewidth]{
\begin{tabular}{cc}
\textsc{A.  Average Handle Time} & \textsc{B.  Chats Per Hour}\\ 
\includegraphics[scale=0.5]{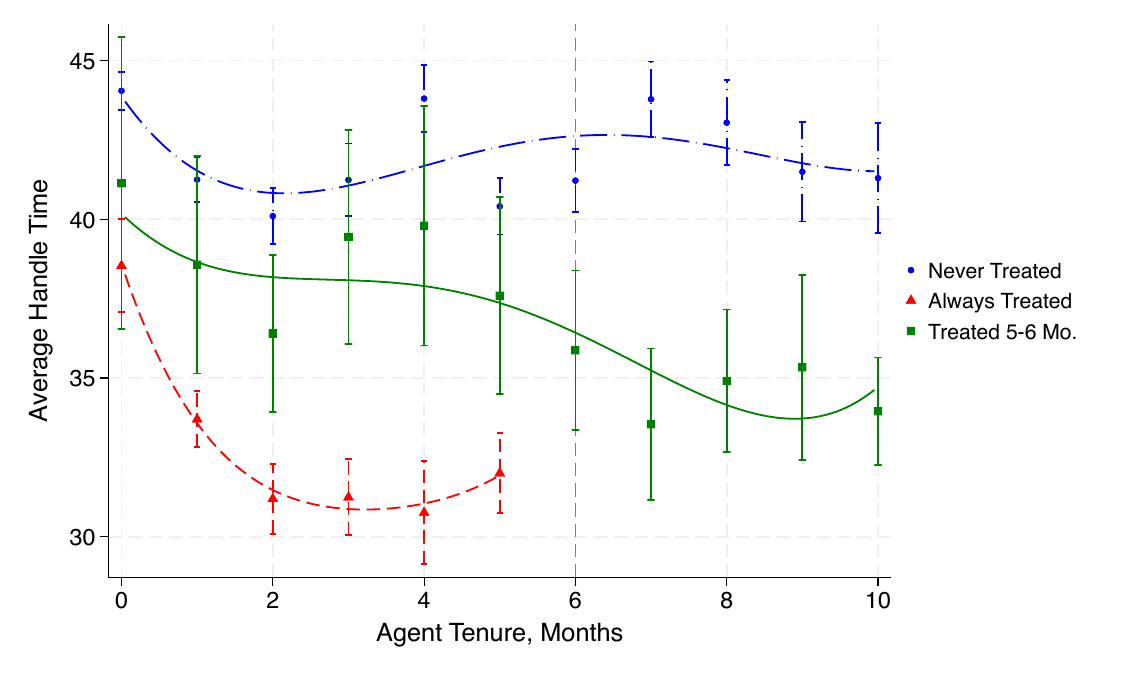} &\includegraphics[scale=0.5]{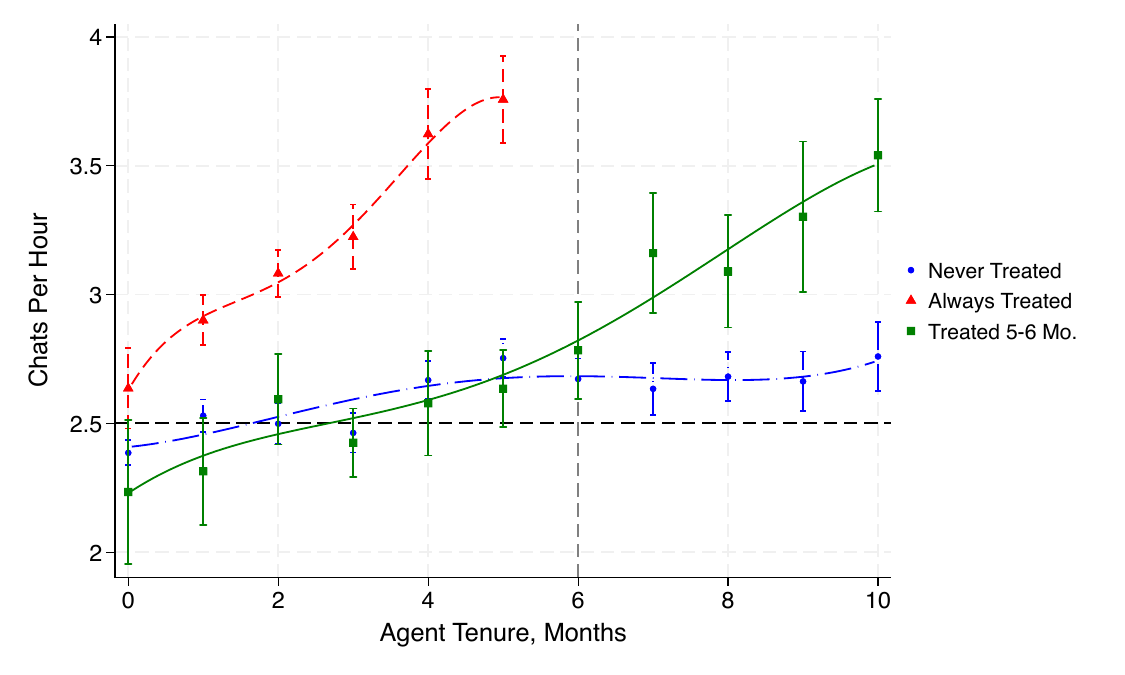} 
\\
\textsc{C. Resolution Rate} & \textsc{D.  Customer Satisfaction}\\ 
\includegraphics[scale=0.5]{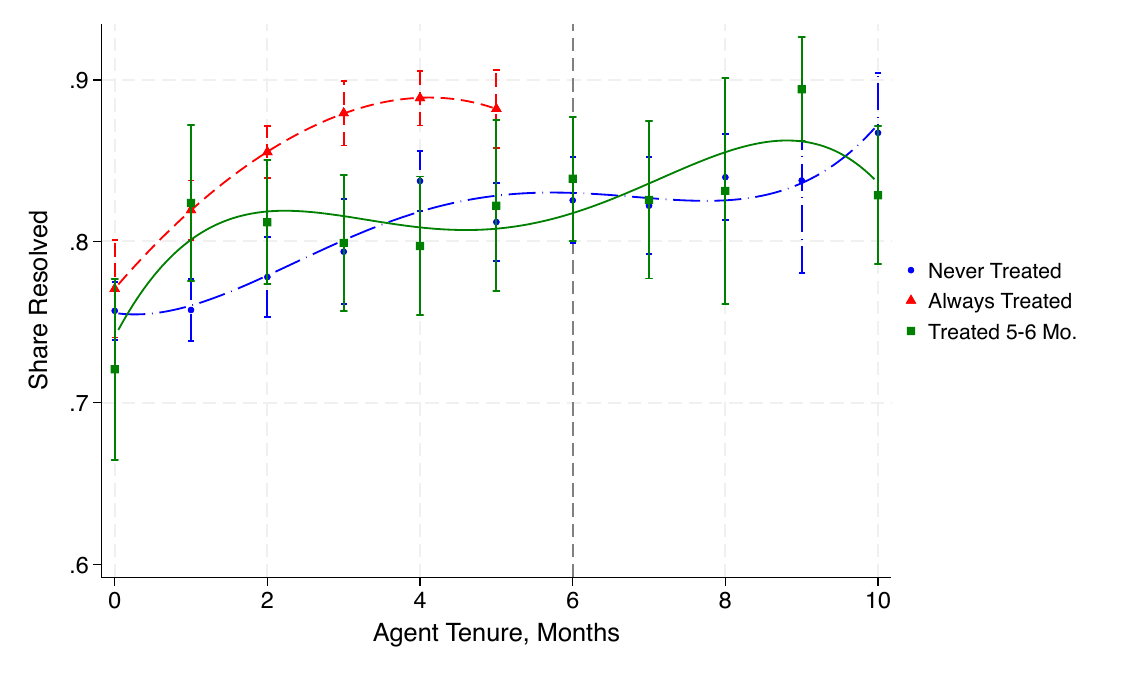} &\includegraphics[scale=0.5]{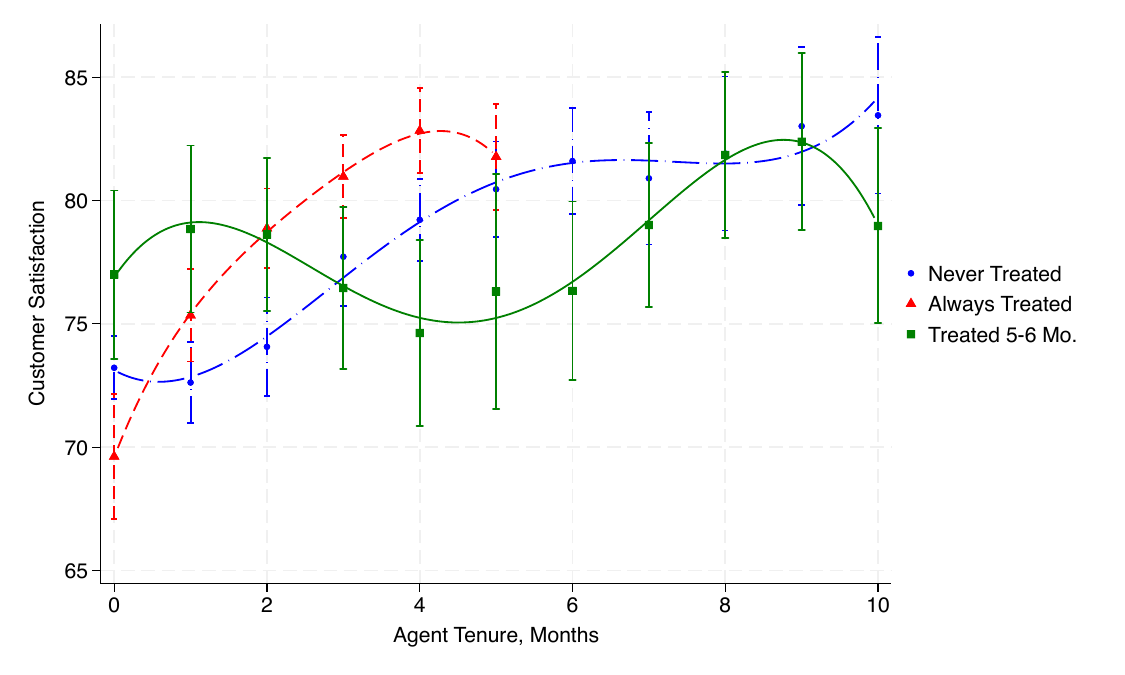} 
\\
\end{tabular}
}

\label{afig:learning_bycohort}
\end{center}
\end{figure}

\begin{footnotesize} 
\begin{singlespace}
\noindent \textsc{Notes}:  These figures plot the experience curves of three groups of agents over their tenure, the x-axis, against five measures of productivity and performance. The short-dashed red lines plot the performance of always-treated agents, those who are start work in their first month with the AI and always have access to the AI suggestions. The long-dashed blue line plots agents who are never treated. The solid green line plots agents who spend their first four months of work without the AI model, and gain access to the AI during their fifth month on the job. All panels include 95\% confidence intervals and observations are at the agent-month level. 
\end{singlespace}
\end{footnotesize}

\clearpage
\begin{figure}[ht!]
\begin{center}
\captionsetup{justification=centering}
\caption{\textsc{Figure \ref{afig:meanrev}: Resolutions per Hour over Time}}

\makebox[\linewidth]{
\includegraphics[scale=0.9]{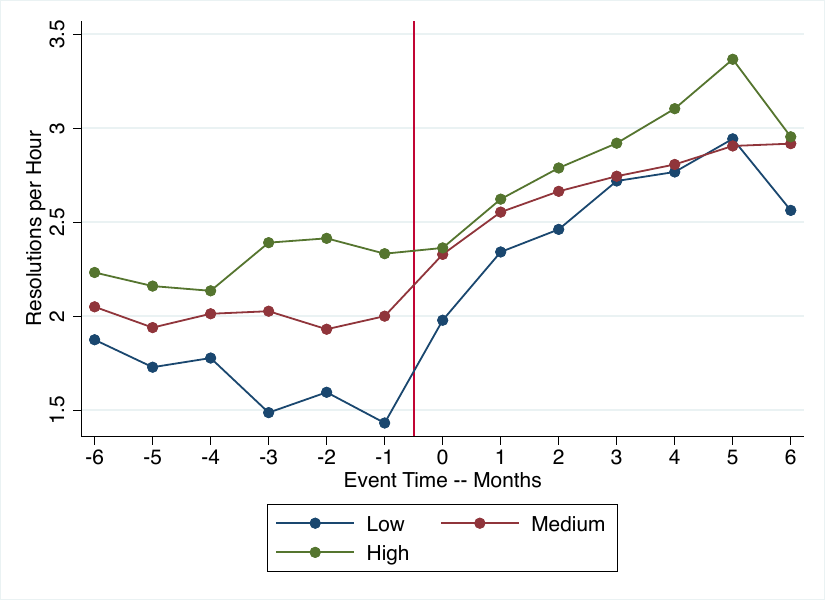} 
}

\label{afig:meanrev}
\end{center}
\end{figure}

\begin{footnotesize} 
\begin{singlespace}
\noindent \textsc{Notes}:  This graph depicts the evolution of average resolutions per hour for agents following the implementation of AI assistance. The graph segments agents into three groups based on their skill level at the time of AI deployment. The green line represents the highest-performing third of agents, those in the top tercile of the skill index. The red line illustrates the progress of agents in the middle tercile, while the blue line tracks those in the bottom tercile, representing the lowest-skilled third at the time of treatment. Agents are categorized based on their skill index at the time of AI implementation. For details on the skill index construction, refer to Appendix Section \ref{asec:key_vars}. 
\end{singlespace}
\end{footnotesize}





\clearpage
\begin{figure}[ht!]
\begin{center}
\captionsetup{justification=centering}
\caption{\textsc{Figure \ref{fig_aiym}: Productivity Impacts by Adoption Cohort}}

\vspace{15pt}
\makebox[\linewidth]{
\begin{tabular}{c}
\textsc{A. Resolutions Per Hour}\\
\includegraphics[scale=0.45]{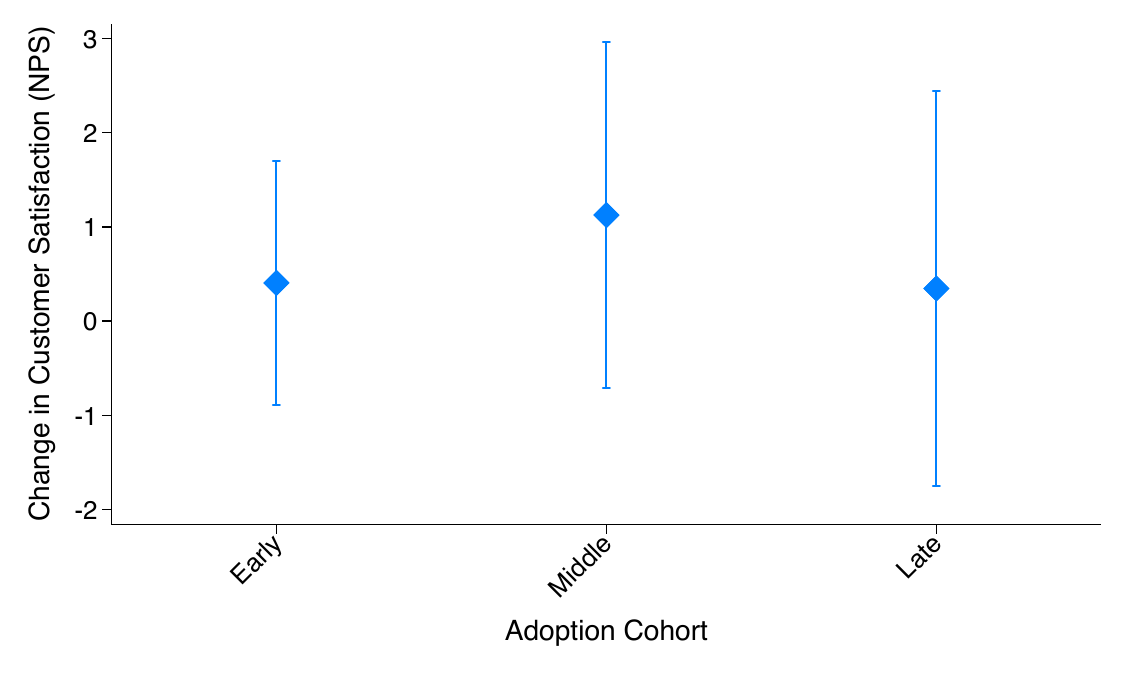} \\
\end{tabular}
}
\vspace{10pt}\\
\makebox[\linewidth]{
\begin{tabular}{cc}
\textsc{B. Average Handle Time} & \textsc{C.  Chats Per Hour}\\
\includegraphics[scale=0.4]{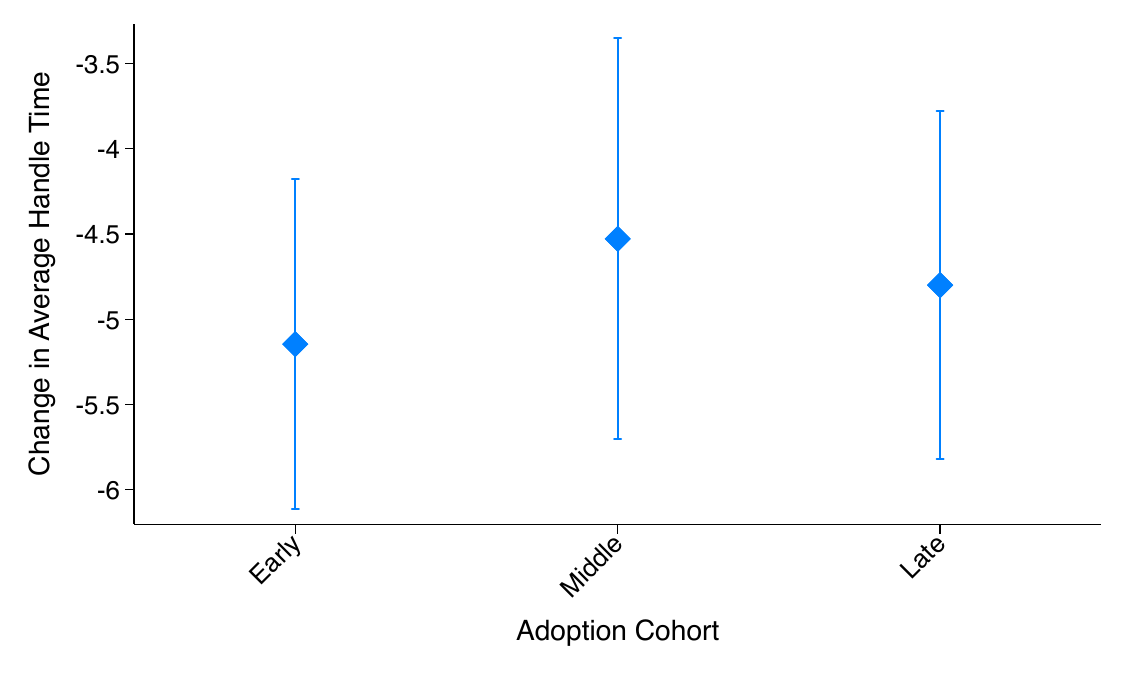} &\includegraphics[scale=0.4]{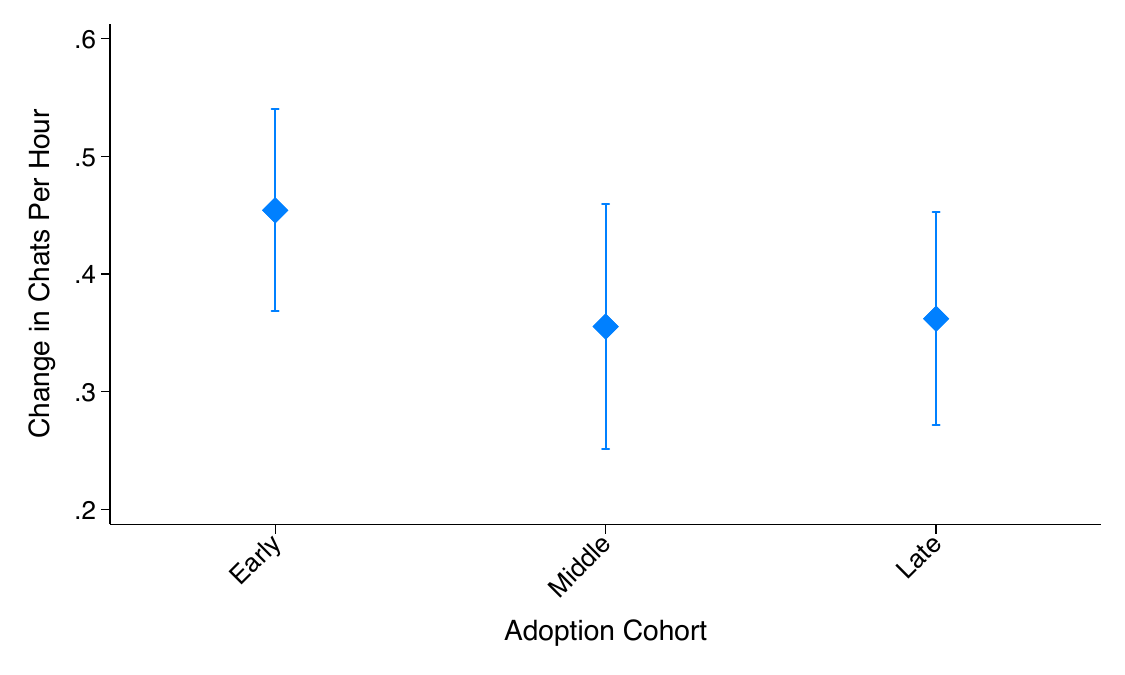} 
\\
\textsc{D. Resolution Rate} & \textsc{E. Customer Satisfaction (NPS)}\\
\includegraphics[scale=0.4]{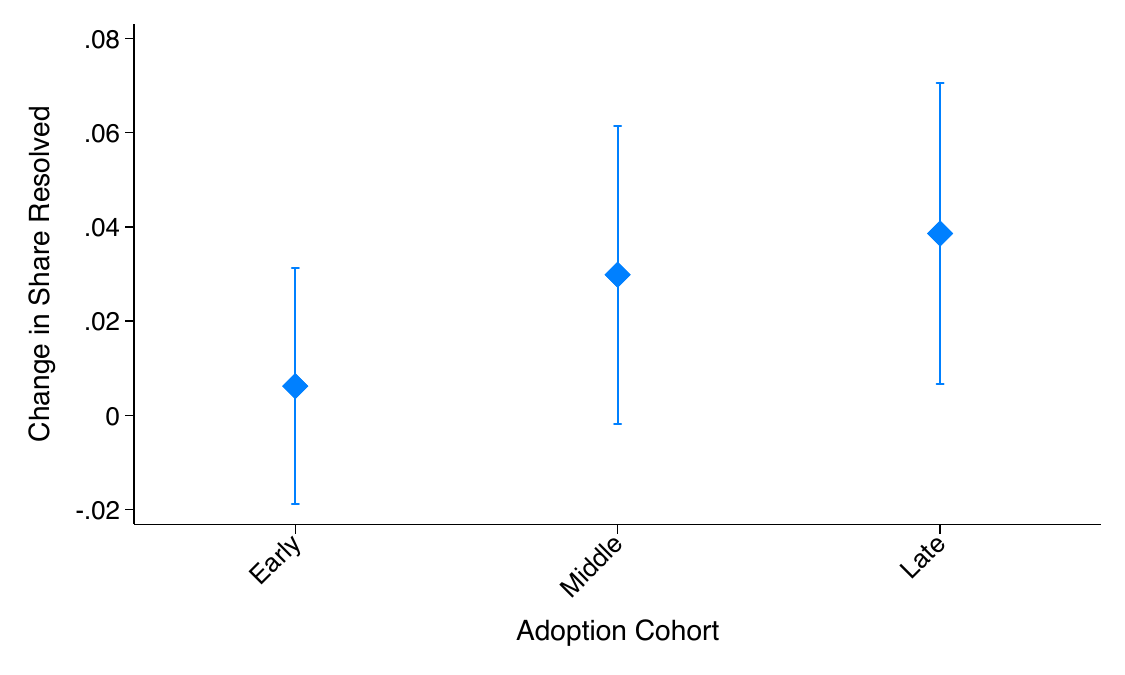} &\includegraphics[scale=0.4]{figures/15Aug2024/DD_tnps_cresta_byaiym.pdf} 
\end{tabular}
}
\label{fig_aiym}
\end{center}
\end{figure}
\begin{footnotesize} 
\begin{singlespace}
\noindent \textsc{Notes}: These figures plots the treatment effect of AI assistance on various outcomes for workers who received access to the AI model in the early, middle, or later period of the rollout.  Because the AI model is periodically updated with new data and pushed to all onboarded workers no more frequently than once a month, we focus on productivity outcomes in the first month after AI adoption in order to compare workers using earlier and later versions of the model. Data are at the agent-month, regression specification follows Equation 1 and we include chat year-month, agent, and months of tenure fixed effects. Robust standard errors are clustered at the agent level. 
\end{singlespace}
\end{footnotesize}

\clearpage
\section{Adherence to AI suggestions}\label{asec:adherence}


\clearpage
\begin{figure}[ht!]
\begin{center}
\captionsetup{justification=centering}
\caption{\textsc{Figure \ref{fig:adherence_decomposed}: Variation in AI Adherence}}
\makebox[\linewidth]{
\begin{tabular}{c}
\includegraphics[scale=0.75]{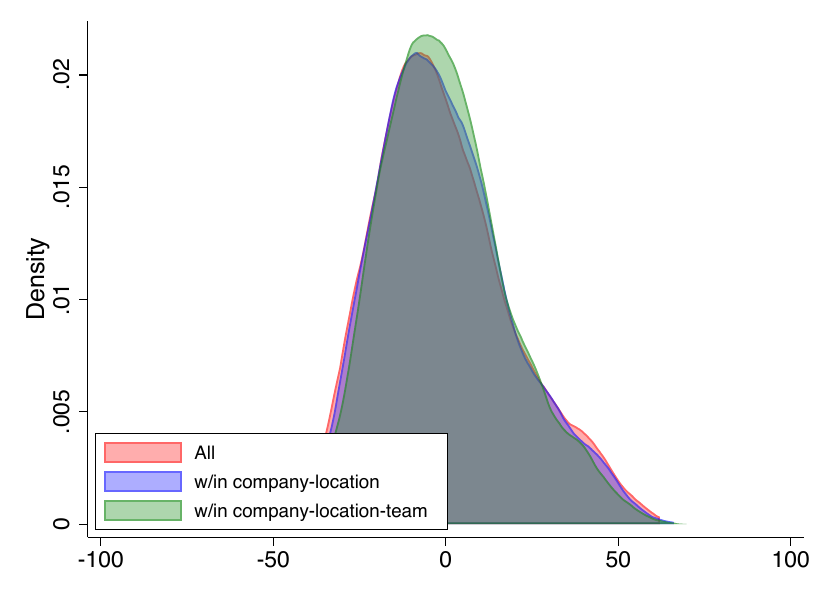} \\
\end{tabular}
}

\label{fig:adherence_decomposed}
\end{center}
\end{figure}

\begin{footnotesize} 
\begin{singlespace}
\noindent \textsc{Notes}: This figure plots the distribution of AI adherence, averaged at the agent-month level, weighted by the log of the number of AI recommendations for that agent-month. ``All'' refers to the overall distribution of unadjusted adherence rates.  ``within company-location'' plots residual adherence after adjusting for company and location fixed effects, and ``within company-location-team'' plots residuals adjusted for company, location, and team fixed effects.   
\end{singlespace}
\end{footnotesize}

\clearpage
\begin{figure}[ht!]
\begin{center}
\captionsetup{justification=centering}
\caption{\textsc{Figure \ref{afig:dd_byreceptivity}: Heterogeneity of AI Impact by Initial AI Adherence, Additional Outcomes}}
\makebox[\linewidth]{
\begin{tabular}{cc}
\textsc{A. Average Handle Time} & \textsc{B.  Chats Per Hour}\\
\includegraphics[scale=0.45]{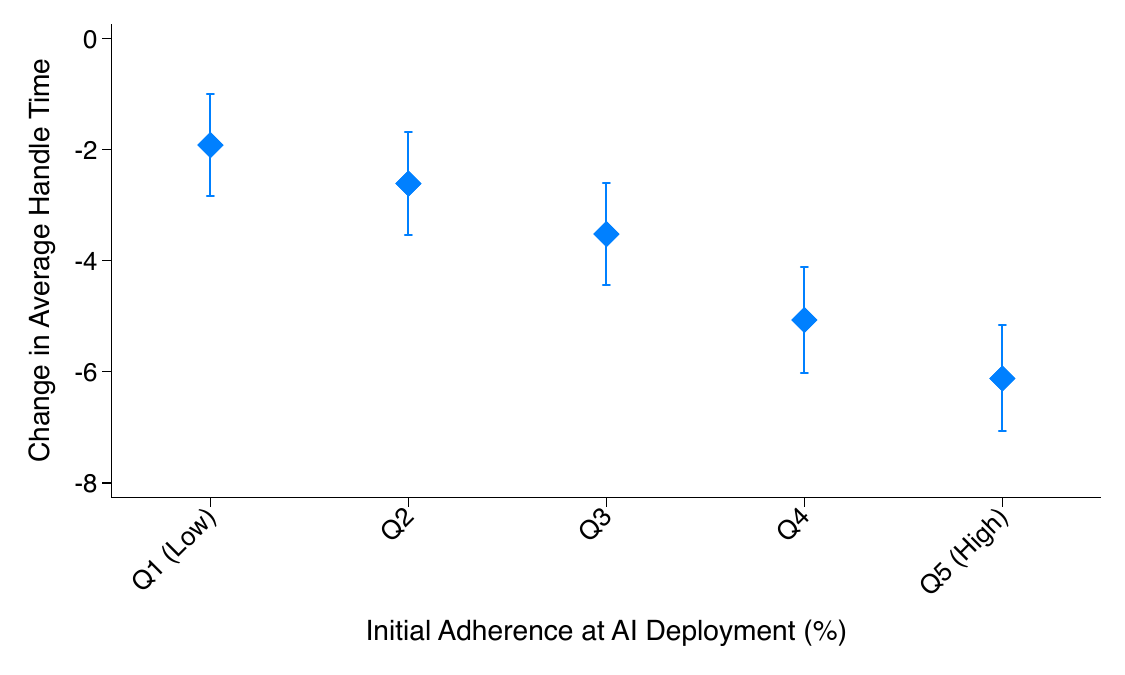} &\includegraphics[scale=0.45]{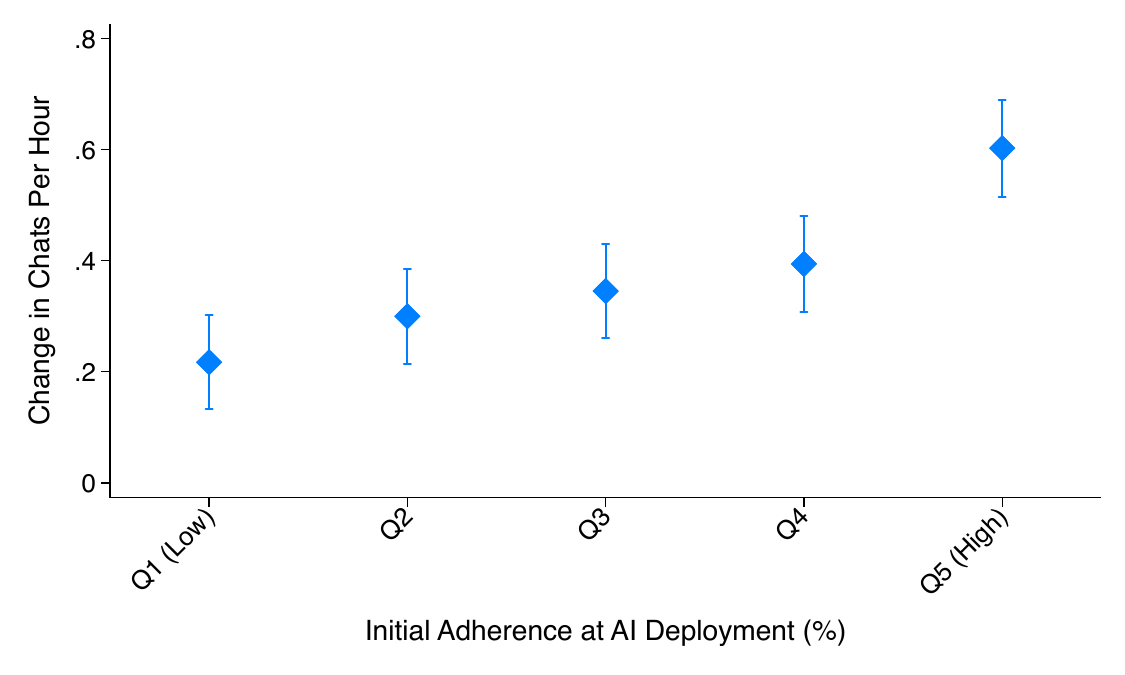} 
\\
\textsc{C. Resolution Rate} & \textsc{D.  Customer Satisfaction (NPS)}\\
\includegraphics[scale=0.45]{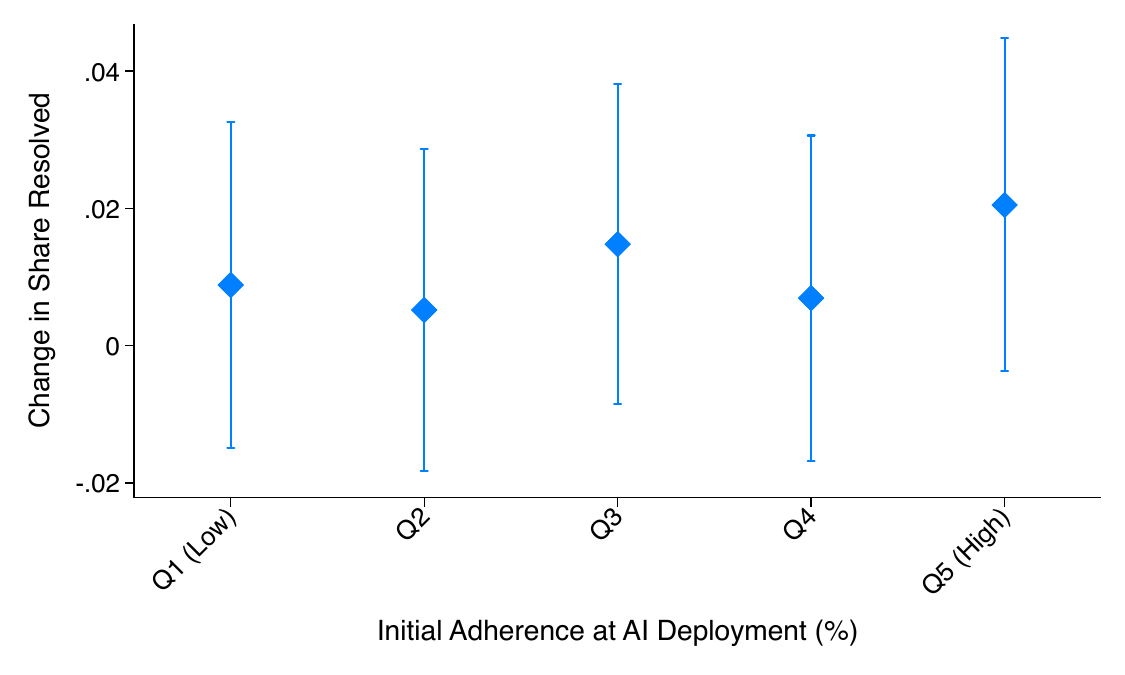} &\includegraphics[scale=0.45]{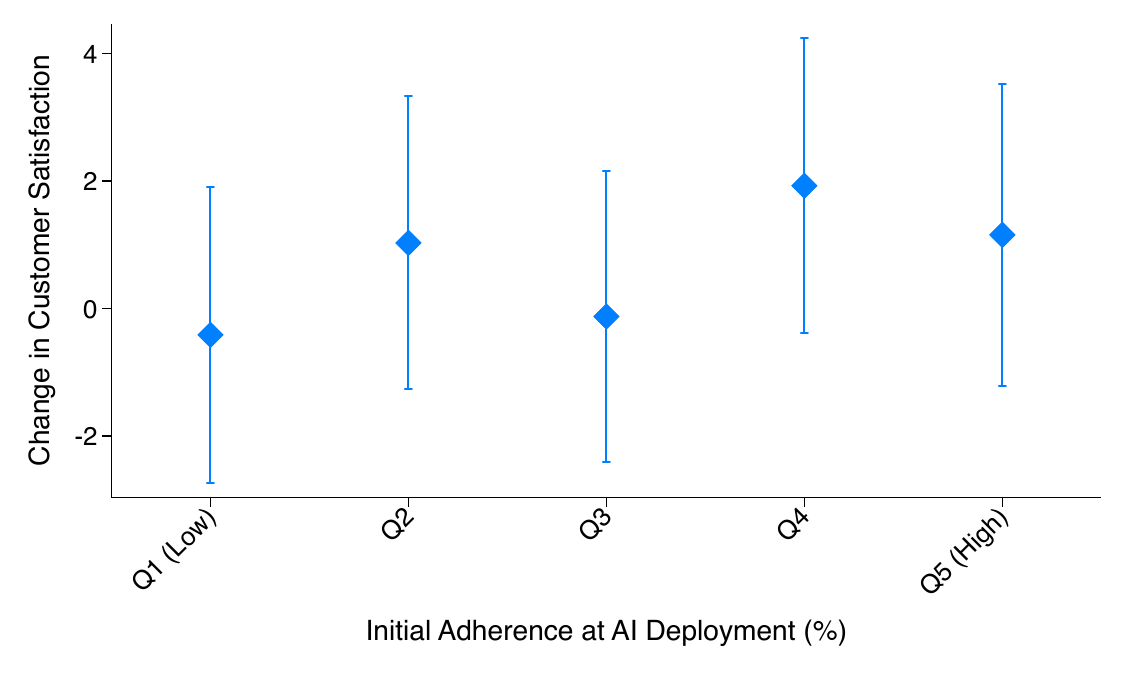} 
\\
\end{tabular}
}

\label{afig:dd_byreceptivity}
\end{center}
\end{figure}

\begin{footnotesize} 
\begin{singlespace}
\noindent \textsc{Notes}: These figures plot the impact of AI model deployment on additional measures of performance by quintile of initial adherence, the share of AI recommendations followed in the first month of treatment. Panel A plots the average handle time or the average duration of each technical support chat. Panel B graphs chats per hour, or the number of chats an agent can handle per hour (including working on multiple chats simultaneously). Panel C plots the resolution rate, the share of chats successfully resolved, and Panel D plots NPS, or net promoter score, is an average of surveyed customer satisfaction. Data is at the agent-level and all regressions include agent and chat year-month, months of agent tenure, with more details in Appendix section \ref{asec:specifications}.
\end{singlespace}
\end{footnotesize}

\clearpage
\begin{table}[ht!]
\begin{center}
                \caption{\textsc{Table \ref{atab:heteroeffects_adh}: Heterogeneity of AI Impact by Initial AI Adherence, Resolutions per Hour}}
                  \vspace{20pt}        
\scalebox{1}{\makebox[\linewidth]{{
\def\sym#1{\ifmmode^{#1}\else\(^{#1}\)\fi}
\begin{tabular}{l*{1}{l}}
\toprule
                &\multicolumn{1}{c}{(1)}\\
                &\multicolumn{1}{c}{By Adherence at AI}\\
\midrule
Q1 (Lowest Adherence)&0.213\sym{***}\\
                &(0.043)         \\
\addlinespace
Q2              &0.233\sym{***}\\
                &(0.042)         \\
\addlinespace
Q3              &0.293\sym{***}\\
                &(0.042)         \\
\addlinespace
Q4              &0.309\sym{***}\\
                &(0.042)         \\
\addlinespace
Q5 (Highest Adherence)&0.432\sym{***}\\
                &(0.043)         \\
\midrule
Year Month FE   &Yes         \\
Agent FE        &Yes         \\
Agent Tenure FE        &Yes         \\
DV Mean         &2.176         \\
Observations    &12,295         \\
\bottomrule
\multicolumn{2}{l}{\footnotesize Standard errors in parentheses}\\
\multicolumn{2}{l}{\footnotesize \sym{*} \(p<0.10\), \sym{**} \(p<0.05\), \sym{***} \(p<0.01\)}\\
\end{tabular}
}
}}
  \label{atab:heteroeffects_adh}
\end{center}
\end{table}

\begin{singlespace}
\footnotesize
\footnotesize
\noindent \textsc{Notes}:  This table presents the results of difference-in-difference regressions estimating the impact of AI model deployment on resolutions per hour by initial adherence.  Column 1 estimates the impact of AI access by initial adherence quintile, including agent, year-month and months of agent tenure fixed effects.  Observations are at the agent-month level and all standard errors are clustered at the agent level. We include chat year-month, agent and months of agent tenure fixed effects. 
\end{singlespace}


\clearpage
\begin{figure}[ht!]
\begin{center}
\captionsetup{justification=centering}
\caption{\textsc{Figure \ref{fig:receptivity_overtime}: AI Adherence over Time}}
\makebox[\linewidth]{
\begin{tabular}{c}
\textsc{A.  By Adherence at AI Model Deployment} \\
\includegraphics[scale=0.5]{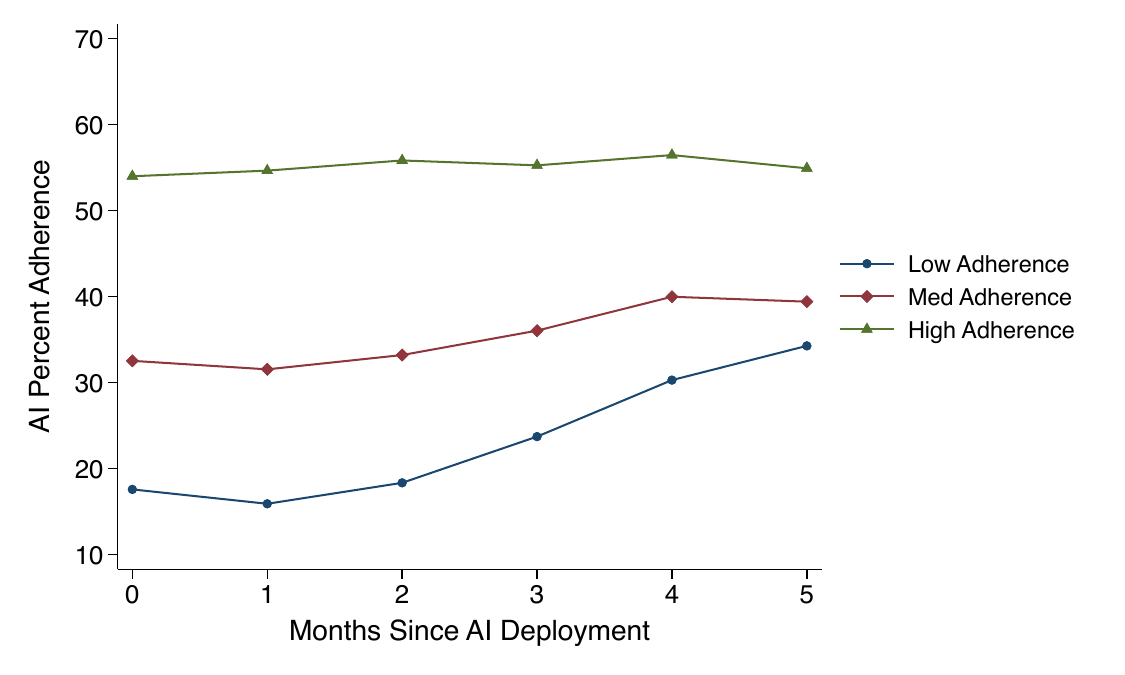} \\

\textsc{B.  By Agent Tenure at AI Model Deployment} \\
\includegraphics[scale=0.5]{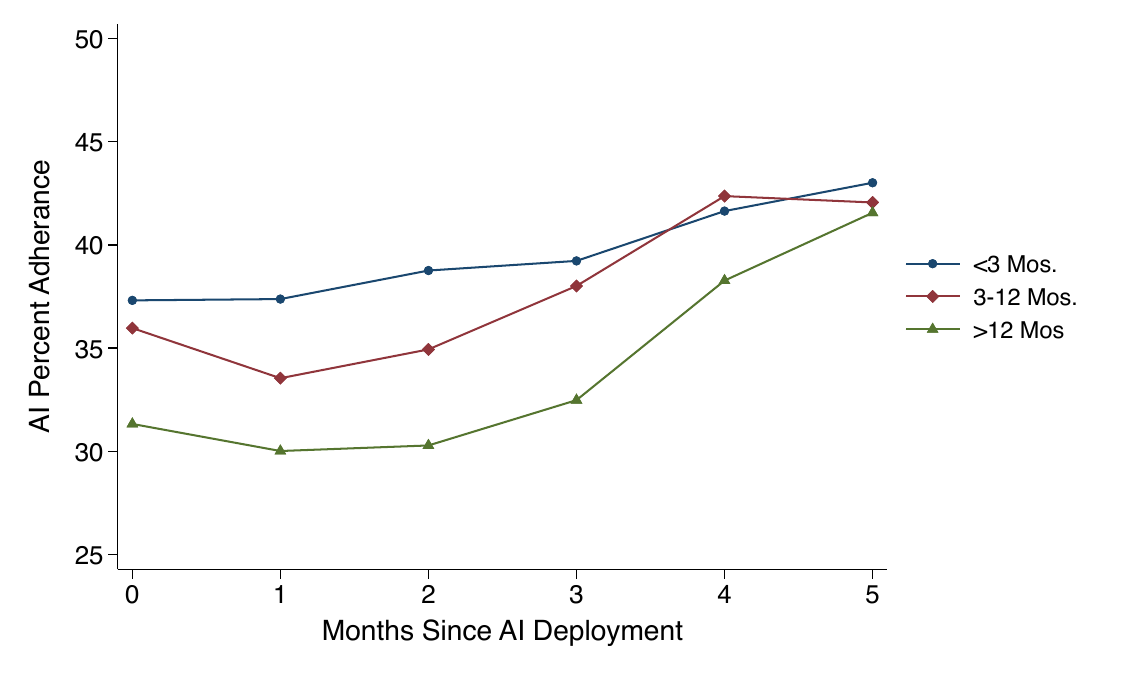} \\

\textsc{C.  By Agent Skill at AI Model Deployment} \\
\includegraphics[scale=0.5]{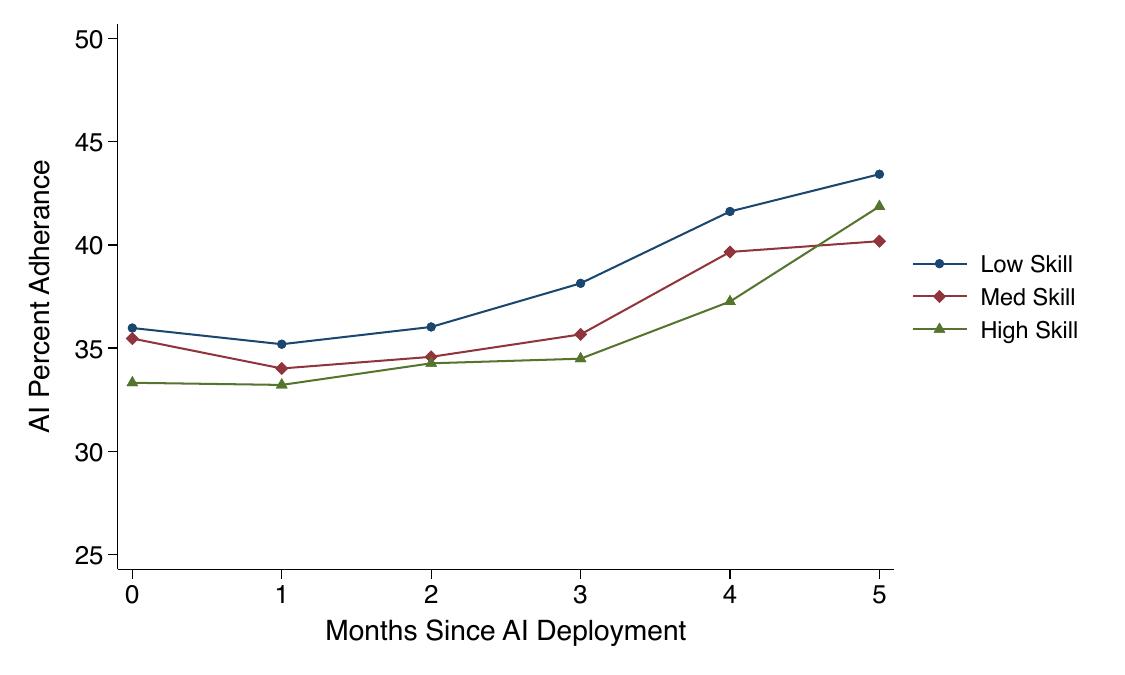} \\

\end{tabular}
}
\label{fig:receptivity_overtime}
\end{center}
\end{figure}

\begin{footnotesize} 
\begin{singlespace}
\noindent \textsc{Notes}:  This figure plots the share of AI suggestions followed by agents as a function of the number of months each agent has had access to the AI model.  In Panel A, we divide agents into terciles based on their adherence to AI suggestions in the first month.  In Panel B, we divide agents into groups based on their tenure at the firm at the time of AI model deployment.  In Panel C, we divide workers into terciles of pre-deployment productivity as defined by our skill index. 
\end{singlespace}
\end{footnotesize}

\clearpage
\begin{figure}[ht!]
\begin{center}
\captionsetup{justification=centering}
\caption{\textsc{Figure \ref{fig:receptivity_overtime_agentfe}: Within-agent AI Adherence over Time}}
\makebox[\linewidth]{
\begin{tabular}{c}
\textsc{A.  By Adherence at AI Model Deployment} \\
\includegraphics[scale=0.5]{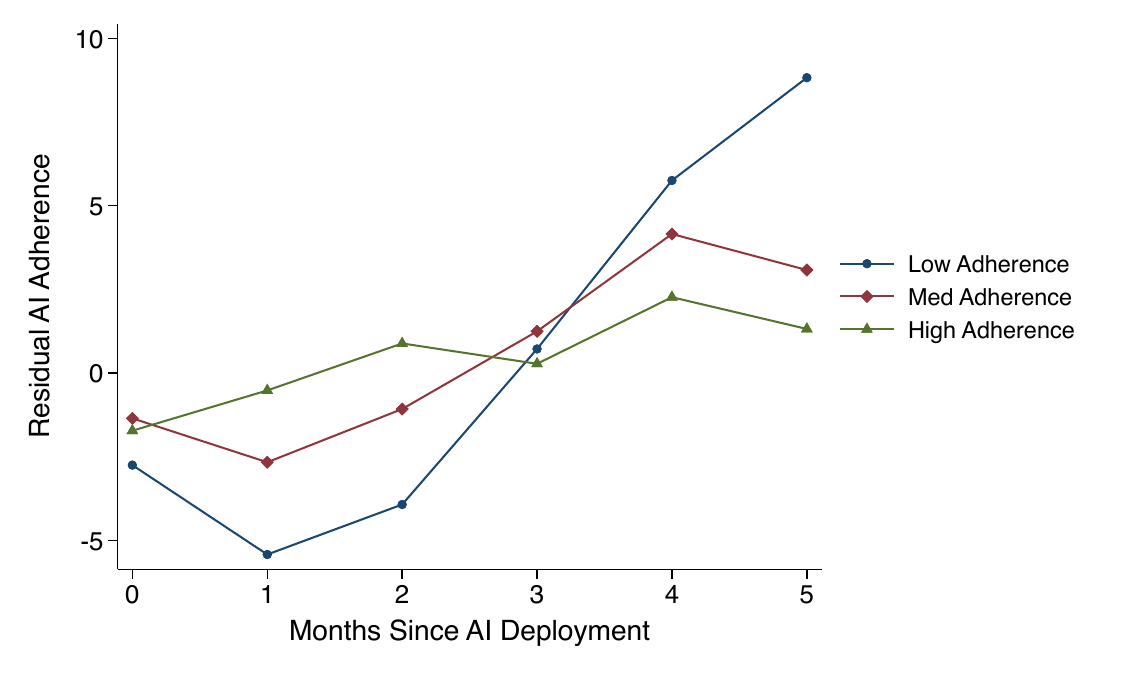} \\

\textsc{B.  By Agent Tenure at AI Model Deployment} \\
\includegraphics[scale=0.5]{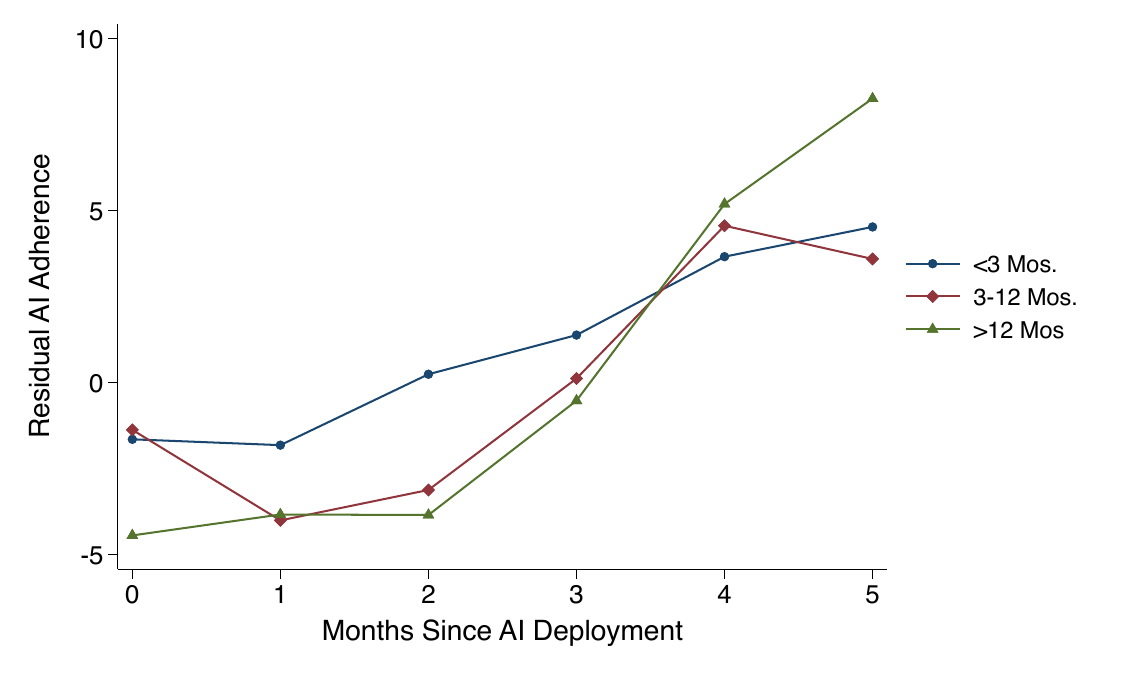} \\

\textsc{C.  By Agent Skill at AI Model Deployment} \\
\includegraphics[scale=0.5]{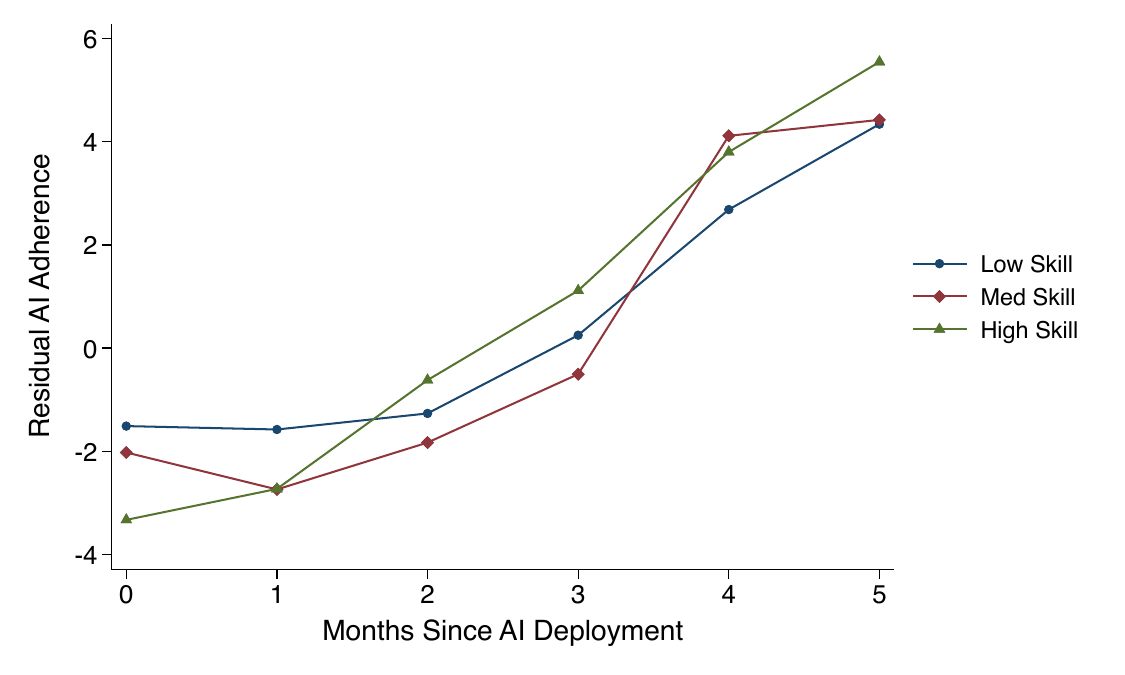} \\

\end{tabular}
}
\label{fig:receptivity_overtime_agentfe}
\end{center}
\end{figure}

\begin{footnotesize} 
\begin{singlespace}
\noindent \textsc{Notes}:  This figure plots the residualized percentage of AI suggestions followed by agents as a function of the number of months each agent has had access to the AI model, after controlling for agent level fixed effects.  In Panel A, we divide agents into terciles based on their adherence to AI suggestions in the first month.  In Panel B, we divide agents into groups based on their tenure at the firm at the time of AI model deployment.  In Panel C, we divide workers into terciles of pre-deployment productivity as defined by our skill index.   
\end{singlespace}
\end{footnotesize}

\clearpage
\section{Worker Learning}\label{asec:learning}


\clearpage
\begin{figure}[ht!]
\begin{center}
\captionsetup{justification=centering}
\caption{\textsc{Figure \ref{afig:outage_example}: Sample AI Outage}}
\makebox[\linewidth]{
\includegraphics[scale=0.85]{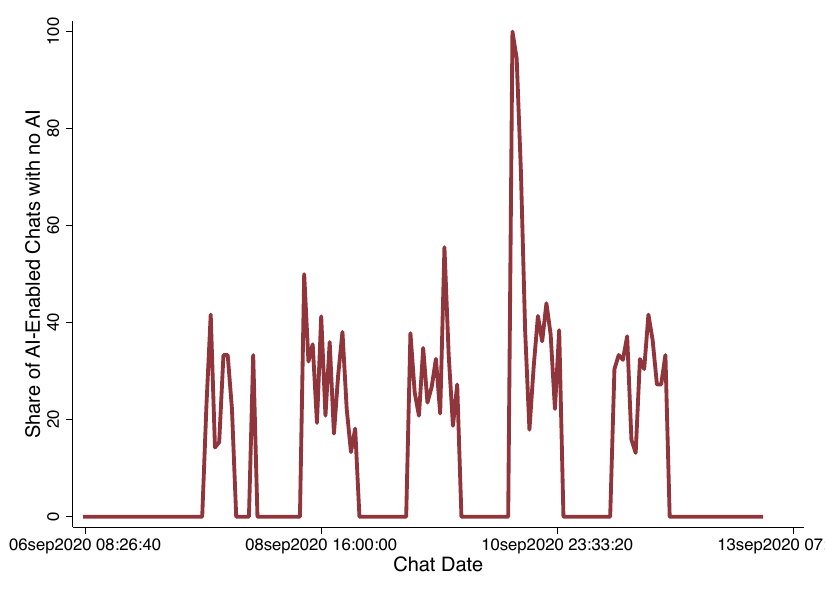} 
}

\label{afig:outage_example}
\end{center}
\end{figure}

\begin{footnotesize} 
\begin{singlespace}
\noindent \textsc{Notes}: This figure plots the share of post-treatment chats with no AI suggestions during a period of a documented software outage.   
\end{singlespace}
\end{footnotesize}

\clearpage

\begin{table}[ht!]
\begin{center}
                \caption{\textsc{Table \ref{atab:ES_outages}: Chat Duration during AI System Outages}}
                  \vspace{20pt}        
\scalebox{.55}{\makebox[\linewidth]{{
\def\sym#1{\ifmmode^{#1}\else\(^{#1}\)\fi}
\begin{tabular}{l*{4}{l}}
\toprule
                &\multicolumn{1}{c}{(1)}&\multicolumn{1}{c}{(2)}&\multicolumn{1}{c}{(3)}&\multicolumn{1}{c}{(4)}\\
                &\multicolumn{1}{c}{Non Outage}&\multicolumn{1}{c}{AI Outages}&\multicolumn{1}{c}{\shortstack{Outages \\ High Receptivity}}&\multicolumn{1}{c}{\shortstack{Outages \\ Low Receptivity}}\\
\midrule
lead10          &0.684         &2.085\sym{***}&3.391\sym{***}&0.284         \\
                &(0.550)         &(0.581)         &(0.880)         &(1.590)         \\
\addlinespace
lead9           &0.549         &2.126\sym{***}&4.056\sym{***}&1.003         \\
                &(0.529)         &(0.551)         &(0.890)         &(1.388)         \\
\addlinespace
lead8           &-0.222         &1.410\sym{***}&3.239\sym{***}&0.340         \\
                &(0.479)         &(0.504)         &(0.771)         &(1.299)         \\
\addlinespace
lead7           &-0.302         &1.267\sym{***}&2.354\sym{***}&1.025         \\
                &(0.470)         &(0.486)         &(0.798)         &(1.141)         \\
\addlinespace
lead6           &0.874\sym{*}  &2.204\sym{***}&4.081\sym{***}&2.210\sym{**} \\
                &(0.463)         &(0.468)         &(0.786)         &(1.079)         \\
\addlinespace
lead5           &0.439         &1.444\sym{***}&2.953\sym{***}&1.736\sym{*}  \\
                &(0.389)         &(0.378)         &(0.591)         &(0.885)         \\
\addlinespace
lead4           &0.415         &1.052\sym{***}&2.222\sym{***}&1.258\sym{*}  \\
                &(0.336)         &(0.321)         &(0.471)         &(0.703)         \\
\addlinespace
lead3           &0.135         &0.343         &1.314\sym{***}&0.218         \\
                &(0.305)         &(0.297)         &(0.482)         &(0.627)         \\
\addlinespace
lead2           &-0.146         &-0.209         &0.723\sym{*}  &-0.869\sym{*}  \\
                &(0.248)         &(0.247)         &(0.411)         &(0.469)         \\
\addlinespace
lag0            &-4.105\sym{***}&-3.471         &-5.236         &9.410\sym{***}\\
                &(0.270)         &(3.713)         &(3.732)         &(3.564)         \\
\addlinespace
lag1            &-4.523\sym{***}&-2.420         &-6.699\sym{*}  &0.000         \\
                &(0.313)         &(3.242)         &(4.052)         &(.)         \\
\addlinespace
lag2            &-4.785\sym{***}&0.767         &-0.201         &0.000         \\
                &(0.335)         &(3.112)         &(3.858)         &(.)         \\
\addlinespace
lag3            &-4.424\sym{***}&-2.851         &-2.677         &6.005         \\
                &(0.333)         &(2.663)         &(3.915)         &(4.099)         \\
\addlinespace
lag4            &-4.225\sym{***}&-2.423\sym{*}  &-4.011\sym{**} &4.851         \\
                &(0.350)         &(1.442)         &(2.015)         &(3.663)         \\
\addlinespace
lag5            &-4.084\sym{***}&0.838         &0.695         &7.871         \\
                &(0.380)         &(1.744)         &(2.042)         &(5.904)         \\
\addlinespace
lag6            &-4.300\sym{***}&-7.799\sym{***}&-8.574\sym{***}&-2.303         \\
                &(0.388)         &(1.836)         &(2.168)         &(3.167)         \\
\addlinespace
lag7            &-4.445\sym{***}&-8.494\sym{***}&-9.532\sym{***}&-2.161         \\
                &(0.393)         &(1.635)         &(1.687)         &(3.209)         \\
\addlinespace
lag8            &-3.804\sym{***}&-4.672\sym{*}  &-2.163         &-8.059\sym{***}\\
                &(0.411)         &(2.647)         &(2.471)         &(3.023)         \\
\addlinespace
lag9            &-3.628\sym{***}&-8.705\sym{***}&-8.842\sym{***}&0.318         \\
                &(0.440)         &(1.660)         &(1.930)         &(5.166)         \\
\addlinespace
lag10           &-3.609\sym{***}&-4.386\sym{**} &-4.486\sym{**} &-0.233         \\
                &(0.501)         &(1.720)         &(2.040)         &(4.273)         \\
\addlinespace
lag11           &-3.305\sym{***}&-2.779\sym{**} &-6.857\sym{**} &0.539         \\
                &(0.486)         &(1.380)         &(3.252)         &(4.003)         \\
\addlinespace
lag12           &-3.479\sym{***}&-1.966         &-4.630         &-8.939\sym{**} \\
                &(0.517)         &(3.003)         &(4.248)         &(3.510)         \\
\midrule
Year Month FE   &Yes         &Yes         &Yes         &Yes         \\
Agent FE        &Yes         &Yes         &Yes         &Yes         \\
Agent Tenure FE &Yes         &Yes         &Yes         &Yes         \\
DV Mean         &41.982         &41.982         &41.982         &41.982         \\
R-squared       &.12         &.104         &.113         &.083         \\
Observations    &2,969,371         &1,829,527         &1,220,852         &323,810         \\
\bottomrule
\multicolumn{5}{l}{\footnotesize Standard errors in parentheses}\\
\multicolumn{5}{l}{\footnotesize \sym{*} \(p<0.10\), \sym{**} \(p<0.05\), \sym{***} \(p<0.01\)}\\
\end{tabular}
}
}}
  \label{atab:ES_outages}
\end{center}
\end{table}

\begin{singlespace}
\footnotesize
\noindent \textsc{Notes}: This table shows the event study coefficients corresponding to the figures in \ref{fig:ES_outage}. Column 1 shows the impact of AI access on call duration during non-outage periods, Column 2 shows the impact when the AI system is experiencing an outage. Column 3 shows the effects on agents who are in the top tercile of AI adherence, while Column 4 shows the impacts on agents in the lowest tercile of adherence. Observations for these OLS regressions are at the agent-chat level and robust standard errors are clustered at the agent level. All specifications include agent, chat year-month, and months of agent tenure fixed effects. 
\end{singlespace}
\normalsize

\clearpage
\section{Topic Heterogeneity}\label{asec:topic}


\clearpage

\begin{footnotesize} 
\begin{singlespace}
\noindent \textsc{Notes}: This figure shows the distribution of chats by common topics in our data. 
\end{singlespace}
\end{footnotesize}

\clearpage
\begin{table}[ht!]
\begin{center}
\caption{\textsc{Table \ref{atab:dd_main_robust}: Impact of AI on Chat Duration, by Topic Commonality}}
\vspace{20pt}        
 \scalebox{1}{\makebox[\linewidth]{\begin{tabular}{lcc} \hline
 & (1) & (2) \\
VARIABLES & \shortstack{Call Duration \\ Aggregate Problem Frequency} & \shortstack{Call Duration \\ Within-Agent Frequency }\\ \hline
 &  &  \\
Q1 (Most Common Overall) & -0.107*** &  \\
 & (0.00717) &  \\
Q2 & -0.115*** &  \\
 & (0.00724) &  \\
Q3 & -0.136*** &  \\
 & (0.00742) &  \\
Q4 (Least Common Overall) & -0.118*** &  \\
 & (0.00842) &  \\
 Q1 (Most Common within Agent) &  & -0.100*** \\
 &  & (0.00927) \\
Q2 &  & -0.103*** \\
 &  & (0.00740) \\
Q3 &  & -0.119*** \\
 &  & (0.00700) \\
Q4 (Least Common within Agent) &  & -0.142*** \\
 &  & (0.00708) \\
 &  &  \\
Observations & 2,089,995 & 2,089,995 \\
R-squared & 0.120 & 0.122 \\
Year Month FE & Yes & Yes \\
Agent FE & Yes & Yes \\
Agent Tenure FE & Yes & Yes \\
 DV Mean & 0.916 & 0.915 \\ \hline
\multicolumn{3}{c}{ Robust standard errors in parentheses} \\
\multicolumn{3}{c}{ *** p$<$0.01, ** p$<$0.05, * p$<$0.10} \\
\end{tabular}
}}
   \label{atab:dd_main_robust}
\end{center}
\end{table}

\begin{singlespace}
\footnotesize
\noindent \textsc{Notes}: This table shows the impact of AI model deployment on call duration by frequency of the chat topic. Regression include controls for chat year-month, agent and months of agent tenure. Data is at the chat level and robust standard errors are clustered at the agent level. Appendix section \ref{asec:key_vars} describes construction of topics and the regression specifications. 
\end{singlespace}
\normalsize

\clearpage
\section{Language Fluency}\label{asec:fluency}


\clearpage
\begin{figure}[ht!]
\begin{center}
\captionsetup{justification=centering}
\caption{\textsc{Figure \ref{afig:englishfluency1}: Distribution of Language Skills}}	

\makebox[\linewidth]{
    \begin{tabular}{c}
    \textsc{A.  Comprehensibility Score} \\
    \includegraphics[scale=0.85]{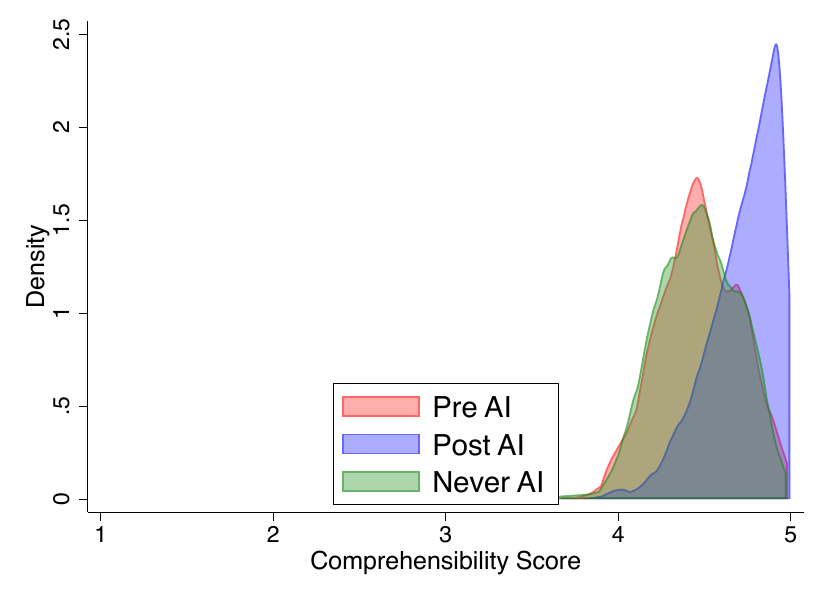} \\
    \textsc{B.  Native Fluency Score}\\
    \includegraphics[scale=0.85]{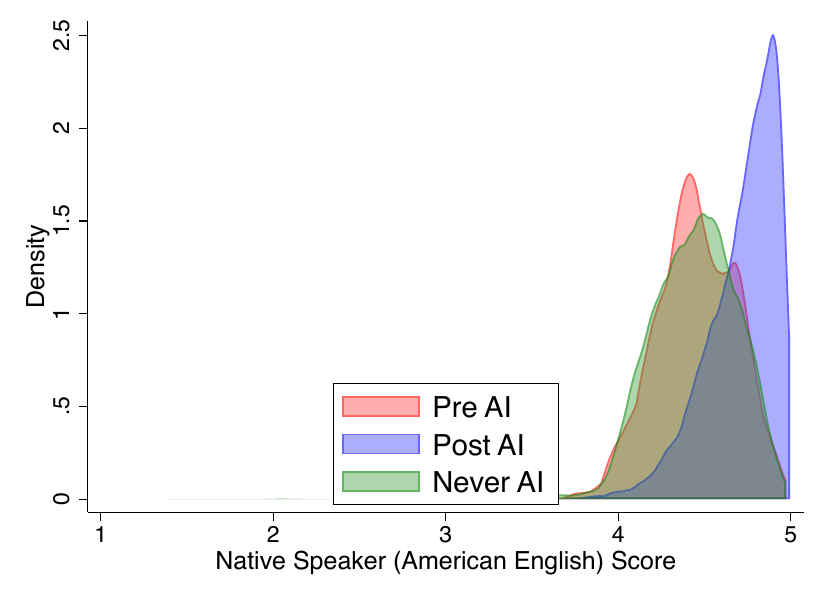}\\
    \end{tabular}
}
\label{afig:englishfluency1}

\end{center}
\end{figure}

\begin{footnotesize} 
	\begin{singlespace}
		\noindent \textsc{Notes}: These figures show the distributions of comprehensibility and native fluency scores. We split this sample into agent-month observations for agents who eventually receive access to the AI system before deployment (``Pre AI''), after deployment (``Post AI''), and for agent-months associated with agents who never receive access (``Never AI'').  Appendix Section \ref{asec:key_vars} describes construction of the language scores.  
	\end{singlespace}
\end{footnotesize}

\clearpage
\begin{table}[ht!]
\begin{center}
                \caption{\textsc{Table \ref{atab:languageDiD}: AI Impact on Language Fluency and Comprehensibility}}
                  \vspace{20pt}        
\scalebox{1}{\makebox[\linewidth]{\begin{tabular}{lcccc} \hline
 & (1) & (2) & (3) & (4) \\
VARIABLES & Native Fluency & Comprehensibility & Native Fluency & Comprehensibility \\ \hline
 &  &  &  &  \\
Post AI X Ever Treated & 0.250*** & 0.241*** &  &  \\
 & (0.00851) & (0.00767) &  &  \\
Post AI X Ever Treated X US = 1 &  &  & 0.159*** & 0.134*** \\
 &  &  & (0.0225) & (0.0208) \\
Post AI X Ever Treated X Philippines = 1 &  &  & 0.251*** & 0.234*** \\
 &  &  & (0.00684) & (0.00672) \\
 &  &  &  &  \\
Observations & 12,772 & 12,772 & 11,271 & 11,271 \\
R-squared & 0.791 & 0.754 & 0.733 & 0.751 \\
Year Month FE & Yes & Yes & Yes & Yes \\
Agent FE & Yes & Yes & Yes & Yes \\
Agent Tenure FE & Yes & Yes & Yes & Yes \\
 DV Mean & 4.564 & 4.597 & 4.592 & 4.608 \\ \hline
\multicolumn{5}{c}{ Robust standard errors in parentheses} \\
\multicolumn{5}{c}{ *** p$<$0.01, ** p$<$0.05, * p$<$0.10} \\
\end{tabular}
}}
  \label{atab:languageDiDh}
\end{center}
\end{table}

\begin{singlespace}
\footnotesize
\footnotesize
\noindent \textsc{Notes}:  This table presents the results of difference-in-difference regressions estimating the impact of AI model deployment on native fluency and comprehensibility.  Observations are aggregated to the agent-month level and regressions include chat year-month, agent and months of agent tenure fixed effects. In Columns 1 and 2, robust standard errors are clustered at the agent level and at the agent location level in Columns 3 and 4. Appendix Section \ref{asec:key_vars} describes construction of the language scores. 
\end{singlespace}

\clearpage
\section{Experience of Work}\label{asec:expofwork}


\clearpage
\begin{figure}[ht!]
	\begin{center}
		\captionsetup{justification=centering}
		\caption{\textsc{Figure \ref{afig:expwork}: Experience of Work}}		
		\makebox[\linewidth]{
			\begin{tabular}{cc}
				\textsc{A.  Customer Sentiment, Histogram} & 	\textsc{B.  Agent Sentiment, Histogram}\\
				\includegraphics[scale=0.4]{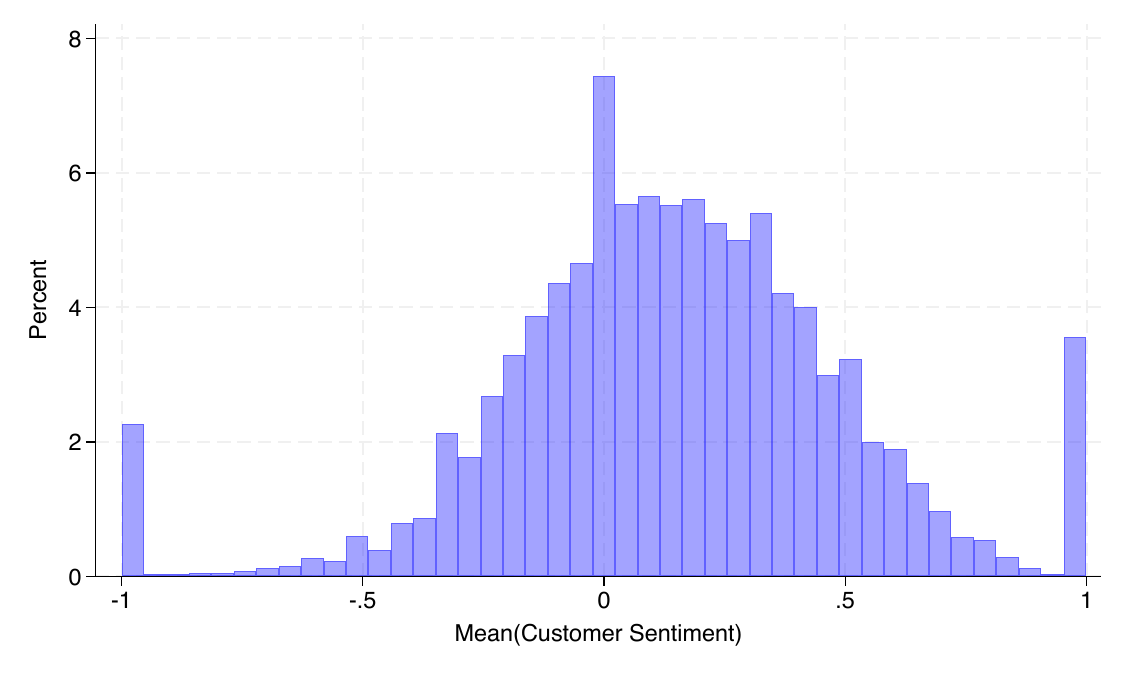} & 	\includegraphics[scale=0.4]{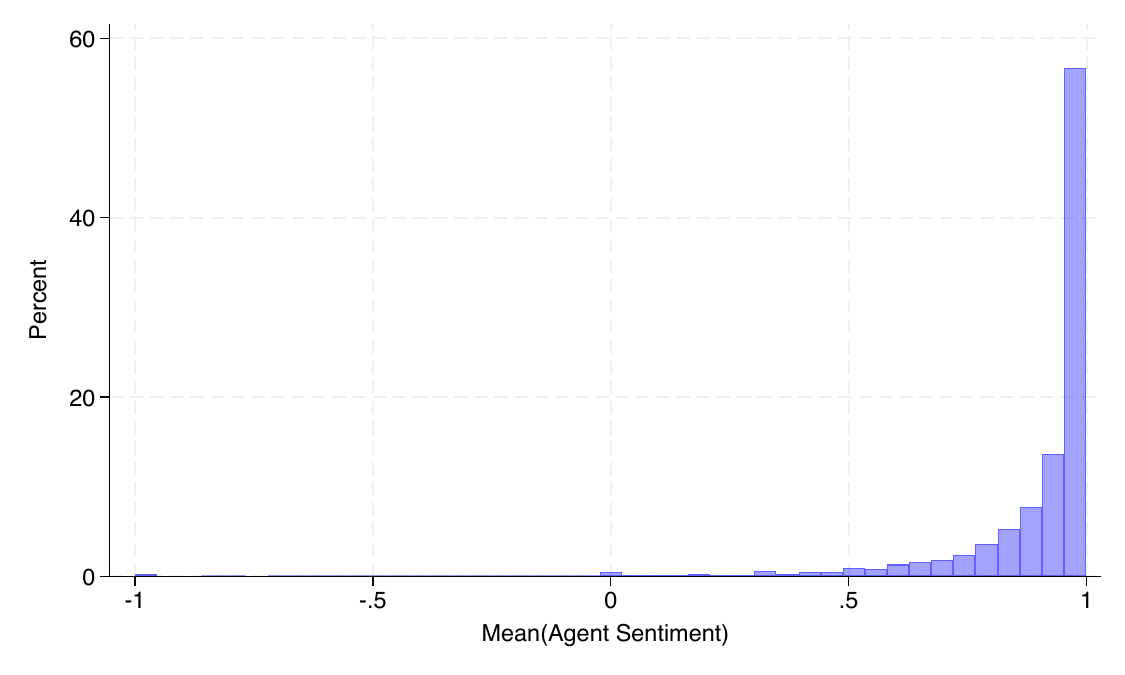} \\
				\textsc{C.  Customer Sentiment, by Pre-AI Skill} & 	\textsc{D.   Customer Sentiment, by Pre-AI Tenure }\\
	\includegraphics[scale=0.4]{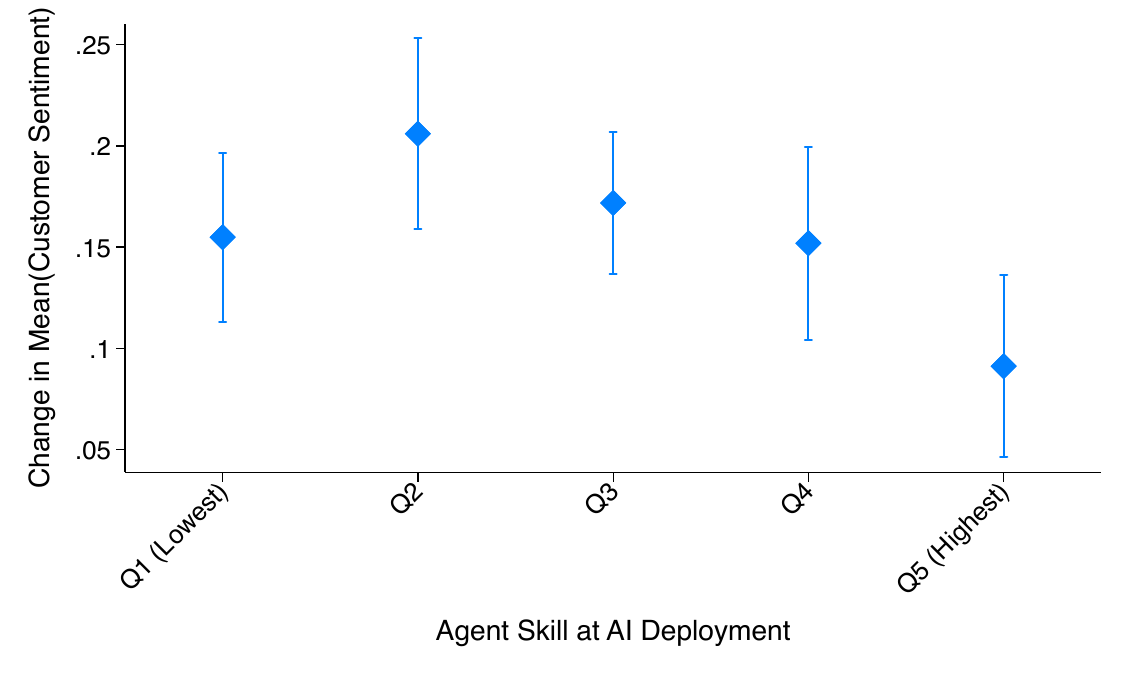} & 	\includegraphics[scale=0.4]{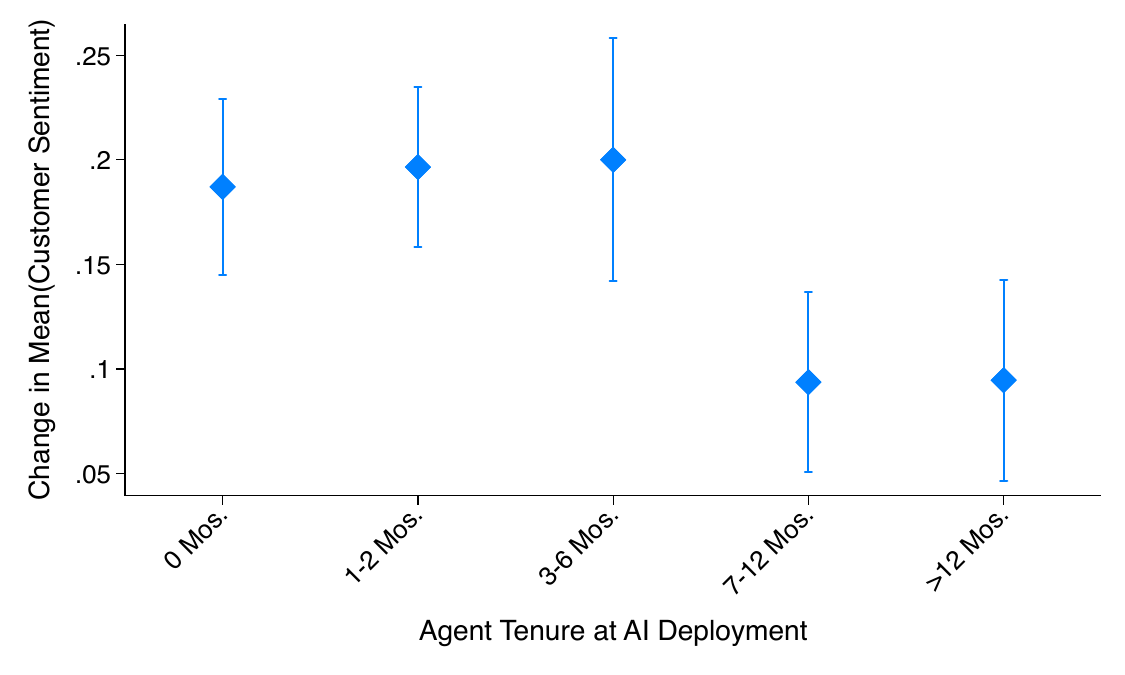} \\
 	\textsc{E.  Manager Assistance, by Pre-AI Skill } & \textsc{F.  Manager Assistance, by Pre-AI Tenure}\\
				\includegraphics[scale=0.4]{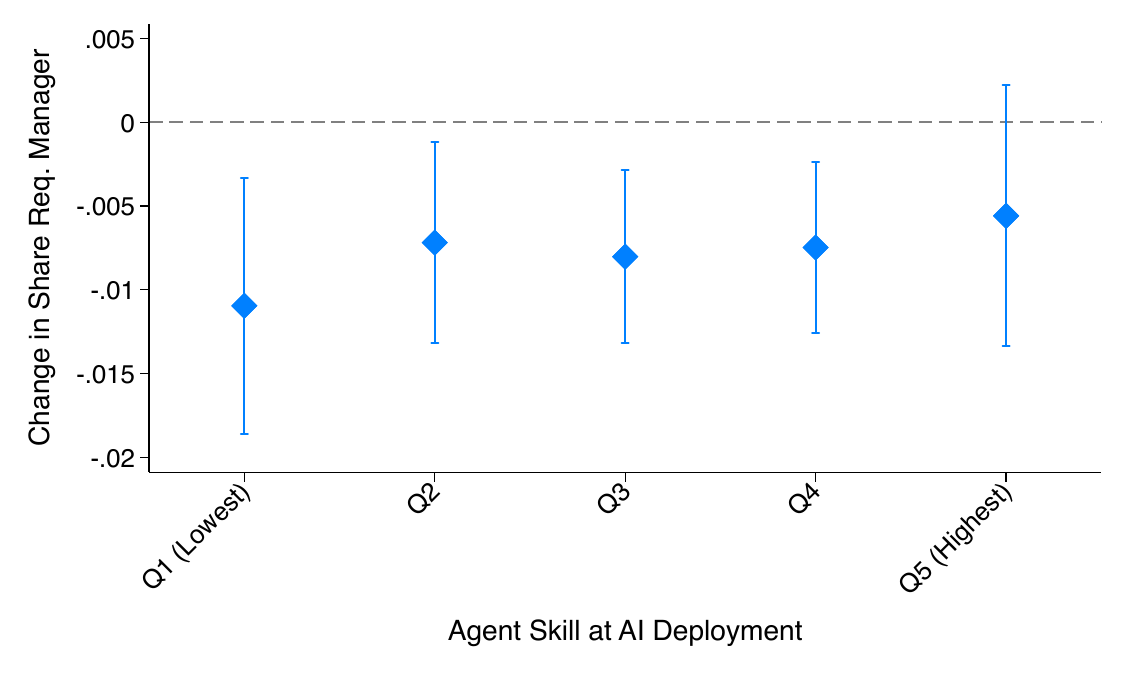} & \includegraphics[scale=0.4]{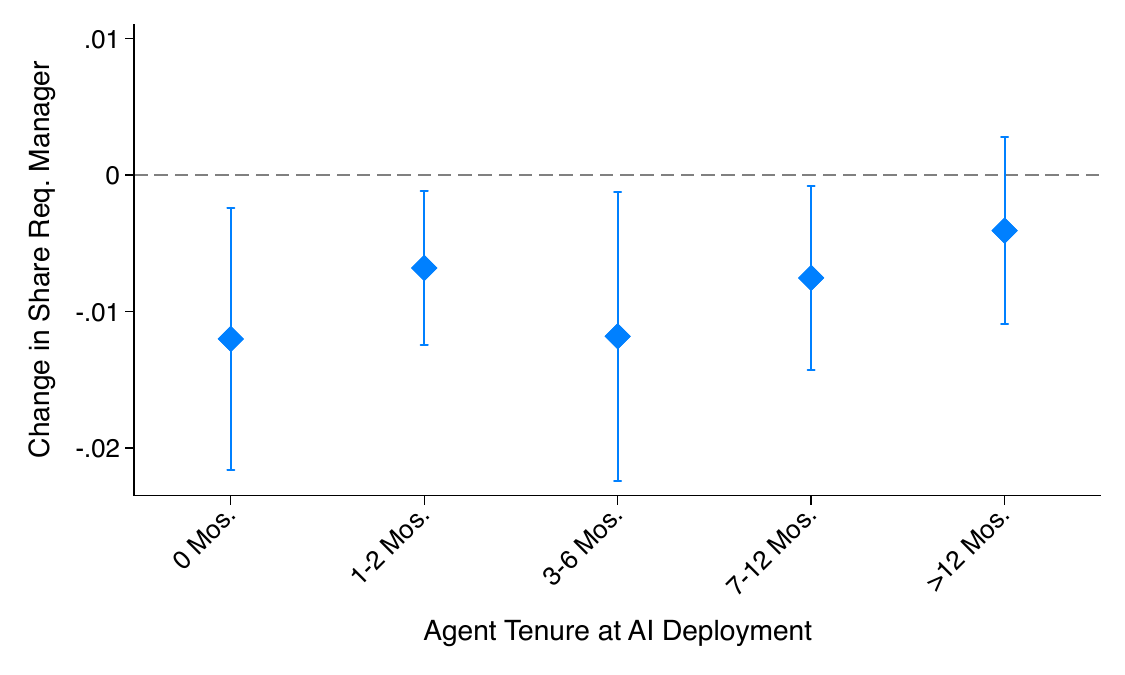} \\
			\end{tabular}
		}
		
		\label{afig:expwork}
	\end{center}
\end{figure}

\begin{footnotesize} 
	\begin{singlespace}
		\noindent \textsc{Notes}: Each panel of this figure illustrates the impact of AI model deployment on aspects of the experience of work. Panel A shows average customer sentiment, while Panel B shows average agent sentiment. Sentiment is measured using SiEBERT, a fine-tuned checkpoint of a RoBERTA, an English language transformer model. Panel C plots the impacts of AI on customer sentiment by agent ex-ante productivity and Panel D plots the effects by agent tenure when the AI is deployed. Panels E and F show the effects of AI on customer requests for manager assistance, by pre-AI agent skill and in by pre-AI agent tenure. Observations for the regression are at the agent-month level, robust standard errors are clustered at the agent level, and regressions include controls for agent, chat year-month and months-of-tenure. 
	\end{singlespace}
\end{footnotesize}

\clearpage
\begin{table}[ht!]
\begin{center}
                \caption{\textsc{Table \ref{atab:attrition}: Attrition}}
                  \vspace{20pt}        
\scalebox{1}{\makebox[\linewidth]{\begin{tabular}{lccc} \hline
 & (1) & (2) & (3) \\
VARIABLES & Leaves this Month & Leaves this Month & Leaves this Month \\ \hline
 &  &  &  \\
Post AI X Ever Treated & -0.0868*** &  &  \\
 & (0.0130) &  &  \\
 0 Mos. &  & -0.0952*** &  \\
 &  & (0.0162) &  \\
1-2 Mos. &  & -0.121*** &  \\
 &  & (0.0167) &  \\
3-6 Mos. &  & -0.0850*** &  \\
 &  & (0.0190) &  \\
7-12 Mos. &  & -0.0803*** &  \\
 &  & (0.0165) &  \\
>12 Mos. &  & -0.0306 &  \\
 &  & (0.0234) &  \\
Q1 (Low Skill) &  &  & -0.0655*** \\
 &  &  & (0.0148) \\
Q2 &  &  & -0.0864*** \\
 &  &  & (0.0134) \\
Q3 &  &  & -0.0741*** \\
 &  &  & (0.0124) \\
Q4 &  &  & -0.0936*** \\
 &  &  & (0.0146) \\
Q5 (High Skill) &  &  & -0.0531*** \\
 &  &  & (0.0145) \\
 &  &  &  \\
Observations & 17,902 & 17,902 & 17,902 \\
R-squared & 0.206 & 0.206 & 0.206 \\
Year Month FE & Yes & Yes & Yes \\
Location FE & Yes & Yes & Yes \\
Agent Tenure FE & Yes & Yes & Yes \\
Agent Company FE & Yes & Yes & Yes \\
 DV Mean & 0.288 & 0.288 & 0.288 \\ \hline
\multicolumn{4}{c}{ Robust standard errors in parentheses} \\
\multicolumn{4}{c}{ *** p$<$0.01, ** p$<$0.05, * p$<$0.10} \\
\end{tabular}
}}
  \label{atab:attrition}
\end{center}
\end{table}

\begin{singlespace}
\footnotesize
\noindent \textsc{Notes}:  This table presents the results of difference-in-difference regressions estimating the impact of AI model deployment on our main measure of productivity, resolutions per hour, the number of technical support problems resolved by an agent per hour (resolutions/hour).  Post AI X Ever Treated captures the impact of AI model deployment on resolutions per hour. Column 1 includes agent geographic location and year-by-month fixed effects.  Columns 2 and 3 include agent-level fixed effects, and Column 3, our preferred specification described by Equation \ref{eq:dd_main}, also includes fixed effects that control for months of agent tenure.  Observations for this  regression are at the agent-month level and all standard errors are clustered at the agent level. Section \ref{sec:rollout} describes the AI rollout procedure.
\end{singlespace}
\normalsize

\end{appendix}

\end{document}